\PassOptionsToPackage{usenames,dvipsnames}{xcolor}

\documentclass[twocolumn,resetfootnote]{aastex701}

\usepackage{subcaption}
\usepackage{amsmath} 
\usepackage{longtable}

\journalinfo{Published in ApJS}

\begin{document}

\title{J-HERTz: J-PLUS Heritage Exploration of Radio Targets at z $<$ 5}
 
\author[orcid=0000-0003-1294-2583,sname={Fern\'andez}]{D. Fern\'andez Gil}
\affiliation{Centro de Estudios de F\'isica del Cosmos de Arag\'on (CEFCA), 1 Plaza San Juan, 44001 Teruel, Spain}
\email[show]{dfernandez@cefca.es, davfergil98@gmail.com}  

\author[orcid=0000-0001-9490-899X,sname={Fern\'andez}]{J.\,A. Fern\'andez-Ontiveros}
\affiliation{Centro de Estudios de F\'isica del Cosmos de Arag\'on (CEFCA), 1 Plaza San Juan, 44001 Teruel, Spain}
\affiliation{Unidad Asociada CEFCA--IAA, CEFCA, Unidad Asociada al CSIC por el IAA, Plaza San Juan 1, 44001 Teruel, Spain}
\email{} 

\author[orcid=0000-0002-5743-3160,sname={L\'opez-Sanjuan}]{C. L\'opez-Sanjuan}
\affiliation{Centro de Estudios de F\'isica del Cosmos de Arag\'on (CEFCA), 1 Plaza San Juan, 44001 Teruel, Spain}
\affiliation{Unidad Asociada CEFCA--IAA, CEFCA, Unidad Asociada al CSIC por el IAA, Plaza San Juan 1, 44001 Teruel, Spain}
\email{}

\author[orcid=0009-0006-1307-7703,sname={Arizo-Borillo}]{F. Arizo-Borillo}
\affiliation{Centro de Estudios de F\'isica del Cosmos de Arag\'on (CEFCA), 1 Plaza San Juan, 44001 Teruel, Spain}
\email{}

\author[orcid=0000-0003-4922-5131,sname={del Pino}]{A. del Pino}
\affiliation{Instituto de Astrof\'isica de Andaluc\'ia (CSIC), Apartado 3004, 18080 Granada, Spain}
\email{}

\author[orcid=0000-0002-4237-5500,sname={Hern\'an-Caballero}]{A. Hern\'an-Caballero}
\affiliation{Centro de Estudios de F\'isica del Cosmos de Arag\'on (CEFCA), 1 Plaza San Juan, 44001 Teruel, Spain}
\affiliation{Unidad Asociada CEFCA--IAA, CEFCA, Unidad Asociada al CSIC por el IAA, Plaza San Juan 1, 44001 Teruel, Spain}
\email{}

\author[orcid=0000-0002-6696-7834,sname={Lumbreras-Calle}]{A. Lumbreras-Calle}
\affiliation{Centro de Estudios de F\'isica del Cosmos de Arag\'on (CEFCA), 1 Plaza San Juan, 44001 Teruel, Spain}
\email{}

\author[orcid=0000-0002-5864-7195,sname={Rahna}]{P. T. Rahna}
\affiliation{Centro de Estudios de F\'isica del Cosmos de Arag\'on (CEFCA), 1 Plaza San Juan, 44001 Teruel, Spain}
\email{}

\author[orcid=0000-0001-8823-4845,sname={Sobral}]{David Sobral}
\affiliation{Departamento de F\'isica, Faculdade de Ci\^encias, Universidade de Lisboa, Edif\'icio C8, Campo Grande, PT1749-016 Lisbon, Portugal}
\affiliation{BNP Paribas Corporate \& Institutional Banking, Torre Ocidente Rua Galileu Galilei, 1500-392 Lisbon, Portugal}
\email{}

\author[orcid=0000-0003-3135-2191,sname={V\'azquez Rami\'o}]{H. V\'azquez Rami\'o}
\affiliation{Centro de Estudios de F\'isica del Cosmos de Arag\'on (CEFCA), 1 Plaza San Juan, 44001 Teruel, Spain}
\affiliation{Unidad Asociada CEFCA--IAA, CEFCA, Unidad Asociada al CSIC por el IAA, Plaza San Juan 1, 44001 Teruel, Spain}
\email{}

\author[orcid=0000-0002-2573-2342,sname={Cenarro}]{A. J. Cenarro}
\affiliation{Centro de Estudios de F\'isica del Cosmos de Arag\'on (CEFCA), 1 Plaza San Juan, 44001 Teruel, Spain}
\affiliation{Unidad Asociada CEFCA--IAA, CEFCA, Unidad Asociada al CSIC por el IAA, Plaza San Juan 1, 44001 Teruel, Spain}
\email{}

\author[orcid=0000-0002-9026-3933,sname={Mar\'in-Franch}]{A. Mar\'in-Franch}
\affiliation{Centro de Estudios de F\'isica del Cosmos de Arag\'on (CEFCA), 1 Plaza San Juan, 44001 Teruel, Spain}
\affiliation{Unidad Asociada CEFCA--IAA, CEFCA, Unidad Asociada al CSIC por el IAA, Plaza San Juan 1, 44001 Teruel, Spain}
\email{}

\author[orcid=0000-0003-2953-3970
,sname={Angulo}]{R. E. Angulo}
\affiliation{Donostia International Physics Centre (DIPC), Paseo Manuel de Lardizabal 4, 20018 Donostia-San Sebasti\'an, Spain}
\affiliation{IKERBASQUE, Basque Foundation for Science, 48013, Bilbao, Spain}
\email{}

\author[orcid=0000-0003-3656-5524,sname={Ederoclite}]{A. Ederoclite}
\affiliation{Centro de Estudios de F\'isica del Cosmos de Arag\'on (CEFCA), 1 Plaza San Juan, 44001 Teruel, Spain}
\affiliation{Unidad Asociada CEFCA--IAA, CEFCA, Unidad Asociada al CSIC por el IAA, Plaza San Juan 1, 44001 Teruel, Spain}
\email{}

\author[sname={Crist\'obal-Hornillos}]{D. Crist\'obal-Hornillos}
\affiliation{Centro de Estudios de F\'isica del Cosmos de Arag\'on (CEFCA), 1 Plaza San Juan, 44001 Teruel, Spain}
\email{}

\author[orcid=0000-0003-1477-3453,sname={Dupke}]{R. A. Dupke}
\affiliation{Observat\'orio Nacional - MCTI (ON), Rua Gal. Jos\'e Cristino 77, S\~ao Crist\'ov\~ao, 20921-400 Rio de Janeiro, Brazil}
\affiliation{University of Michigan, Department of Astronomy, 1085 South University Ave., Ann Arbor, MI 48109, USA}
\affiliation{University of Alabama, Department of Physics and Astronomy, Gallalee Hall, Tuscaloosa, AL 35401, USA}
\email{}

\author[orcid=0000-0001-5471-9166,sname={Hern\'andez-Monteagudo}]{C. Hern\'andez-Monteagudo}
\affiliation{Instituto de Astrof\'isica de Canarias, La Laguna, 38205, Tenerife, Spain}
\affiliation{Departamento de Astrof\'isica, Universidad de La Laguna, 38206, Tenerife, Spain}
\email{}

\author[orcid=0009-0002-6052-8723,sname={Moles}]{M. Moles}
\affiliation{Centro de Estudios de F\'isica del Cosmos de Arag\'on (CEFCA), 1 Plaza San Juan, 44001 Teruel, Spain}
\email{}

\author[sname={Sodr\'e Jr.}]{L. Sodr\'e Jr.}
\affiliation{Instituto de Astronomia, Geof\'isica e Ci\^encias Atmosf\'ericas, Universidade de S\~ao Paulo, 05508-090 S\~ao Paulo, Brazil}
\email{}

\author[sname={Varela}]{J. Varela}
\affiliation{Centro de Estudios de F\'isica del Cosmos de Arag\'on (CEFCA), 1 Plaza San Juan, 44001 Teruel, Spain}
\email{}

\begin{abstract}
We introduce \textsc{J-HERTz} (J-PLUS Heritage Exploration of Radio Targets at $z < 5$), a new multi-wavelength catalog that combines optical narrow-band photometry from J-PLUS, infrared observations from \textit{WISE}, and deep low-frequency radio data from LoTSS for nearly half a million sources across 2100\,deg$^2$ of the northern sky. Key innovations of \textsc{J-HERTz} include Bayesian neural network classifications for 390,000 galaxies, 31,000 quasars, and 20,000 stars, along with significantly improved photometric redshifts for 235,000 galaxies compared to previous J-PLUS DR3 and LoTSS DR2 estimates. We identify 831 candidate Galactic radio stars, which, if confirmed, would constitute a significant addition to the number of radio-emitting stars identified to date. Among radio-loud galaxies with spectroscopic observations, $\gtrsim$20\% lack Seyfert or LINER signatures, indicating a substantial population of optically quiescent radio galaxies, in agreement with previous works. Spectral energy distribution fitting of their host galaxies using J-PLUS photospectra reveals systematically low specific star formation rates, consistent with quenched stellar populations. \textsc{J-HERTz} thus provides a powerful dataset to exploit radio-optical synergies, enabling studies that span from the origin of stellar radio emission to the AGN life cycle and the role of jet activity in shaping host galaxy evolution.
\end{abstract}


\section{Introduction}\label{sect:Introduction}
Radio astronomy has made substantial advances since the first radio emission coming from the center of our galaxy was discovered \citep{Jansky1933}. In 90 years, radio continuum surveys have improved from angular resolutions of several degrees to sub-arcsecond scales thanks to the development of radio interferometry, and from frequencies of a few hundred MHz to a full coverage of the radio window up to millimeter wavelengths. This advancement came hand in hand with the technical development that enabled the construction of larger and more efficient antennas, as well as improved backend systems with more sensitive receivers and faster correlators. Some of the most important radio continuum surveys in the last decades include the NVSS\footnote{NRAO (National Radio Astronomy Observatory) VLA Sky Survey.} \citep{Condon1998} at 1.4 GHz, the Cambridge 3C Radio Survey \citep{Bennett1962} at 178 MHz, the Very Large Array (VLA) FIRST\footnote{Faint Images of the Radio Sky at Twenty-Centimeters.} Survey \citep{Becker1994} at 1.5 GHz, the Green Bank GB6 catalog \citep{Gregory1996} at 4.85 GHz, the RACS\footnote{Rapid ASKAP Continuum Survey} \citep{Hale2021} at 888 MHz and the SUMSS\footnote{Sydney University/Molonglo Sky Survey.} \citep{Mauch2013} at 843 MHz.

Because radio emission is not scattered or absorbed by dust grains and Compton thick structures are transparent to radio frequencies, this range yields complementary information to optical observations, tracing free-free and synchrotron processes, and the Rayleigh-Jeans tail of thermal emission, which are not accessible at other wavelengths \citep{Condon2016,Padovani2016}. Thus, radio surveys are indispensable for the study of a wide array of astronomical sources, including radio galaxies, quasars, ultra luminous starburst galaxies, Active Galactic Nuclei (AGN), planetary nebulae, supernova remnants, pulsars, and variable stars.

\begin{figure*}[t]
  \includegraphics[width=\textwidth]{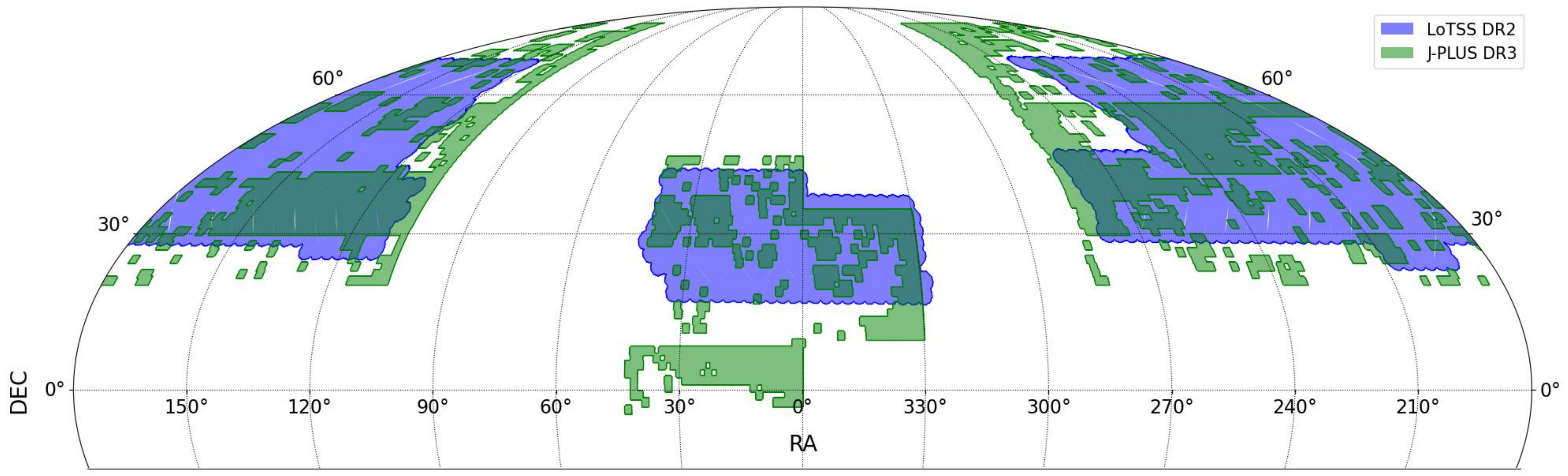}
  \caption{Sky coverage of the 2100 deg$^2$ common area of J-PLUS DR3 (green shaded area) and LoTSS DR2 (blue shaded area).}\label{fig:sky_coverage}
\end{figure*}

Of particular interest for this work are low-frequency radio surveys ($\nu < 500 \, \text{MHz}$), which are sensitive to older synchrotron emission and therefore probe aged and extended radio structures where self-absorption effects are not significant, such as halos, relics, and jet lobes \citep{Melrose1985,Ferrari2008}. Some of the most notable low-frequency surveys are the VLA Low-frequency Sky Survey (VLSS) at 74 MHz \citep{Cohen2007}, the 8C radio survey at 38 MHz \citep{Rees1990}, the GaLactic and Extragalactic All-sky MWA survey (GLEAM) at 72-231 MHz \citep{Wayth2015,Hurley2017} and the Westerbork Northern Sky Survey (WENSS) at 330 MHz \citep{deBruyn2000}. These cover a wide sky area with 0.5--4.5 arcmin resolution and flux density limits of 18--1000 mJy at $5\sigma_\mathrm{rms}$. A breakthrough in sensitivity and sky coverage was achieved with the LOw Frequency ARray \citep[LOFAR; ][]{Haarlem2013} Two-metre Sky Survey \citep[LoTSS; ][]{Shimwell2022}. LoTSS DR2 covers the 120--168 MHz range over $\sim$ 5\,700 deg$^2$ of the northern sky at 6 arcsec resolution, reaching a flux limit of $\sim$0.4 mJy for compact sources and a sensitivity of $\sim$0.25 mJy/beam at $5\sigma_\mathrm{rms}$, providing an eightfold increase in the surface density of radio sources compared to previous works. This is complemented by the ongoing 42–66 MHz LOFAR Low Band Antenna Sky Survey \citep[LoLSS; ][]{deGasperin2021} and the LOFAR Decametre Sky Survey (LoDSS) at 14–30 MHz. Some remarkable results are the discovery of a $\sim$ 7 Mpc giant radio galaxy \citep{Oei24} or the search for stellar radio emission as a possible indicator of exoplanet interaction with M dwarfs \citep{Callingham2021}.

The real potential of radio surveys is unleashed when they are combined with other frequencies that offer complementary information for a comprehensive understanding of astronomical sources. For instance, optical, infrared (IR), and ultraviolet observations enable the measurement of photometric and spectroscopic redshift, stellar masses, and star formation rates (SFRs) of galaxies. They also allow estimates the bolometric luminosity and radio-loudness of quasi-stellar objects (QSOs), as well as the mass and variability of stars. In particular, optical spectroscopy of radio sources provides more precise determination of the origin of the radio emission and the nature of the host. Examples of such multi-wavelength studies include the cross-match between SDSS\footnote{Sloan Digital Sky Survey \citep{York2000,Stoughton2002,Strauss2002}.}, NVSS, and FIRST by \citet{Best2005} and the identification of NVSS sources within the 6dF Galaxy Survey \citep{Mauch2007}, which together show that radio-loud AGN preferentially reside in massive galaxies. In this regard, the co-evolution of SMBHs and galaxies has been investigated by \citet{Heckman2014}, who combined radio, IR, optical, and X-ray data, and by \citet{Ching2017}, who analyzed the LARGESS\footnote{Large Area Radio Galaxy Evolution Spectroscopic Survey.} sample of FIRST sources with SDSS spectroscopy. These works revealed that the majority of radio AGN are Low-Excitation Radio Galaxies (LERGs), associated with jet activity from massive black holes in elliptical galaxies. By contrast, High-Excitation Radio Galaxies (HERGs) are associated with less massive, radiatively efficient black holes typically found in bluer, star-forming galaxies. In a follow-up study with the GAMA\footnote{Galaxy And Mass Assembly.} survey, \citet{Ching2017b} showed that luminous LERGs are preferentially associated with denser environments and galaxy groups. More recently, \citet{Arnaudova2024} combined LoTSS with spectroscopically confirmed QSOs from SDSS DR14 \citep{Abolfathi2018} to show that radio-loud QSOs tend to exhibit higher SFRs than their radio-quiet counterparts.

On the other hand, photometric surveys allows efficient access to extensive, large-area samples without the need for targeted observations. This enables photometric redshift (photo-$z$) estimates, albeit with lower precision and potential contamination by outliers. To amend this, the use of narrow-band optical observations provides, among other, improved photo-z estimations by exploiting spectral features in galaxies, essential for clustering studies \citep{Molino2019}, and precise stellar parameter determinations \citep{Yuan2023}. Additionally, narrow-band filters offer detailed information on the spectral energy distribution (SED) and the emission lines for a better characterization of the star formation main sequence and SFR density in galaxies \citep{VilellaRojo2021}, the discovery of high-z Lyman $\alpha$ QSOs \citep{Spinoso2020}, and the identification of large samples of extreme emission-line galaxies \citep{Lumbreras2022}. The catalog presented in this study exploits the advantages of narrow-band photometry from the Javalambre Photometric Local Universe Survey \citep[J-PLUS DR3;][]{Cenarro19,LopezSanJuan24} with low-frequency radio counterparts from the LOFAR Two-Metre Sky Survey (LoTSS DR2), and IR photometry from WISE. J-PLUS filters are specifically designed to provide accurate stellar population parameters and low-redshift photo-$z$ estimates. While other broad-band surveys such as DESI Legacy Imaging Surveys \citep{Dey2019} offer deeper and wider coverage, J-PLUS adds the ability to resolve spectral features and provides improvements in the photo-$z$ and SFR estimations of high signal-to-noise sources, particularly when prominent spectral features such as bright emission lines or the $4000$\,\AA \ break are present.

The resulting product is the J-HERTz catalog (J-PLUS Heritage Exploration of Radio Targets at z $<$ 5), a public, multi-wavelength resource covering $\sim$2100 deg$^2$ to characterize nearly half a million sources in the northern sky. J-HERTz provides a uniform, cross-matched data set with probabilistic Bayesian-neural-network classifications for galaxies, QSOs, and stars; improved photo-$z$ estimates that leverage the J-PLUS narrow bands; and a radio-loudness indicator based on LOFAR/WISE flux ratios, together with SED-based stellar masses and SFR for the galaxy population. These elements enable population-wide studies spanning galaxy evolution, AGN duty cycles and feedback, and the demographics of radio sources across astrophysical classes --\,including a large sample of candidate radio-emitting stars\,-- within a single, coherent framework. These capabilities will significantly improve with the upcoming J-PAS survey \citep[][Vázquez-Ramió, in prep.]{Bonoli2021}, including 56 narrow bands and deeper observations with respect to J-PLUS by $\sim1.5$ mag., which will be incorporated in future releases of J-HERTz. Throughout this paper, we adopt a cosmology with $H_0 = 70$~km\,s$^{-1}$\,Mpc$^{-1}$, $\Omega_m = 0.3$, and $\Omega_\Lambda = 0.7$

\section{A radio-optical catalog for the Northern Hemisphere}\label{sect:A radio-optical catalog for the Northern Hemisphere}

The combination of J-PLUS and LoTSS entails a perfect compromise between sky coverage, resolution, and depth. Their common area reaches $\sim$2100 deg$^2$ of the northern sky (Fig.~\ref{fig:sky_coverage}) and their angular resolution is of the same order (1 arcsec vs. 6 arcsec), while reaching a similar depth down to flux densities of $\sim$ 0.1 mJy. Both surveys contain cross-identifications in WISE, allowing for an extension to IR wavelengths. Furthermore, J-PLUS contains both broad and narrow bands, providing more accurate properties of the optical counterparts.

\subsection{Catalog description}\label{sect:catalog description}
The J-HERTz catalog contains $489\,897$ objects found in J-PLUS DR3 and LoTSS DR2. The available information for each object includes IDs, optical fluxes and magnitudes, radio fluxes and sizes, WISE band magnitudes, photometric and spectroscopic redshifts and source classification. The whole catalog is available in the online version of this publication and in Zenodo at doi:\url{10.5281/zenodo.17651891}. Additionally, a reduced version with the object IDs and some of the value-added parameters is accessible through the J-PLUS DR3 Data Access Services\footnote{\url{https://archive.cefca.es/catalogues/jplus-dr3}} and can be queried through TAP or ADQL services. All the included parameters are listed in the Appendix. We show the distribution of some of the main properties in Section \ref{sec:properties} and discuss them in Section \ref{sec:Discussion}.

\begin{table}[h!]
\caption{J-PLUS photometric system.}
\centering
\begin{tabular}{cccc}
\hline
\hline
Filter & Central & FWHM \\
                & Wavelength [\AA]            & [\AA]   \\
\hline
u     & 3485  & 508  \\
J0378 & 3785  & 168   \\
J0395 & 3950  & 100    \\
J0410 & 4100  & 200     \\
J0430 & 4300  & 200  \\
g     & 4803  & 1409  \\
J0515 & 5150  & 200    \\
r     & 6254  & 1388    \\
J0660 & 6600  & 138   \\
i     & 7668  & 1535   \\
J0861 & 8610  & 400     \\
z     & 9114  & 1409     \\ \hline
\end{tabular}
\label{tab:filters}
\end{table}

\subsection{J-PLUS DR3}
The J-PLUS DR3 catalog covers about $\sim 3200\,\rm{deg^{2}}$ of the northern sky, containing $\sim$47 million objects between galaxies, QSOs and stars. The observations are done at the Observatorio Astrofísico de Javalambre \citep{Cenarro14} with the Javalambre Auxilary Survey Telescope of 83\,cm diameter (JAST80) and T80Cam, a 9.2k $\times$ 9.2k pixels CCD that provides a $2\deg^2$ field of view and a pixel scale of $0.55$ arcsec pix$^{-1}$ \citep{MarinFranch15}. 
It has 12 filters described in Table \ref{tab:filters}, a combination of 5 broad and 7 medium or narrow bands, reaches a depth of AB $\sim$ 21.8 mag at the r band at 5$\sigma$ with a 3 arcsec aperture and a $1.2$ arcsec full width at half maximum (FWHM) point spread function. J-PLUS provides flux densities and magnitudes in its 12 available bands, allowing for an accurate characterization of stars in our galaxy as well as 2D photo-spectral information for resolved galaxies in the local Universe \citep{Rahna2025} and accurate photo-z estimates for moderately bright extra galactic sources. For our catalog and this work we use their \texttt{AUTO} photometry otherwise stated, which takes an automatically fitted ellipse from \texttt{SExtractor}. J-PLUS archives also provide cross-matches with other catalogs like Sloan Digital Sky Survey (SDSS) DR12 \citep{Alam2015}, catWISE2020 \citep{Marocco2020} and {\it Gaia} DR3 \citep{Gaia2023}.

\subsection{LoTSS DR2}
LoTSS DR2 \citep{Shimwell2022} is a low-frequency radio catalog that provides fluxes at 144 MHz with a 6 arcsec resolution, which is particularly useful to detect wide area radio emission with lower surface brightness. It is based on LOFAR observations and covers 27\% of the Northern Hemisphere sky ($\sim$ 5700 deg$^2$), containing around 4.4 million radio objects. It includes optical and/or IR identifications in unWISE \citep{Lang2014} and DESI Legacy Imaging Surveys DR9 for 85\% of the objects using a combination of likelihood-ratio methods, algorithmic optical identification and extensive visual inspection \citep{Hardcastle2023}. The catalog provides fitted magnitudes from WISE as well as other optical and UV catalogs, estimated radio emission sizes, photometric redshifts, and spectroscopic redshifts from SDSS DR16 \citep{Ahumada2020}, the first
data release of the DESI spectroscopic survey \citep{DESI2025} and the Hobby-Eberly Telescope Dark Energy Experiment (HETDEX) DR1 \citep{Davis2023} when available. Using the most reliable redshifts, they also compute the linear sizes, radio luminosities, stellar masses and rest-frame magnitudes.

\begin{figure*}
\centering

\begin{subfigure}{0.8\textwidth}
  \includegraphics[width=\textwidth]{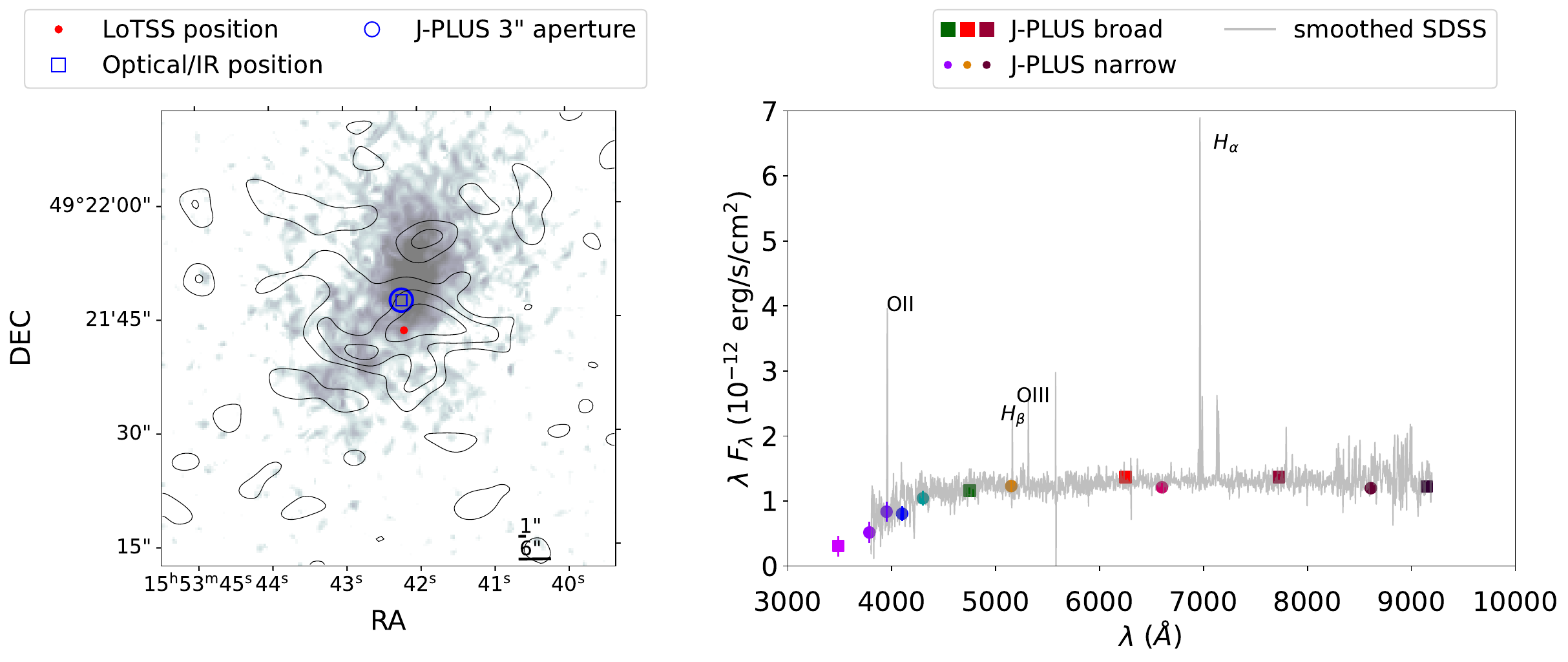}
  \caption{Star-forming galaxy with strong H$\alpha$ emission. \texttt{TILE\_ID} = 93594, \texttt{NUMBER} = 2529 and z = 0.06.}
  \label{fig:radio_ratio_example_1}
\end{subfigure}

\vspace{0.2cm}

\begin{subfigure}{0.8\textwidth}
  \includegraphics[width=\textwidth]{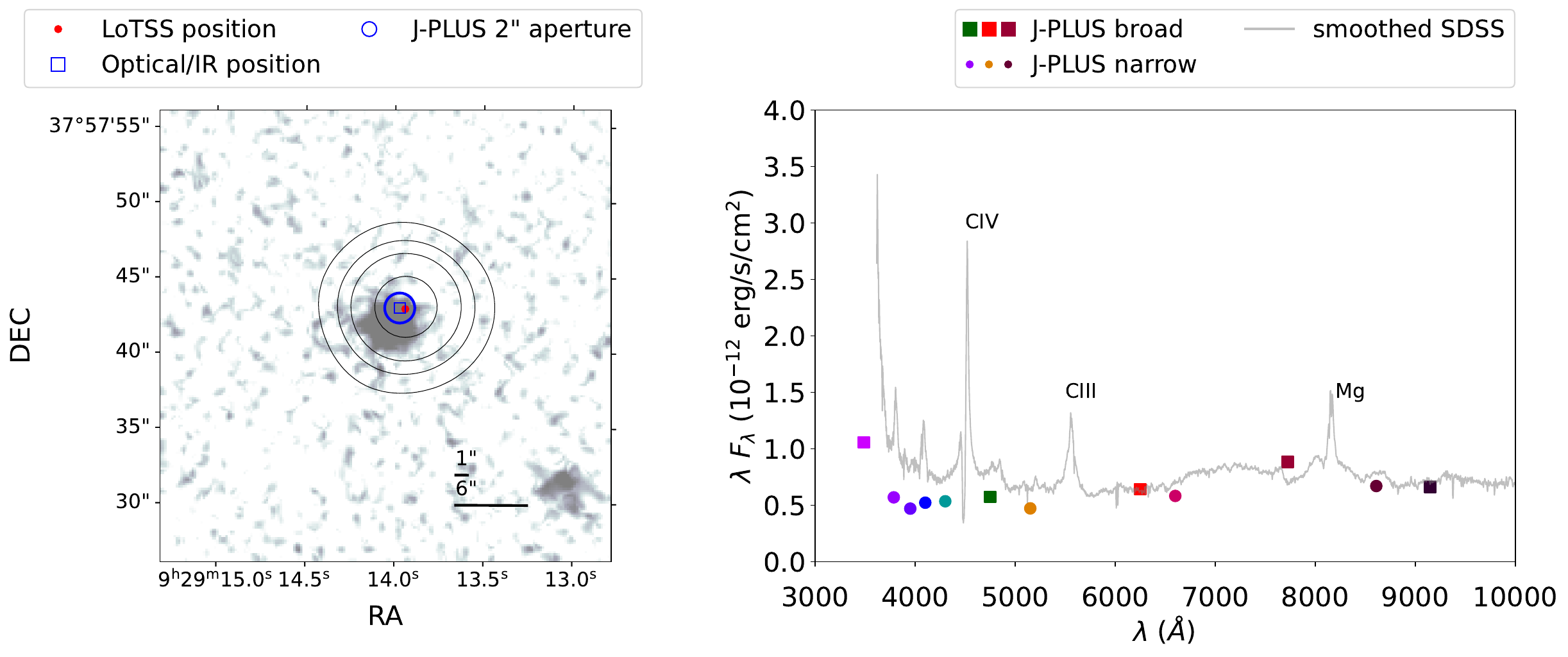}
  \caption{QSO with \ion{C}{3}] and \ion{C}{4} emission lines and a blue steep AGN continuum. \texttt{TILE\_ID} = 89037, \texttt{NUMBER} = 12178 and z = 1.92.}
  \label{fig:radio_ratio_example_2}
\end{subfigure}

\vspace{0.2cm}

\begin{subfigure}{0.8\textwidth}
  \includegraphics[width=\textwidth]{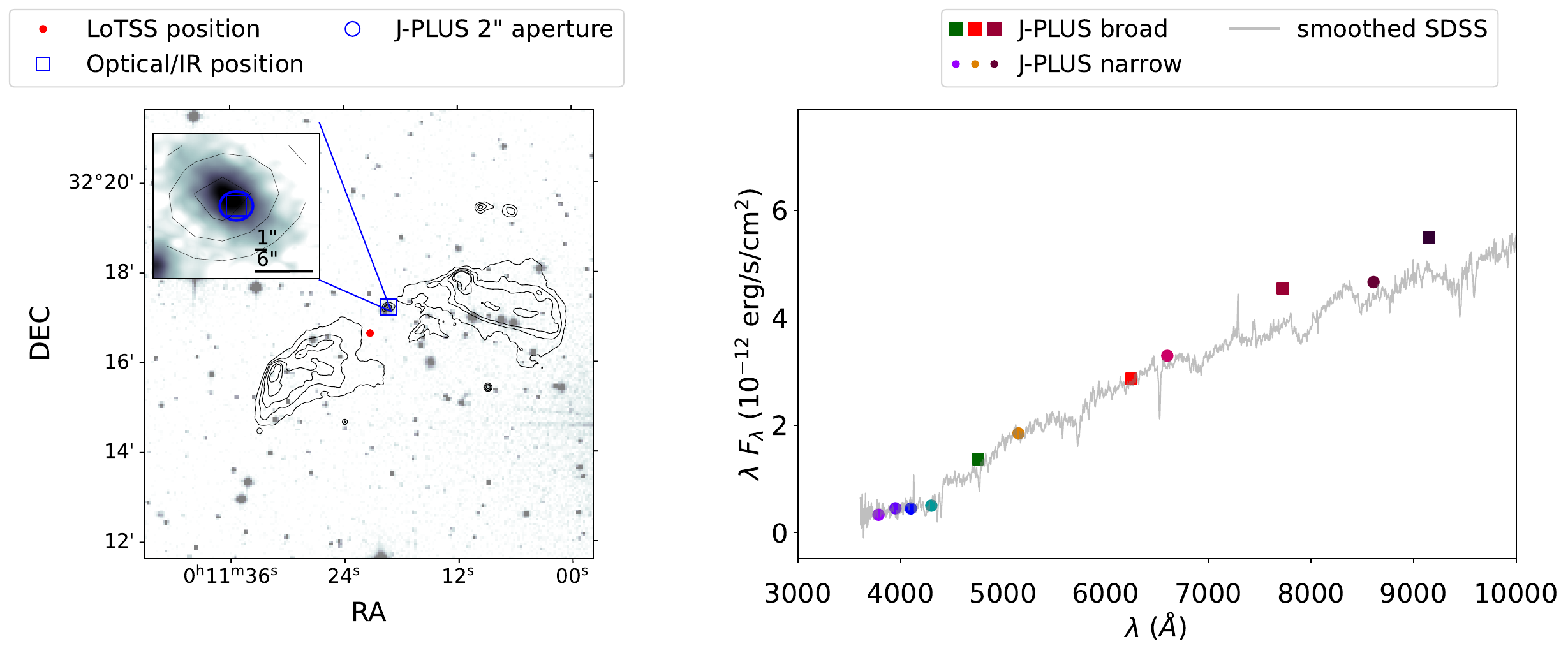}
  \caption{Radio galaxy with 10 arcmin. extended jets. The continuum drop at blue wavelengths indicates old stellar populations. \texttt{TILE\_ID} = 102551, \texttt{NUMBER} = 8756 and z = 0.11.}
  \label{fig:radio_ratio_example_3}
\end{subfigure}

\caption{Three objects included in J-HERTz, with their unique J-PLUS
identification and their SDSS DR18 \citep{Almeida2023} spectroscopic redshifts. Left panels: optical image from J-PLUS in gray scale, with the J-PLUS and LoTSS positions (blue square and red dot respectively), and the radio contours (black lines) from LoTSS drawn at 1, 2, 3, 5 and 7$\sigma$. The blue circle is the J-PLUS aperture selected to match the size of the SDSS fiber in each case (either 2 arcsec for BOSS or 3 arcsec for SDSS). The 1 arcsec and 6 arcsec lines show the resolutions of J-PLUS and LoTSS respectively, when visible. Right panels: Fluxes and error bars measured in the 12 J-PLUS filters with the specified aperture as the colored symbols (squares for broad and circles for narrow bands, most error bars are smaller than the symbol size) and their SDSS DR18 spectra in gray, smoothed and re-scaled to the J-PLUS flux in the $r$ band.}
\label{fig:optical_radio_collage}
\end{figure*}

\subsection{Radio-optical cross-match}
We use the optical/IR identifications of LoTSS provided by \citet{Hardcastle2023}. In their work, they first cross-match the DESI Legacy Imaging Surveys to the unWISE catalog using a 2 arcsec radius to determine an optical/IR position, which is then assigned to a radio source in LoTSS DR2. We do not cross-match these identifications with the J-PLUS positions directly, but rather with the catWISE and SDSS identifications that J-PLUS provides. As most of the common objects in J-PLUS and LoTSS have identifications in WISE, this ensures that they are cross-matched using the same survey as reference and maximizes genuine matches. We first cross-match a subsample of the positions with a high tolerance (5 arcsec) and analyze the separations. The median is 0.23 arcsec, indicating a good agreement in most positions, while the 90th and 98th percentiles are 1.26 arcsec and 3.24 arcsec, respectively. We decide to use a 1 arcsec tolerance for the cross-match, which is slightly conservative to avoid spurious associations but still keeps a high completeness of genuine matches. We obtain 489,897 unique objects in total, which correspond to $\sim$80\% of the LoTSS DR2 \citet{Hardcastle2023} identifications with a S/N $>$ 5 in the common area. The angular distances between the SDSS/catWISE identifications and the original J-PLUS positions are provided in the \texttt{AngDist} parameter, and $\sim$92\% of them are smaller than 2 arcsec. Fig.~\ref{fig:optical_radio_collage} shows three examples of objects contained in the catalog, including J-PLUS 3 arcsec photometry, LoTSS radio emission and SDSS optical spectra. Panel (a) shows a star-forming galaxy, as can be seen from the strong H$H\alpha$ emission line. Panel (b) presents a QSO with \ion{C}{3}] and \ion{C}{4} emission lines and a very strong blue continuum coming from the AGN. Finally, panel (c) is a radio galaxy whose jets extend 10 arcmin and has a drop in emission at blue wavelengths, indicating the presence of an old stellar population. There is an overall good agreement between the optical SED of J-PLUS and the optical spectra of SDSS of the examples, with small deviations that may be caused by different calibrations, PSF models and sky subtractions, or due to some lost flux outside of the SDSS fiber.

\subsection{Object classification}
We use the recently developed Bayesian Artificial Neural Networks for the Javalambre
Observatory Survey (BANNJOS) machine learning-based algorithm \citep{delPino2024} to classify sources into galaxies, QSOs, and stars. This algorithm uses Bayesian Artificial Neural Networks to provide a PDF (Probability Distribution Function) and correlations between the three classes. It uses the full photometric information provided by J-PLUS in different apertures, as well as the morphology and other external properties such as the proper motions from {\it Gaia} and the IR bands from WISE. Note that this classification refers to the optical/IR morphological and spectral properties of the counterparts and is not intended to classify the radio properties of the active nucleus. This method improves upon previous works because it includes uncertainties and correlations, allowing for a more precise filtering of the results according to the user's preferences for accuracy and completeness. 

\begin{figure}[!ht]
  \centering
  \includegraphics[width=\columnwidth]{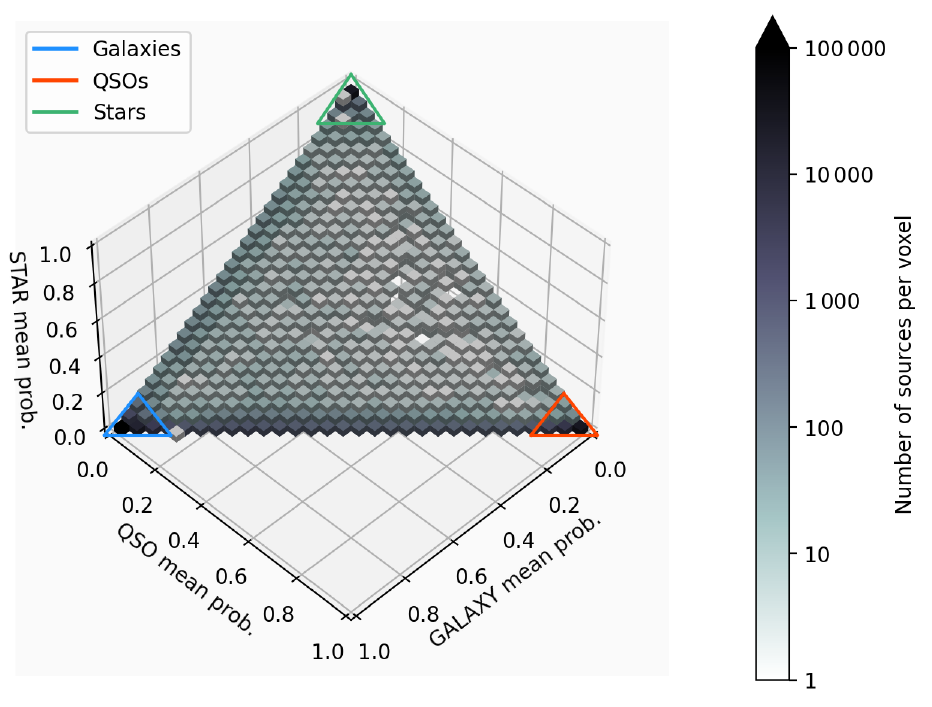}
  \caption{Mean probabilities of being a galaxy, QSO or star for all objects in J-HERTz according to the BANNJOS \citep{delPino2024} classification. Each voxel is colored according to the number of points inside. The voxels inside the blue, orange and green contours contain objects classified as galaxies, QSOs and stars with a mean probability above 0.9, which correspond to over 90\% of the catalog and are the ones we consider for the rest of this work.}\label{fig:prob_class_dist}
\end{figure}

\begin{figure*}[!t]
  \includegraphics[width=\textwidth]{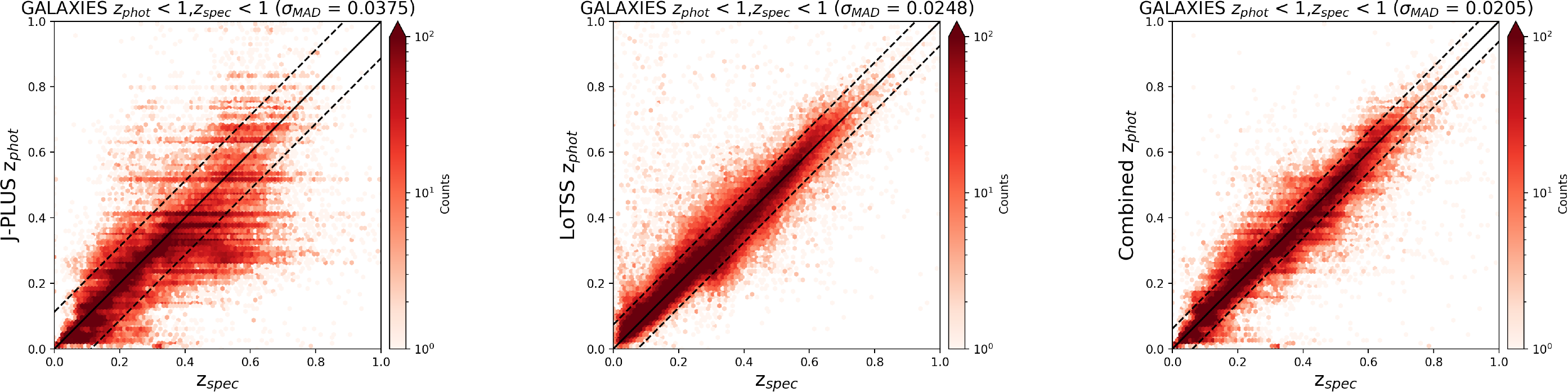}
  \caption{Comparison between different photo-z values in J-PLUS (left), LoTSS \citep[][center]{Duncan2022}, and our combined  estimate from J-PLUS and LoTSS conflation (right), with the spectroscopic z taken from SDSS, DESI or HETDEX, when available, for sources classified as galaxies with both photometric and spectroscopic z $<$ 1. The solid black line represents the one-to-one relation, and the dashed lines are drawn at $\pm 3\sigma_{\text{MAD}}$ (the sigma is indicated at the top of each panel). The conflation improves the accuracy and precision of both original photo-zs.}\label{fig:photo_z_comparison}
\end{figure*}

BANNJOS provides the full PDF, its mean, standard deviation and various percentiles of each of the three classes for every object. Fig.~\ref{fig:prob_class_dist} shows the distribution of the mean probabilities of each class for all objects. $\sim$91\% of objects are classified into one of the three classes with a mean probability above 0.9, and there are no objects with mean probabilities above 0.9 in more than one class. Using that value as threshold for classification keeps a high completeness while ensuring the classification is accurate. With this criterion, we obtain 393,023 galaxies, 31,234 QSOs and 21,617 stars. In this work, we additionally remove 72,098 galaxies with J-PLUS magnitude $r$ $>$ 21, as their photometric uncertainties are higher, but we keep them in the released catalog.

\subsection{Radio emission sizes and fluxes}\label{sec:Source sizes}
When the source has a clear, continuous radio emission, \cite{Hardcastle2023} estimate the angular size with a Gaussian fit, using the resolution criterion described \citep{Shimwell2022}, where they use a integrated-to-peak flux ratio and a signal-to-noise threshold. When the source has several composite radio components they try to estimate the angular size from the largest dimension of the convex perimeter encompassing all components. A small number of sources also have manual size measurements, and others have a ``flood-fill size'' measurement, where they use the individual components as the starting point and only include pixels above the local noise level. All sources whose size is not measured with a Gaussian fit are taken to be resolved. There will be a higher resolution release in the future that will allow to better constrain compact sources. Using the \texttt{Resolved} flag in their catalog, there are $\sim$450k unresolved and only $\sim$40k resolved radio sources in our sample, but their criteria is very restrictive and only considers the most ideal cases as resolved. We decide to include all sources in our catalog and our analysis, but we only consider as valid the size measurements of sources with \texttt{Resolved} = True, which we will refer to simply as \texttt{Resolved} radio sources from now on. We include this flag in our final catalog as well. LoTSS provides the peak flux density as well as the total flux within the estimated angular size of a source. 

\subsection{Redshifts}
There are available spectroscopic redshifts for $\sim$34\% of our sources, mainly from SDSS and DESI, and photometric redshift estimates based on broad bands for $\sim$95\% of sources in LoTSS and narrow bands for $\sim$97\% of sources in J-PLUS. \citet{Hernan2024} demonstrate that photo-z accuracy can be improved when the probability distributions from a deeper wide-band survey are combined with those from a shallower narrow-band survey. This method outperforms alternatives such as fitting all the photometry together or using a weighted average of point estimates. 

LoTSS DR2 includes spectroscopic redshifts from SDSS DR16, the first data release of the DESI spectroscopic survey, or HETDEX DR1 when available. We update or add new SDSS spectroscopic redshifts from the latest DR18 performing a 2 arcsec cross-match between the SDSS positions and the J-PLUS optical/IR identifications in catWISE and SDSS. When the spectroscopic redshift from SDSS/DESI is above 5, or has a difference with the photometric redshift of that same survey larger than 0.2, it is marked with the \texttt{unreliable\_spec\_z} flag. LoTSS also provides photometric redshift estimates for the sample with optical identifications in the DESI Legacy Imaging Surveys DR8 taken from \citet{Duncan2022}. The latter uses Gaussian Mixture Models (GMMs) to separate the parameter space into different regions that are trained and tested separately with the \texttt{GPz} code \citep{Almosallam2016}. They include a `best z', which always contains the spectroscopic redshift if available, or the photo-z estimate if flagged as reliable. If none of these are available, it is left empty. The J-PLUS DR3 photometric system includes essential broad and narrow bands that enable precise photo-z measurements for nearby galaxies with $z < 0.2$ \citep{Molino2019}. This accuracy is achieved through continuous coverage of the 4\,000\,\AA \ break and the detection of specific emission lines. The photo-z estimates are obtained the same way as with the miniJPAS catalog \citep{Bonoli2021}, which is described in detail in \citet{Hernan2021}, but adapted to the properties of the J-PLUS data. In short, they use a modified version of the L\textsc{e}P\textsc{hare} code \citep{Arnouts1999} and input 50 synthetic galaxy templates generated with the Code Investigating GALaxy Emission \citep[CIGALE;][]{Boquien2019}. They search in the range of 0 $<$ z $<$ 1 with a resolution of 0.001 to prioritize physically plausible solutions, as high-redshift objects in J-PLUS are expected to be dominated by QSOs due to brightness limits. They apply a prior derived from galaxy counts in the VIMOS VLT Deep Survey \citep[VVDS][]{LeFevre2005} and obtain final redshift PDFs for all sources regardless of their morphological classification to avoid biases, albeit only galaxy templates are used. Therefore, the J-PLUS photo-zs should be primarily trusted only for sources classified as galaxies. 

Both J-PLUS and LoTSS provide PDFs for the photo-z estimates. In the latter, the GMM method assumes the PDF is Gaussian by definition, where the mean is the photo-z estimate and the standard deviation its error. In J-PLUS, the PDF can have an arbitrary shape and even be multi-modal. The photo-z estimate is the value with the maximum $P(z)$ and its error is obtained from the 68th confidence interval. By leveraging the datasets from both surveys, we can enhance precision by conflating their PDFs to derive a new combined photo-z estimation, following \citet{Hernan2024}. We only use the estimates of objects classified as galaxies and photo-z $<1$ in both surveys, as the J-PLUS photo-z estimates are computed using lower redshift galaxy templates. Among these, the percentage of sources with a photo-z $1\sigma$ confidence interval below 0.05 increases from 33\% in J-PLUS and 56\% in LoTSS to 72\% in the new combined photo-z. The number of combined photo-zs that have smaller errors than either survey is $\sim$235,000. In cases with available spectroscopic redshifts, Fig.~\ref{fig:photo_z_comparison} shows the comparison with the different photo-zs for galaxies with both photo and spectroscopic redshifts below $z < 1$. To quantify the improvement in the redshift determinations, the following statistical diagnostics are introduced:

\begin{itemize}

    \item $\Delta z_i$: the relative error in $z_{\text{phot,i}}$, defined as
    \begin{equation}
        \Delta z_i = \frac{z_{\text{phot,i}} - z_{\text{spec,i}}}{1 + z_{\text{spec,i}}}
    \end{equation}

    \item MAPE: the median absolute percentage error, defined as
    \begin{equation}
        \text{MAPE} =  \text{median} \left(100 |\Delta z_i|\right) \, \%
    \end{equation}
    \item $\sigma_{MAD}$: the scaled median absolute deviation, equivalent to the standard deviation $\sigma (\Delta z_i)$ for a normal distribution, but less sensitive to outliers, defined as
    \begin{equation}
    \sigma_{\text{MAD}} = 1.48 \times \text{median}|\Delta z_i - \text{median}(\Delta z_i)|
    \end{equation}
    \item $\eta$: the percentage of outliers in a sample (objects with more than 0.03 relative error), defined as
    \begin{equation}
    \eta = 100\frac{N\left(|\Delta z_i| > 0.03\right)}{N_{total}} \, \%
    \end{equation}

\end{itemize}
where $i$ denotes an individual value, and the median runs through all of them. The values obtained for the original and the combined photo-zs are shown in Table~\ref{tab:stats}, both for the whole sample shown in the plots and for sources with photo-z $<$ 0.1 and r $<$ 17.

\begin{table}[ht!]
\caption{Median Absolute Percentage Error (MAPE), Median Absolute Deviation ($\sigma_{{MAD}}$) and fraction of outliers ($\eta$)} for the original and combined photo-z.\label{tab:stats}
\centering
\begin{tabular}{ c|c|c|c } 
   &  \text{MAPE} &  $\sigma_{\text{MAD}}$ & $\eta$ \\
\hline\\[-0.3cm]
 \text{J-PLUS photo-z}  & 2.67 \% & 0.0375 & 46.9 \% \\
 \text{LoTSS photo-z}   & 1.68 \% & 0.0248 & 28.7 \% \\
 \text{Combined photo-z} & 1.41 \% & 0.0205 & 25.7 \% \\
 \hline\\[-0.3cm]
\text{J-PLUS photo-z $<$ 0.2, r $<$ 17}  & 0.94 \% & 0.0132 & 15.6 \%\\
 \text{LoTSS photo-z $<$ 0.2, r $<$ 17}   & 1.22 \% & 0.0182 & 17.4 \%\\
 \text{Combined photo-z $<$ 0.2, r $<$ 17} & 0.77 \% & 0.0110 & 9.9\%\\
\hline
\end{tabular}
\end{table}

Although J-PLUS narrow-band photo-zs have more outliers and are outperformed by the DESI Legacy Surveys broad-band photo-zs in the overall sample, the median relative error of the narrow-band photo-zs is comparable and quite low. For brighter (and higher signal-to-noise) objects, J-PLUS yields better results, which reflects the precision of the estimates. Remarkably, the combined photo-z performs better across all metrics for both samples, which justifies the combination of the DESI Legacy Surveys photo-zs with the J-PLUS narrow-band photo-zs.

For our catalog, we defined a final best redshift, \texttt{z\_best}, which corresponds, in the following order, to:
\begin{enumerate}
    \item The latest SDSS DR18 \citep{Almeida2023} spectroscopic redshift, if available, without warnings and \texttt{unreliable\_spec\_z} = 0.
    \item The SDSS, DESI or HETDEX spectroscopic redshift that LoTSS provides without warnings and \texttt{unreliable\_spec\_z} = 0.
    \item LoTSS photo-z if the source is classified as a QSO and LoTSS  \texttt{flag\_qual} = 1.
    \item The combined photo-z if the source is classified as a galaxy, both surveys have a photo-z, the LoTSS photo-z $<$ 1.
    \item J-PLUS photo-z if the source is classified as a galaxy, the J-PLUS \texttt{ODDS} $>$ 0.5 and the J-PLUS photo-z is not near the boundary of 1.
    \item LoTSS photo-z if the source is classified as a galaxy and the LoTSS \texttt{flag\_qual} = 1 .  
\end{enumerate}
where J-PLUS \texttt{ODDS} is an indicator of the confidence in $z_{\text{phot}}$ and is defined as the probability that $|\Delta z|$ is smaller than 0.03. LoTSS \texttt{flag\_qual} selects sources with reliable redshifts, reasonable uncertainty, minimal contamination from nearby sources, low star-likelihood and free from imaging artifacts. We provide the \texttt{z\_source} column, which specifies the origin of the redshift in each case, and can be one of the following: `Combined zphot', `J-PLUS zphot', `LoTSS zphot', `LoTSS zspec', and `SDSS DR18 zspec'. If none of the listed conditions are fulfilled, \texttt{z\_best} and \texttt{z\_source} are left blank. Roughly 34\% of total sources in the catalog have a spectroscopic best redshift, and 92\% of the galaxies have a best combined photo-z. We also provide an \texttt{incompatible\_spec\_class} flag, which indicates when the SDSS/DESI spectroscopic class is different from the BANNJOS class (more information in Section \ref{sec:Stranger_things}). Our \texttt{z\_best} together with the cosmology described at the end of Section \ref{sect:Introduction} is used to estimate the radio linear size from the estimated angular size and the radio luminosity ($L_{144}$) from the total source flux on the assumption of a spectral index $\alpha$ = 0.7.

\begin{figure*}[!t]
  \centering
  \includegraphics[width=\textwidth]{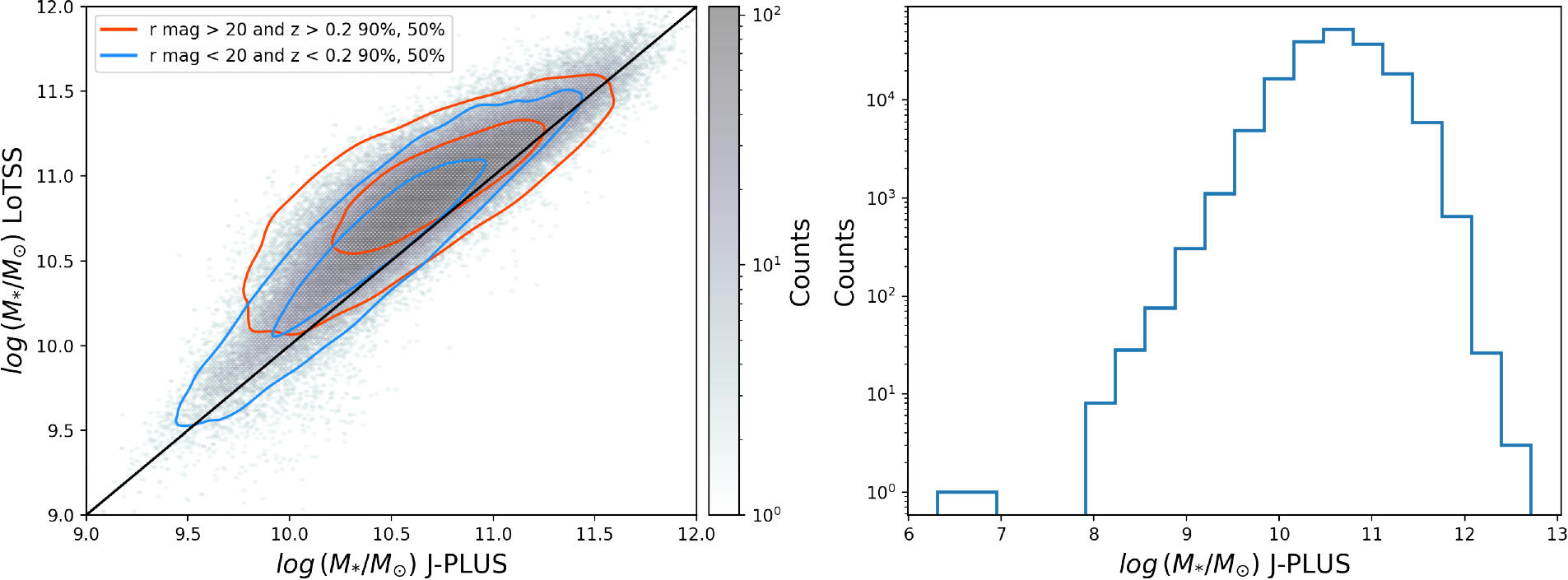}
  \caption{Left: LoTSS stellar masses taken from \citet{Hardcastle2023} adjusted for our \texttt{best\_z} against J-PLUS stellar masses computed by CIGALE \citep{Arizo2025} for galaxies with r $<$ 21 and low uncertainty (see text). The orange contours contain 90\% and 50\% of galaxies with r mag above 20 and redshift above 0.2 (faint), while the blue contours contain 90\% and 50\% of galaxies with r mag below 20 and redshift below 0.2 (bright). Right: Distribution of the J-PLUS stellar mass of galaxies with r $<$ 21 and low uncertainty. The most extreme outliers were adjusted with redshifts that differed considerably from the original J-PLUS photo-z and should not be trusted.}\label{fig:stellar_mass_galaxies}
\end{figure*}

\subsection{Radio-loudness}
Traditionally, AGN are divided into radio-loud and radio-quiet classes. The former correspond to the kinetic mode of black hole activity, where most of the energy output is dominated by powerful jets \citep{Merloni2007}. On the other hand, radio-quiet sources are in the radiative mode, dominated by the accretion disk, although the origin of their radio emission is not well established yet \citep{Chen2023}. There are various radio-loudness definitions in the literature, e.g., based on the radio-to-optical \citep{Kellermann1989} or radio-to-X-ray \citep{Terashima2003} ratios. Here we introduce a new radio-loudness definition, based on the flux ratio between 144 MHz and WISE W2 ($4.6\, \rm{\micron}$), to analyze the distribution of star-forming galaxies and AGN in our sample. This ratio, discussed in this work for the first time, is particularly convenient for identifying radio-loud AGN, including optically quiescent radio galaxies with past jet activity, because the low-frequency radio flux is sensitive to older and less energetic synchrotron emission. On the other hand, the IR flux probes the presence of hot dust associated with the active nucleus or star formation. In star-forming galaxies this ratio would be dominated by the star-formation activity and the associated contribution from supernova remnants, while in radio-quiet QSOs low ratios are caused by the strong IR dust emission. Radio-loud sources have an extra contribution to the radio fluxes caused by the jet activity, which, at low frequencies typically comes from extended radio structures. This extra contribution distinguishes the galaxies with high radio-loudness ratios (hereafter radio-loud galaxies) from star-forming galaxies and radio-quiet QSOs. Note that although the rest-frame frequencies of this ratio are different for each redshift, our range is not large enough to avoid tracing the synchrotron continuum in radio and the dust emission in the IR. Most galaxies and a large fraction of QSOs are below $z = 2.5$. In this case, the WISE W2 band rest-frequency would correspond to $1.3\, \rm{\micron}$, which is still representative of the AGN hot dust (e.g. \citealt{Elvis1994}). On the other hand, the LoTSS 144 MHz frequency would correspond to approximately 500 MHz, which still probes the synchrotron continuum. Although redshift may introduce a scatter in the radio-to-IR ratios, the latter still provides a meaningful radio-quiet and radio-loud source separation. Additionally, variability may affect the low frequency and mid- to near IR continuum differently, owing to the distinct timescales associated with the jet and hot dust components. A larger variability is expected in the latter due to the faster response to luminosity changes in the accretion disk. Nevertheless, we only expect a minor fraction of IR-variable sources in our sample ($\lesssim$10\%), as suggested by \citet{Sanchez2017} for a sample of AGN at z $<$ 2. Additionally, the radio-loudness distribution covers a range of 7 orders of magnitude, thus significant deviations are only expected for sources with extraordinarily large variability amplitudes.

\subsection{Stellar masses and star formation rates}\label{sec:Stellar masses}
\citet{Arizo2025} obtain stellar masses, SFRs, rest-frame colors and luminosities from J-PLUS extended sources according to the \texttt{sglc\_prob\_star} parameter \citep{LopezSanJuan2019} with the CIGALE SED fitting code. They adopt J-PLUS photometric redshifts, fluxes and flux errors and use \citet{Bruzual2003} stellar populations with \citet{Chabrier2003} initial mass functions with different metallicities. We scale their mass estimates and SFRs (averaged over the last 100 Myrs), initially computed for sources with J-PLUS photo-z estimates up to $z\, \sim \, 0.35$, to our new \texttt{z\_best} using the previously defined cosmology. Approximately $40$\% of the sources in our sample have a stellar mass estimate, with the majority ($\sim$95\%) being galaxies. We discard the values for QSOs, as AGN templates in CIGALE do not provide reliable stellar mass estimates.

These results are compared with LoTSS-based estimates from \citet{Hardcastle2023}, who follow the SED-fitting approach in \citet{Duncan2022} using DESI Legacy Surveys broad-bands and WISE photometry, but introducing photo-z uncertainties. After scaling their masses to our \texttt{z\_best}, both estimates are compared in the left panel of Fig. \ref{fig:stellar_mass_galaxies}. LoTSS values show a positive median offset of 0.18 dex with 0.20 dex scatter for galaxies with reliable estimates and $r<21$. The offset likely arises from differences in flux measurements and apertures, while the scatter reflects photometric uncertainties. Brighter galaxies ($r < 20$ and z $<$ 0.2) show a lower scatter than the median squared error of the J-PLUS mass estimates, meaning that the errors are slightly overestimated. Both offset and scatter are slightly larger for fainter galaxies ($r > 20$ and z $>$ 0.2). Overall, J-PLUS narrow-band derived masses are in good agreement with LoTSS broad-band estimates, confirming the robustness of our approach. The J-HERTz catalog includes stellar masses and SFRs at \texttt{z\_best} from J-PLUS photometry. The right panel of Fig. \ref{fig:stellar_mass_galaxies} shows the distribution of stellar masses, with a few extreme outliers caused by large differences between the original J-PLUS photo-z and the newly obtained \texttt{z\_best}, which mainly happens whtn the spectroscopic class in SDSS/DESI is not a galaxy and can be filtered with the \texttt{incompatible\_spec\_class} flag.

\begin{figure*}[!t]
  \centering
  \includegraphics[width=1\textwidth]{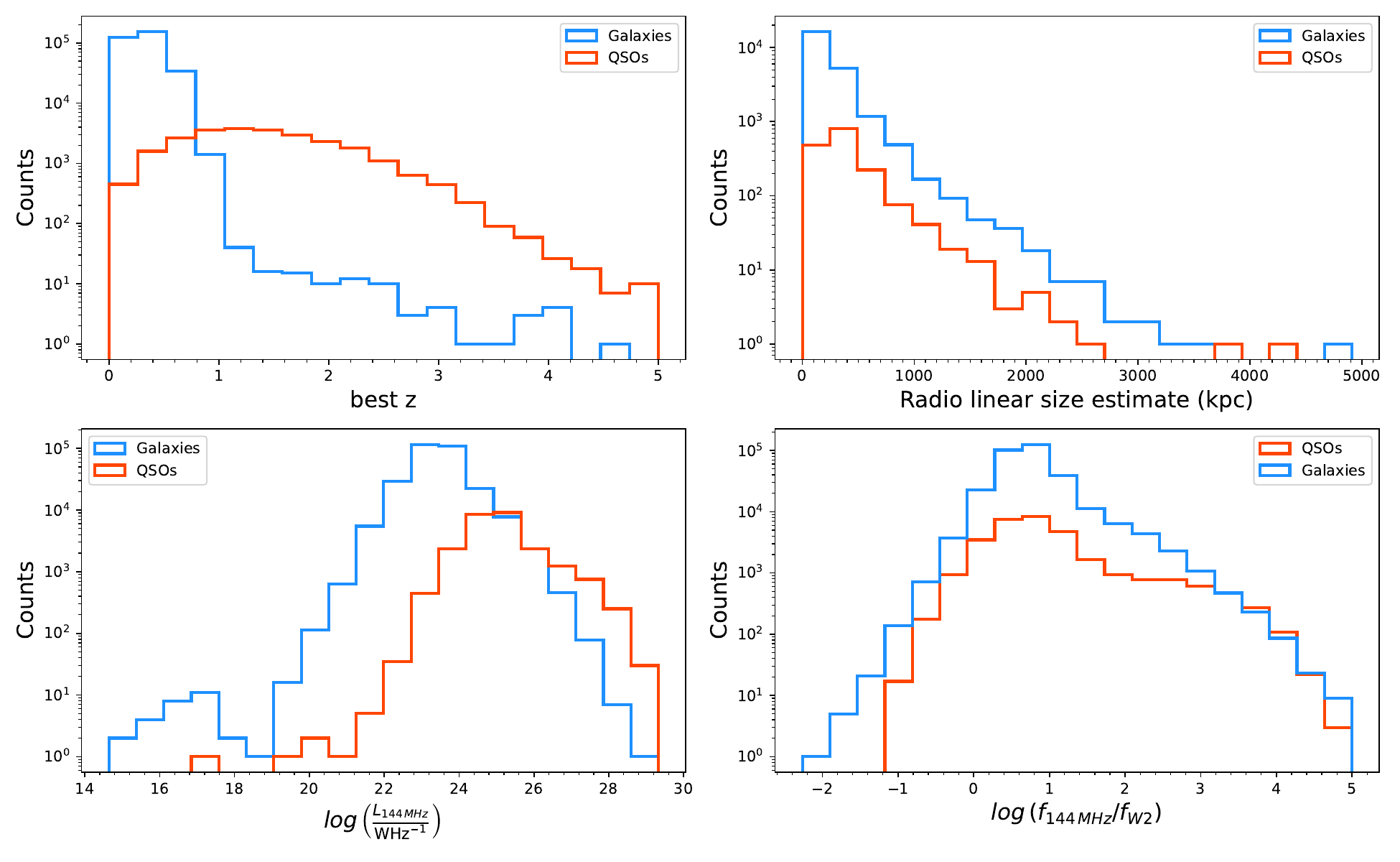}
  \caption{Distribution of the best redshift (upper left), the linear size of the radio emission for sources with an angular size larger than 6 arcsec and classify as resolved in \cite[][upper right]{Shimwell2022}, the radio luminosity (lower left), and the ratio between the radio 144 MHz and the IR W2 band fluxes as radio-loudness (lower right) for QSOs (orange) and galaxies with $r < 21$ mag (blue). Note that stars are not included in these histograms.}\label{fig:radio_properties} 
\end{figure*}

\section{Source properties in J-HERTz}\label{sec:properties}
\subsection{QSOs}\label{sec:QSOs}
In this section we analyze the main properties of the nearly 31k QSOs included in the catalog. These include their best redshift, radio luminosity, radio linear size and radio-loudness, whose distributions are shown in the orange histograms of Fig.~\ref{fig:radio_properties}. The QSO best-z distribution in our sample peaks around $z \sim 1-2$, in agreement with the higher density of active galaxies found at the cosmic noon \citep{MadauDickinson2014}, but it spans a wide range of values. The lowest value with a compatible spectroscopic class is 0.028 from SDSS (WISEA J160443.63+531517.8), although this appears to be a misidentification of H$\alpha$ emission, which actually corresponds to [OII] emission at $z=0.79$. The highest value of $z=7.01$ also comes from SDSS (WISEA J234219.09+270845.9), however spectroscopic redshifts above 5 are unreliable. Cases like either of these two examples (when the spec-z and photo-z of the same spectroscopic survey are very different, or the spec-z is larger than 5) are marked with an \texttt{unreliable\_spec\_z} flag and are not included in the histogram or the \texttt{z\_best} parameter, although they are included in the \texttt{zspec} parameter (see Section \ref{sec:Stranger_things}). The radio luminosities peak at $L_{144 MHz} \sim 10^{25}\, \rm{W\,Hz^{-1}}$, with the majority falling between $10^{22}$ and $10^{29}\,$ $\rm{W\,Hz^{-1}}$. The few outliers with very low luminosities correspond to sources where the \texttt{z\_best} comes from a low spec\_z of an object spectrally classified as a star, and can be avoided with the \texttt{incompatible\_spec\_class} flag. The estimated linear size of the QSO radio emission for \texttt{Resolved} sources with an angular size larger than 6 arcsec mostly falls below 2700 kpc, with the most extended radio source reaching 4.3 Mpc and 20 arcmin \citep[WISEA J093139.04+320400.1; ][]{Kuzmicz2018}. We consider the host galaxies resolved if their major axis ($2 \, \texttt{A\_WORLD}$) is larger than the typical FWHM value of 1.2 arcsec in J-PLUS. By this criterion, approximately 20\% of the QSO host galaxies detected by J-PLUS are resolved. Over 70\% of these have an axis ratio $b/a > 0.8$, indicating a high degree of sphericity typical of elliptical galaxies, which are likely the usual hosts of QSOs. Taking only the cases with a relative error in $g$ and $r$ bands below 10\%, QSOs have a median and dispersion ($g-r$) color of 0.20 $\pm$ 0.12. The radio-loudness of QSOs is not described by a single Gaussian distribution, showing instead a bimodal distribution, i.e. with two local maxima centered around $\text{log} \, (f_{144 \, MHz}$/$f_{W2}) \sim 0.7$ and $\sim2.5$. We explore this further in Section \ref{sec:Radio and IR}.

\subsection{Galaxies}\label{sec:Galaxies}
We have $\sim$390k galaxies in our catalog, $\sim$320k with $r < 21$ mag. The best redshifts and radio properties are shown in Fig.~\ref{fig:radio_properties}. Around 72\% of the best-zs are below 0.5, the lowest belonging to NGC 2976 (a small galaxy in the M81 group) at z = $9 \cdot 10^{-6}$. Only $\sim$0.50\% are above 1, and many of them are spectroscopically classified as QSOs or have unreliable spec-zs, which can be filtered using the \texttt{incompatible\_spec\_class} and \texttt{unreliable\_spec\_z} flags). The radio luminosities peak at $L_{144 MHz} \sim 10^{23.5}\, \rm{W\,Hz^{-1}}$, with most falling between $10^{20}$ and $10^{28}$ $\rm{W\,Hz^{-1}}$ and some lower outliers than can also be filtered with the same flags. The estimated linear size of the galaxies' radio emission for \texttt{Resolved} sources with an angular size larger than 6 arcsec, mainly falls below 3000 kpc, with the most extended radio source reaching 4.9 Mpc and 12 arcmin (WISEA J003623.89+253611.1), and the second largest with 4.4 Mpc and 40 arcmin being 3C 236, one of the largest known giant radio galaxies \citep{Shulevski2019}. Using the same optical resolution criteria as defined in Sect.~\ref{sec:QSOs} for the hosts, approximately 87\% of the galaxies are resolved in J-PLUS, a much higher percentage than the QSOs. This highlights the effectiveness of BANNJOS in using optical morphology to classify sources. Their axis ratio distribution is wider as well, with 40\% of sources having an axis ratio $b/a < 0.8$, indicating more varied morphological types. Taking only the cases with a relative error below 10\% in $r$ and $g$ bands, galaxies have a median and dispersion ($g-r$) color of $0.69 \pm 0.14$, much redder than QSOs, but showing a lot of variation from bluer ($g-r < 0.3$) up to redder ($g-r \sim 1$) colors. The radio-loudness in galaxies appears to have a single distribution with a peak around $\text{log} \, (f_{144 \, MHz}$/$f_{W2}) \sim 0.5$, but it is skewed towards higher values and reaches $\text{log} \, (f_{144 \, MHz}$/$f_{W2}) \sim 5$.

 The right panel of Fig.~\ref{fig:stellar_mass_galaxies} shows the distribution of stellar masses of galaxies (Sect.~\ref{sec:Stellar masses}). We filter for galaxies with $r < 21$ mag, masses with low uncertainty and redshift values with no flags. Most galaxies have a stellar mass between $10^8$ and $10^{12}$ $M_{\odot}$, peaking at $10^{10.5}$ $M_{\odot}$. The few outliers, although passing all the filters, were adjusted from the original values from \citet{Arizo2025} with a new $\texttt{z\_best}$ that differs considerably from the original J-PLUS photo-z used and should not be trusted. For example, the cases with extremely low masses were classified as stars in SDSS and have spectroscopic redshifts compatible with zero, which when used to adjust the original masses resulted in these low values. On the other hand, the sources with high masses were classified as QSOs in SDSS and have much higher spectroscopic redshifts. These problematic cases can be avoided using again the \texttt{incompatible\_spec\_class} flag. In any case, the fraction of galaxies with stellar masses outside $[10^{8},10^{12}] \, M_{\odot}$ is only 0.11\%.

\begin{figure*}[h]
  \centering
  \includegraphics[width=0.9\textwidth]{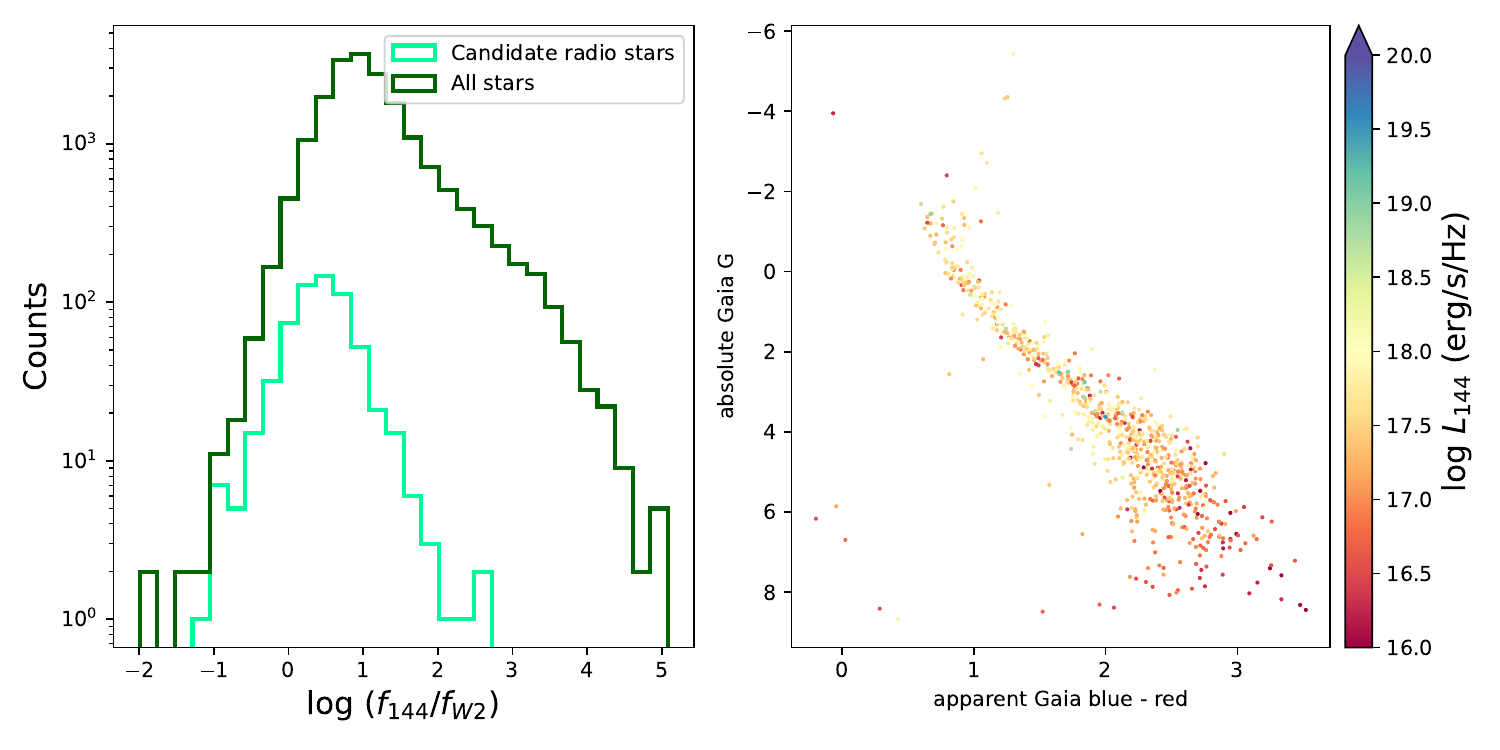}
  \caption{Left: radio-loudness distribution of all the objects classified as stars (dark green) and our genuine radio star candidates (light green). Right: position of our radio star candidates in the Gaia color-magnitude diagram, color coded by their radio luminosity. The G-band is at $\sim$330–1050 nm, the red-band at $\sim$640–1050 nm and the blue-band at $\sim$330–680 nm.}\label{fig:stars_radio_info}
\end{figure*}

\begin{figure*}[!t]
  \centering
  \includegraphics[width=1\textwidth]{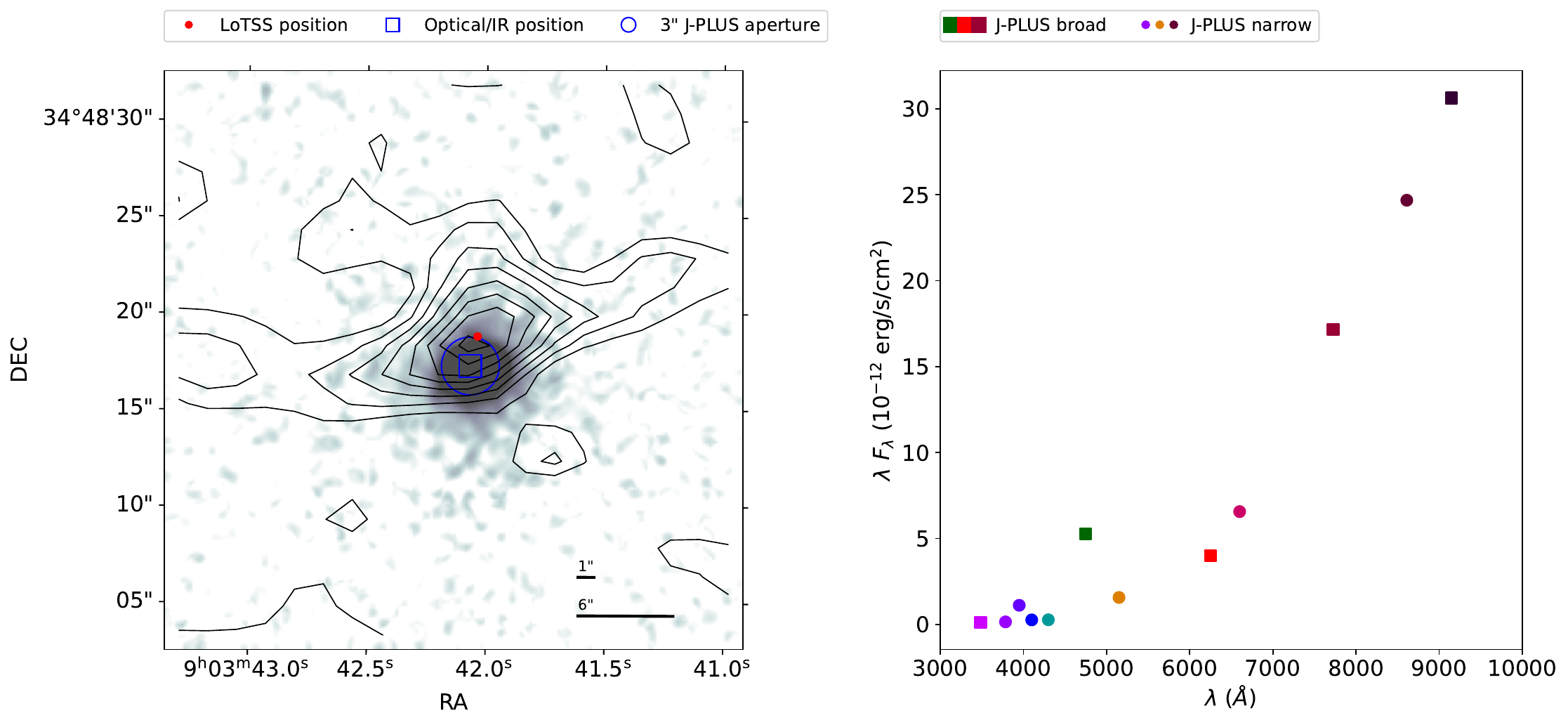}
  \caption{A candidate radio star, also found in the Sydney Radio Star Catalogue \citep[][]{Driessen2024} and among the population of M dwarfs observed at low radio frequencies from \citet{Callingham2019}. It also shows a prominent Ca II emission line from a flare. Its unique J-PLUS identification is \texttt{TILE\_ID} = 87918 and \texttt{NUMBER} = 18792. Left panel: optical image from J-PLUS in gray scale, with the J-PLUS and LoTSS positions (blue square and red dot respectively), and the radio contours (black lines) from LoTSS drawn at values between 0.5 and 5$\sigma$. The blue circle is the J-PLUS 3 arcsec aperture used. Right panel: 3 arcsec aperture fluxes and error bars measured in the 12 J-PLUS filters as the colored symbols (squares for broad and circles for narrow bands). No SDSS spectrum is available for this source.}\label{fig:star_extended_radio}
\end{figure*}

\subsection{Stars}\label{sect:Stars}
Our initial cross-matched sample includes $\sim$21.6k objects classified as stars by BANNJOS with associated radio emission. However, most of these are expected to be spurious associations (\citealt{Callingham2019}), and therefore a series of filters are applied to identify a more reliable set of candidates. First, we remove objects with inconsistent spectroscopic classifications, multiple BANNJOS identifications, or with a nearby galaxy or QSO within the radio emission region (see Sect.~\ref{sec:Stranger_things}). This step eliminates most extended radio sources incorrectly associated with stars. After filtering, we retain $\sim$9k extended radio sources with \texttt{Resolved} = False and only 347 \texttt{Resolved} stars with extended radio emission, likely contaminated by nearby undetected sources. We also recover $\sim$5k compact ($<6^{\prime\prime}$, \texttt{Resolved}=False) sources. This last group constitutes the most promising candidate radio stars. Because genuine radio stars are intrinsically faint (typical $L_\nu \sim 10^6$–$10^{12}$ W Hz$^{-1}$; \citealt{Callingham2021,Driessen2024}), LoTSS flux limits constrain the distance where detection is possible to $\sim$1 kpc of the Sun. We cross-match our compact, filtered stars with Gaia DR3, and retain only sources with reliable parallaxes ($\varpi/\sigma_\varpi > 5$) and distances $d<1$ kpc, using the catalog in \citet{BailerJones2021}. This yields a final sample of 831 stars, with inferred radio luminosities consistent with the expected range. To assess the possibility of chance alignments between radio AGN and optical stars, we follow the likelihood-ratio method of \citet{Callingham2019} and \citet{Sutherland1992}. Compared with offset positions, we find an excess of cross-matches within 5$^{\prime\prime}$ by a factor $\sim$25–40, indicating that the majority of these associations are unlikely to be spurious. The radio-loudness distribution of our candidates is significantly narrower and more radio-quiet than that of the full star sample, consistent with expectations for nearby stars (left panel of Fig. \ref{fig:stars_radio_info}). In the Gaia color–magnitude diagram (right panel of Fig.~\ref{fig:stars_radio_info}), our candidates populate the main sequence and display a luminosity gradient similar to that reported for confirmed radio stars in the Sydney Radio Star Catalogue \citep[Fig.~4 in][]{Driessen2024}. Since that catalogue is dominated by southern-hemisphere objects, only two sources are cross-matched with our sample (one of which is highlighted in Fig.~\ref{fig:star_extended_radio}). Both stars are also included in the sample of M dwarfs observed at low radio frequencies by \citet{Callingham2021}. Although the overlap is small, this agreement provides encouraging support for the reliability of our candidate selection. We therefore propose this final set of 831 sources as candidate radio stars. These are marked with a \texttt{radio\_star\_candidate} = `reliable' in the final catalog, while the rest of the stars have an `unreliable' value. A more detailed characterization is required, however, before any of them can be confirmed as genuine detections. Possible mechanisms to explain the radio continuum emission in these sources include stellar coronal activity \citep{Callingham2021} or magnetic interactions with close-in exoplanets \citep{Pope2021}, although further follow-up will be needed to establish the nature of these objects.

\section{Discussion}\label{sec:Discussion}
In this section we discuss the main properties of galaxies and QSOs, focusing on their radio and IR continuum. Specifically, the results based on improved photo-zs and stellar masses derived with narrow-band photometry are compared with those obtained by \citet{Hardcastle2023,Hardcastle2025} for the larger LoTSS DR2 sample. Additionally, we analyze the radio-loudness properties for the different populations identified by BANNJOS. In the case of galaxies, we discuss how the radio activity relates with the masses and SFRs. Finally, in Section \ref{sec:Stranger_things} we discuss some caveats and user warnings for the catalog.

\begin{figure*}[!t]
  \centering
  \includegraphics[width=1\textwidth]{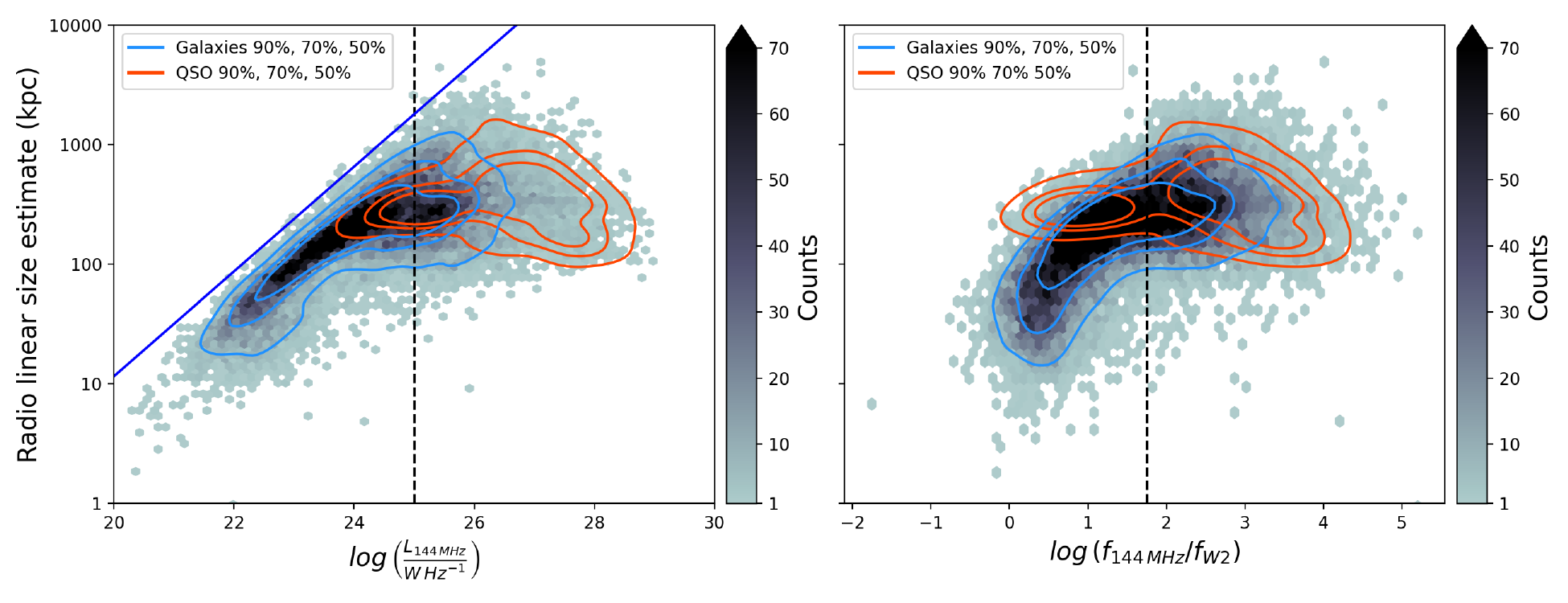}
  \caption{Left: radio linear size estimate against radio luminosity. The black dashed line shows the threshold between FRI and FRII objects \citep[$L_{144 \, MHz} = 10^{25}\, \rm{W\,Hz^{-1}}$;][]{FaranoffRiley1974}. The blue solid line shows the surface brightness limit of LoTSS. Right: radio linear size estimate against radio-loudness. The black dashed line shows the threshold between radio-quiet and radio-loud objects ($\text{log} \, (f_{144 \, MHz}$/$f_{W2}) = 1.76$). In both figures, only the linear sizes of \texttt{Resolved} sources with an angular size larger than 6 arcsec are included. The color map shows the density distribution of all sources, the orange contours show the distribution of QSOs and the blue contours show the distribution of galaxies with r $<$ 21. Each contour contains 90\%, 70\% and 50\% of sources inside.}\label{fig:radio_sizes}
\end{figure*}

\subsection{Radio luminosities, loudness and sizes}\label{sec:Radio and IR}

\begin{figure*}[!t]
  \includegraphics[width=\textwidth]{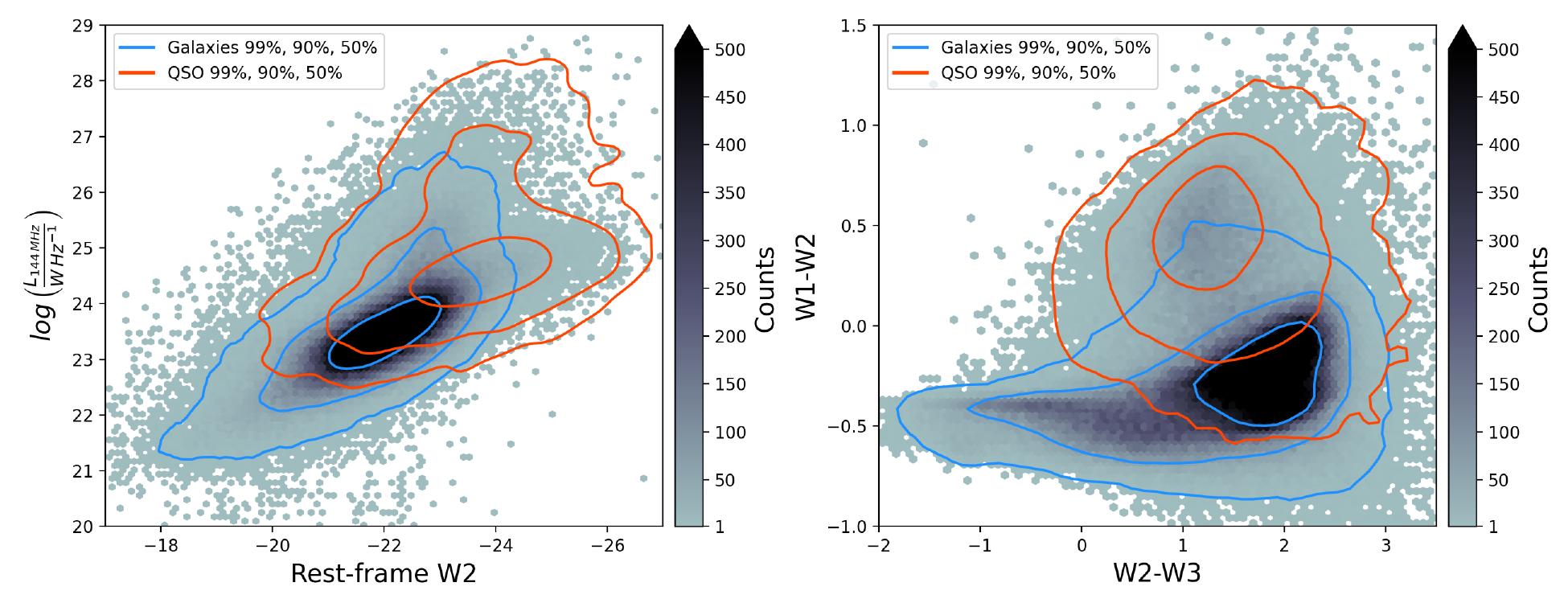}
  \caption{Left: Logarithm of the 144 MHz radio luminosity against WISE W2 absolute magnitude. Right: WISE color-color diagram. In both figures, the color map shows the density distribution of all sources, the orange contours show the distribution of QSOs and the blue contours show the distribution of galaxies with r $<$ 21. Each contour contains 99\%, 90\% and 50\% of sources inside.}\label{fig:radio_WISE_class_comparison}
\end{figure*}

The lower left panel in Fig.~\ref{fig:radio_properties} 
shows that QSOs radio luminosities peak at a higher value when compared to galaxies, possibly due to a selection effect given the different redshift distributions of the two populations. Limiting the comparison only to QSOs with \texttt{z\_best} $<$ 1, they peak at a much closer value of $10^{24}\, \rm{W\,Hz^{-1}}$, although this remains slightly higher than for galaxies. Only $\sim$2\% of the galaxies would fall within the Faranoff-Riley II (FRII) luminosity cut \citep[$L_{144 \, MHz} > 10^{25}\, \rm{W\,Hz^{-1}}$;][]{FaranoffRiley1974}, while that percentage is $\sim$40\% for QSOs. The left panel of Fig.~\ref{fig:radio_sizes} illustrates the relationship between radio luminosity and estimated radio linear sizes for \texttt{Resolved} sources with an angular size larger than 6 arcsec. Many of the galaxies fall in the Faranoff-Riley I (FRI) region, where the radio emission linear size is strongly correlated with the radio luminosity. On the other hand, most QSOs fall in the FRII region and their radio size and luminosity are not strongly correlated. All sources reach more extended radio emission in the FRII region, typically due to larger jets and brighter radio lobes at their edges, but this FRI/FRII threshold does not appear to divide the sample in a meaningful way. Moreover, the lack of low-luminosity extended sources is most likely caused by selection effects, as shown by the solid blue line indicating the surface brightness limit of LoTSS (see fig. 8 in \citealt{Hardcastle2025}). The distribution of sizes and luminosities in our sample is consistent with the radio-luminous AGN population in their work.

The lower right panel of Fig.~\ref{fig:radio_properties} shows that QSOs seem to exhibit a bimodal distribution in radio-to-IR radio-loudness. The majority of sources peak at lower ratios, with an extended tail or possibly a second distribution at higher ratios. This bimodality has been investigated in previous works using radio-to-optical ratios or radio luminosities as radio-loudness indicators, with mixed results. \citet{Kellermann1989,Laor2000,Ivezic2002,Kellermann2016,Balokovic2012,Zhang2021} and \citet{Arnaudova2024} find that the radio-loudness distribution can be best fitted with two Gaussian components, or that there are different properties among the two populations, even if the separation is not a clear division. 
On the other hand, \citet{White2000} and \citet{Cirasuolo2003} find continuous distributions from low to high radio-loudness without a clear gap, but their analysis is limited to samples of a few hundred QSOs. \citet{Mahony2012} find a bimodal distribution for a sample covering a wide range in redshift, which is interpreted as evolutionary effect rather than an intrinsic property of the QSO population.

We perform an analysis using the Bayesian Information Criterion (BIC) and the Akaike Information Criterion (AIC), comparing a 2-normal distribution with different continuous asymmetric distributions such as the log-normal, skewed-normal and exponential-normal. To avoid possible selection effects, we limit the sample to QSOs with \texttt{z\_best} $<$ 1. The 2-normal distribution is favored over the others by both diagnostics, with $\Delta \text{BIC}$ and $\Delta \text{AIC}$ $>$ 500. The best fit for the whole QSO sample provides a Gaussian distribution centered at $\mu_1 \sim 0.69$ with $\sigma_1 = 0.44$, and a second one centered at $\mu_2 \sim 2.55$ with $\sigma_2 = 0.73$. The radio-loudness value at the intersection of these Gaussians is adopted as the threshold between both populations, i.e. at $\text{log} \, (f_{144 \, MHz}$/$f_{W2}) = 1.76$. We consider sources below this threshold as radio-quiet, and radio-loud above it. However, we note that this is not an absolute division, as the tails of both populations overlap. $\sim$5\% of QSOs are radio-loud at low redshifts (z $<$ 0.2) and the fraction increases to $\sim$12\% at z $\sim$ 1, which is very similar to the usual fraction of radio-loud QSOs observed at low redshifts in the literature \citep{Kellermann1989,Ho2002}. The radio-loud fraction reaches a maximum of $\sim$20\% at z = 3.5 – 4.0, but selection effects could come into play at higher redshifts. Moreover, the exact fraction of radio-loud objects will depend on the chosen definition for radio-loudness and, as mentioned before, there is still a debate about whether there is a real bimodality in the radio-loudness distribution of QSOs. 

\begin{figure*}
\centering

\begin{subfigure}{0.48\textwidth}
  \includegraphics[width=\textwidth]{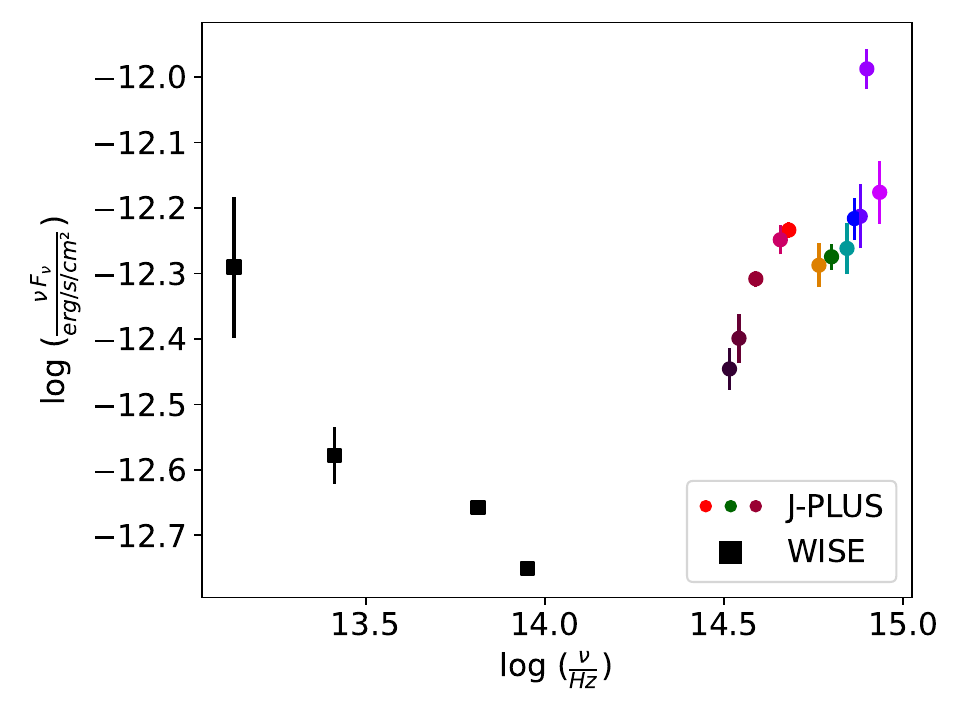}
  \caption{QSO: 101999-21337 and SDSS spec-z = 1.43.}
  \label{fig:SED_example_1}
\end{subfigure}
\hfill
\begin{subfigure}{0.48\textwidth}
  \includegraphics[width=\textwidth]{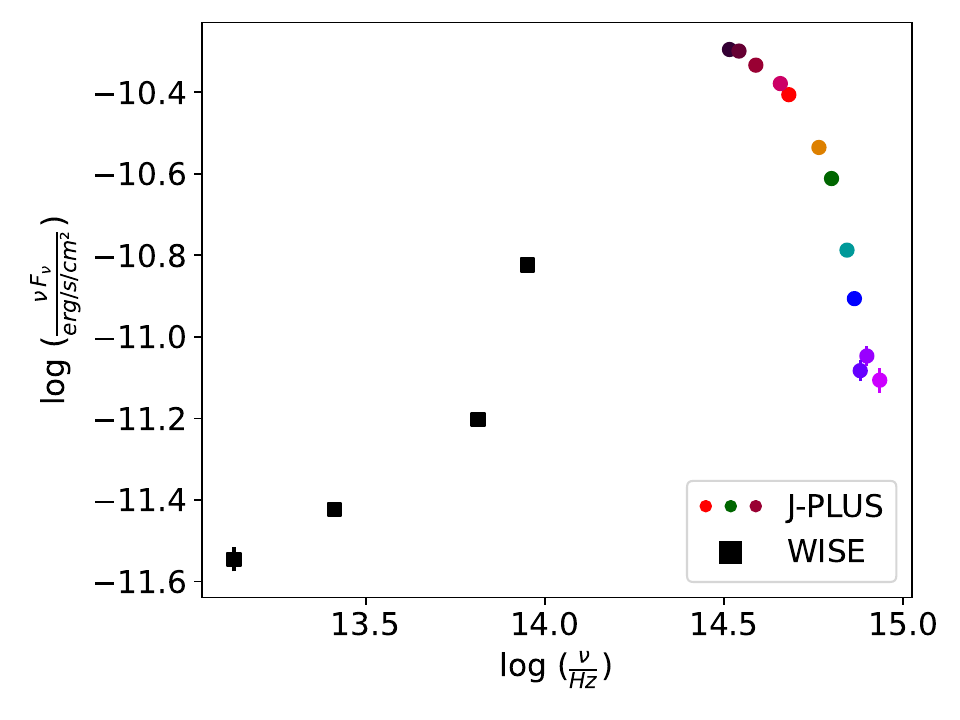}
  \caption{Galaxy: 102528-20171 and combined photo-z = 0.03.}
  \label{fig:SED_example_2}
\end{subfigure}

\caption{Example of the SED of two objects with high radio luminosity ($\log \left( \frac{L_{144 \, \mathrm{MHz}}}{\mathrm{W} \, \mathrm{Hz}^{-1}} \right) > 25$). Their unique J-PLUS \texttt{TILE\_ID}-\texttt{NUMBER} identification and redshift is shown in the captions. Colored symbols and error bars show the 12 J-PLUS band fluxes (squares for broad and circles for narrow bands) with \texttt{AUTO} photometry, while the black squares show the WISE IR fluxes. Some error bars are too small to be visible. Left: object classified as QSO with bluer emission in the optical and relatively high WISE W1--W2 color. Right: galaxy with redder optical emission and relatively low W1--W2 color.}
\label{fig:radio_ratio_both_examples}
\end{figure*}

Galaxies in our sample show radio-loudness values centered around $\log (f_{144,\mathrm{MHz}}/f_{W2}) = 0.5$, without exhibiting the bimodality observed in QSOs. Overall, only $\sim$5\% of galaxies fall into the radio-loud category. Their high radio-to-infrared flux ratios suggest that these nuclei have undergone intense jet activity in the past, consistent with the long timescales traced by low-frequency synchrotron emission. Nevertheless, these galaxies show no clear signatures of ongoing nuclear activity in their J-PLUS photometry or WISE colors. To study them in more detail, we cross-match our radio-loud galaxies with the catalog by \citet{Drake2024}, who provide LoTSS sources with SDSS emission line spectra from the Portsmouth catalog \citep{Thomas2013}, radio-excess probabilities, and spectral classifications of a subset of hosts with full line diagnostics in a BPT diagram \citep{Baldwin1981}. J-HERTz has 3,621 radio-loud galaxies in common with their sample. Following the classification by \citet{Drake2024}, galaxies are identified as Low-Excitation Radio Galaxies \citep[LERGs; ][]{Laing1994,Heckman2014} if they exhibit radio-excess and fall within the LINER, composite, or star-forming regions in the BPT diagram. The sources located below the \citet{Kewley2001} star-forming boundary can be considered as optically quiescent LERGs, i.e. radio-loud galaxies with a weak or negligible contribution from the AGN to the nebular ionized gas. Among the common sample, 12\% of radio-loud galaxies are LERGs falling below the star-forming boundary, 5\% are compatible with HERGs, and 45\% lack a clear classification or miss any of the BPT lines (H$\alpha$, H$\beta$, [\ion{O}{3}]$\lambda 5007$ or [\ion{N}{2}]$\lambda 6583$).  

Nevertheless, the total fraction of optically quiescent LERGs should also include passive galaxies without significant line emission that cannot be classified in the BPT diagram, such as optically inactive radio galaxies. Following the criteria used in \citet{Stasinska2025}, a host is considered inactive if its H$\alpha$ equivalent width is less than 3\AA. Among the radio-loud galaxies without a BPT classification but with measured H$\alpha$ equivalent widths (and continuum S/N $>$ 3), 7\% are found to be optically inactive.  This implies that the total fraction of optically quiescent LERGs is approximately $\gtrsim$20\% of the radio-loud galaxy population. It is interesting to compare this result with the fraction of remnant AGN, i.e. sources in which jet activity at the nucleus has completely ceased, usually identified as radio galaxies with undetected radio cores in deep surveys or with ultra-steep spectral indices. Previous studies have reported upper limits of $\lesssim$5--10\% for the fraction of AGN remnants among radio-loud galaxies \citep{Mahatma2018,Dutta2023,Mostert2023}. This suggests that most optically quiescent LERGs still host an active radio core that is unable to ionize the gas but continues to power the radio lobes. Identifying these systems is crucial for understanding the AGN duty cycle, as they represent a substantial fraction of the radio-loud population ($\gtrsim$20\%) showing negligible or undetectable signs of nuclear activity in the optical, yet with ongoing jet activity that may contribute significantly to the black hole accretion history.

The right panel in Fig.~\ref{fig:radio_sizes} shows the radio-loudness against the radio linear size for \texttt{Resolved} QSOs and galaxies with angular sizes larger than 6 arcsec. Most sources are radio-quiet; QSOs show a bimodal distribution, whereas galaxies show a continuous distribution towards the radio-loud side. 60\% of radio-loud galaxies and 39\% of radio-loud QSOs have extended and resolved radio emission ($>$ 6 arcsec and \texttt{Resolved}), while those percentages are 6\% for radio-quiet galaxies and 2\% for radio-quiet QSOs. The size 1$\sigma$ dispersion of the \texttt{Resolved} sources with angular size larger than 6 arcsec and are 60--450\,kpc for radio-loud galaxies, $\lesssim80$\,kpc for radio-quiet galaxies, 170--640\,kpc for radio-loud QSOs and 230--400\,kpc for radio-quiet QSOs. Radio-loud sources tend to have more extended radio emission. This is expected since jet activity boosts the radio luminosity and produces large radio structures. Nevertheless, many sources on the radio-quiet side also show extended emission. Most of them lie close to the threshold and likely belong to the tail of the radio-loud population with less powerful jets. Among sources that are $3\sigma$ away from the radio-loud mean ($\text{log} \, (f_{144 \, MHz}$/$f_{W2}) < 0.35$) and most probably radio-quiet, there are only 32 QSOs and 224 galaxies with resolved and extended emission. Some of the radio-quiet galaxies with resolved extended radio emission are extended in the optical themselves and the emission is coming from their star formation.

\begin{figure*}[!t]
  \centering
  \includegraphics[width=0.8\textwidth]{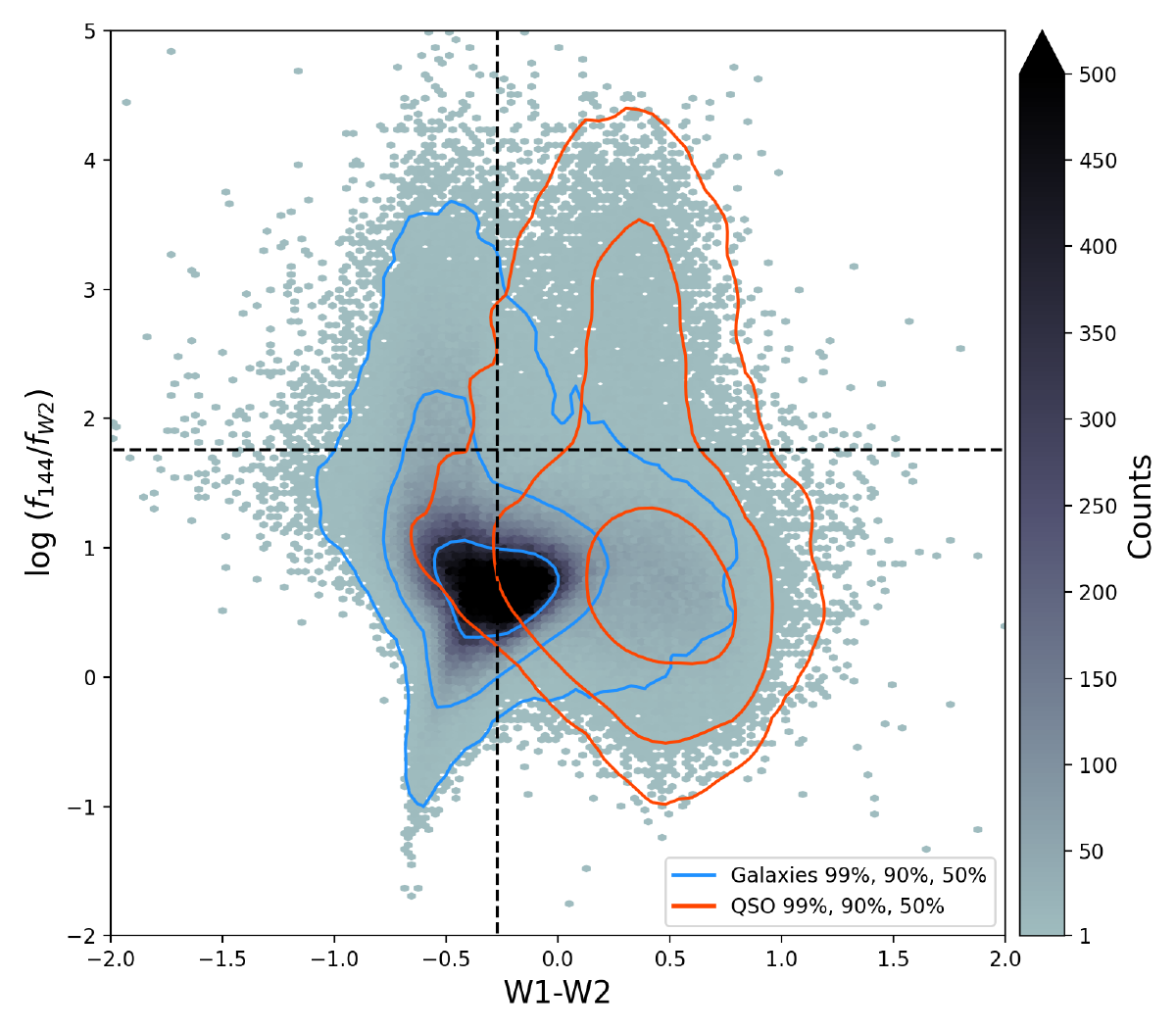}
  \caption{Radio-loudness parameter against WISE W1-W2 colors. The vertical dashed line in the right plot divides sources between blue and red W1-W2 colors at W1-W2 = -0.27. In both figures, the horizontal dashed lines divide sources between radio-quiet and radio-loud at $\text{log} \, (f_{144 \, MHz}$/$f_{W2}) = 1.76$. The color maps show the density distribution of all sources, the orange contours show the distribution of QSOs and the blue contours show the distribution of galaxies with r $<$ 21. Each contour contains 99\%, 90\% and 50\% of sources inside.}\label{fig:WISE_radio_loudness}
\end{figure*}

\subsection{IR properties}

\begin{figure*}[!t]
  \centering
  \includegraphics[width=1\textwidth]{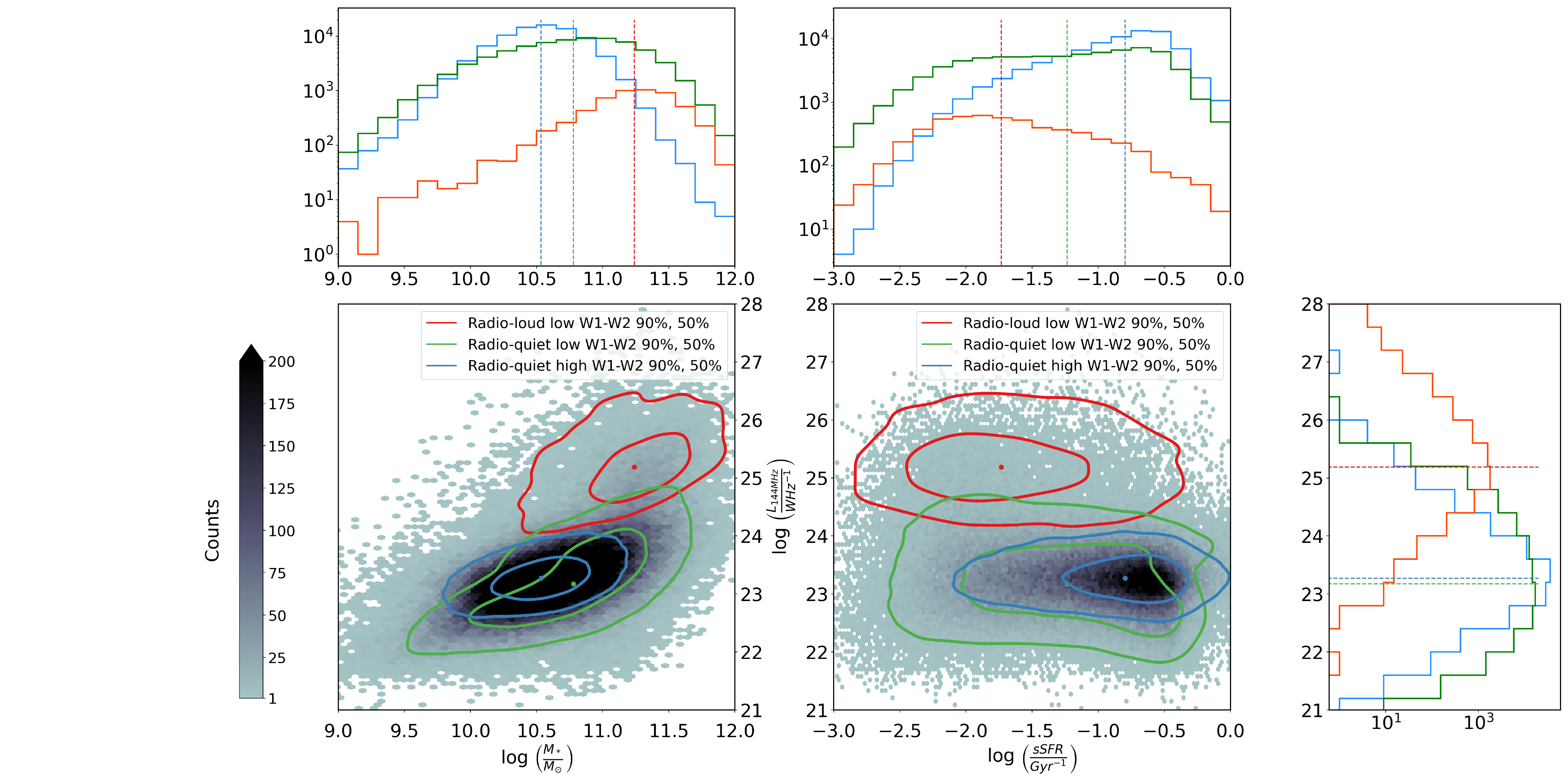}
  \caption{Left: radio luminosity against stellar mass of galaxies. Right: radio luminosity against specific star formation rate (sSFR) averaged over the last 100 Myrs. The color map shows the density distribution of all galaxies. The blue contour contains 90\% and 50\% of all radio-quiet high IR color galaxies, the green contour contains 90\% and 50\% of all radio-quiet low IR color galaxies and the orange contour contains 90\% and 50\% of all the radio-loud galaxies.}\label{fig:stellar_mass_Radio_galaxies}
\end{figure*}

The IR emission provides a powerful diagnostic to distinguish star-forming galaxies from radio-loud AGN. In the radio–IR plane (left panel of Fig.~\ref{fig:radio_WISE_class_comparison}), most galaxies follow the well-known IR–radio relation \citep{Helou1985}, consistent with radio emission from star formation activity (see Fernández Gil et~al., in prep., for a discussion of the SFR-radio relation in J-HERTz galaxies). QSOs, on the other hand, occupy the brightest IR region, characterized by strong dust emission. Among them, only a fraction are radio loud, showing radio excess above the IR–radio relation ($L_\mathrm{144 MHz} \gtrsim 10^{25}\, \rm{W\,Hz^{-1}}$) together with radio-loud galaxies, while most QSOs are radio quiet and overlap in luminosity with the star-forming galaxy population.

The WISE color–color diagram (Fig.~\ref{fig:radio_WISE_class_comparison}, right) further distinguishes these populations. The W1-W2 color traces hot dust heated by AGN activity, while W2-W3 traces warm dust from star-formation. Passive galaxies occupy the bluer region of the diagram, as they show little IR emission in either color, whereas star-forming galaxies show redder W2–W3 values, and AGN/QSOs have the redder W1–W2 colors. One of the main parameters used by BANNJOS to identify QSOs is indeed W1-W2 (e.g. Fig.~\ref{fig:SED_example_1} for a QSO with red W1-W2 color; Fig.~\ref{fig:SED_example_2} for a galaxy with blue W1-W2 color). A similar WISE color–color diagram is shown in \citet{Hardcastle2025}, where star-forming galaxies, radio-excess AGN, and spectroscopically confirmed QSOs occupy regions consistent with those found in our sample, confirming that mid-IR colors are effective in distinguish between different source types.

The radio-loudness versus W1–W2 color plane (Fig.~\ref{fig:WISE_radio_loudness}) highlights these differences. It is divided in 4 regions by two dashed lines: the radio-loud/radio-quiet threshold defined earlier at $\log\,(f_\mathrm{144\,MHz}$/$f_{W2}) = 1.76$, and a blue/red division at W1-W2 = $-0.27$, obtained by fitting two components to the radio-loud galaxies and QSOs, which are separated by their W1-W2 colors. Radio-loud galaxies exhibit relatively blue W1–W2 colors, consistent with LERGs or optically inactive radio galaxies, disfavoring scenarios involving dusty, obscured Type 2 QSOs \citep{Gurkan2014,Mingo2016}. QSOs populate mainly the reddest colors. Most radio-quiet galaxies have blue W1-W2 colors, but a subset shows higher IR colors comparable to those of QSOs. Approximately 48\% of the galaxies with W1-W2 $>$ 0.25 and available spectra are classified as QSOs in SDSS, suggesting that many of them are likely identified as galaxies due to their resolved or extended morphologies. Alternatively, some could be starburst galaxies with intense star formation heating their dust, or systems dominated by older stellar populations that emit strongly in the near-IR. These scenarios will be explored further in the next section.

\begin{figure*}[!t]
  \centering
  \includegraphics[width=0.8\textwidth]{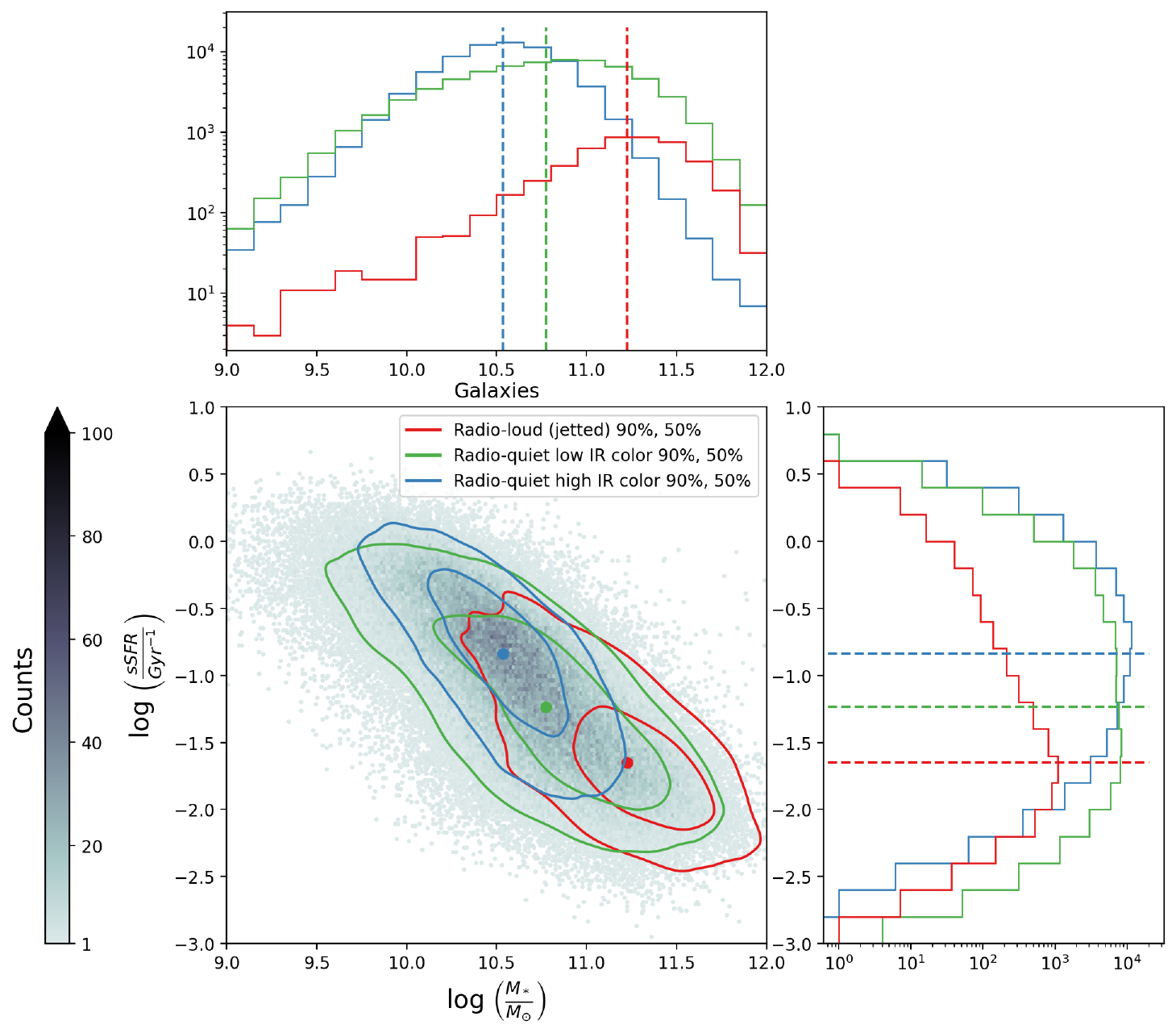}
  \caption{Specific star formation rate (sSFR) averaged over the last 100\,Myrs against stellar mass of galaxies. The color map shows the density distribution of all galaxies. The blue contour contains 90\% and 50\% of all radio-quiet high IR color galaxies, the green contour contains 90\% and 50\% of all radio-quiet low IR color galaxies and the orange contour contains 90\% and 50\% of all the radio-loud galaxies.}\label{fig:stellar_mass_SFR_galaxies}
\end{figure*}

\subsection{Stellar masses and star formation rates of galaxies}
To explore host galaxy properties, we divide the sample into three groups based on their distribution in Fig.~\ref{fig:WISE_radio_loudness}: radio-loud galaxies, radio-quiet galaxies with blue IR colors, and radio-quiet galaxies with red IR colors. Figure~\ref{fig:stellar_mass_Radio_galaxies} shows the relationship between radio luminosity, stellar mass and sSFR for these groups. Radio-quiet galaxies follow a stellar mass–radio luminosity relation similar to the radio-IR correlation, as the IR traces the older stellar population that dominates the stellar mass when no AGN is present. On the other hand, the radio luminosity in radio-loud galaxies is independent of the stellar mass \citep{Hardcastle2023}. Both radio-quiet galaxy populations have similar sSFR values. As mentioned above, red W1-W2 colors observed in some radio-quiet galaxies may result from nuclear activity in a spatially resolved galaxy that was not classified as a QSO by BANNJOS, although some of these sources also reach high sSFR values consistent with starbursts ($12$\% have sSFR $> 0.5\,\mathrm{Gyr}^{-1}$).

\begin{table*}[!t]
\centering
\caption{Median host properties of the three galaxy groups. Uncertainties correspond to the 16th and 84th percentiles. The sSFR values correspond to the 1$\sigma$ dispersion.}
\label{tab:galaxy_groups}
\begin{tabular}{lccc}
\hline\hline
Galaxy type & $\log(M_*/M_\odot)$ & $\log(L_{144\,\mathrm{MHz}}/\mathrm{W\,Hz^{-1}})$ & sSFR [$\mathrm{Gyr^{-1}}$] \\
\hline
Radio-quiet, W1-W2 $> 0.27$ & $10.53^{+0.30}_{-0.33}$ & $23.28^{+0.34}_{-0.30}$ & 0.04–0.42 \\
Radio-quiet, W1-W2 $< 0.27$ & $10.78^{+0.44}_{-0.56}$ & $23.20^{+0.74}_{-0.63}$ & 0.02–0.25 \\
Radio-loud [$\log\,(f_\mathrm{144\,MHz}$/$f_{W2}) > 1.76$] & $11.23^{+0.29}_{-0.39}$ & $25.19^{+0.52}_{-0.45}$ & 0.01–0.07 \\
\hline
\end{tabular}
\end{table*}

The stellar mass, radio luminosity, and sSFR of each group are summarized in Table~\ref{tab:galaxy_groups} and illustrated in Fig.~\ref{fig:stellar_mass_SFR_galaxies}. Radio-quiet galaxies with red IR colors are generally less massive and exhibit enhanced sSFRs, including both starbursts and possibly some cases with AGN contamination, representing the youngest systems in the sequence. Radio-quiet galaxies with blue IR colors show intermediate stellar masses and sSFRs, consistent with systems in transition from active star formation to quiescence, where the growing central black hole may already begin to influence star formation. Finally, radio-loud galaxies are the most massive and have the lowest sSFRs, consistent with quenched star formation in passive, typically elliptical hosts. The overall distribution supports an evolutionary connection between these groups, in which star-forming galaxies evolve through intermediate, partially quenched stages toward a radio-loud phase dominated by low-excitation AGN activity.

\subsection{Stranger things}\label{sec:Stranger_things}

\begin{figure*}[!t]
  \centering
  \includegraphics[width=0.8\textwidth]{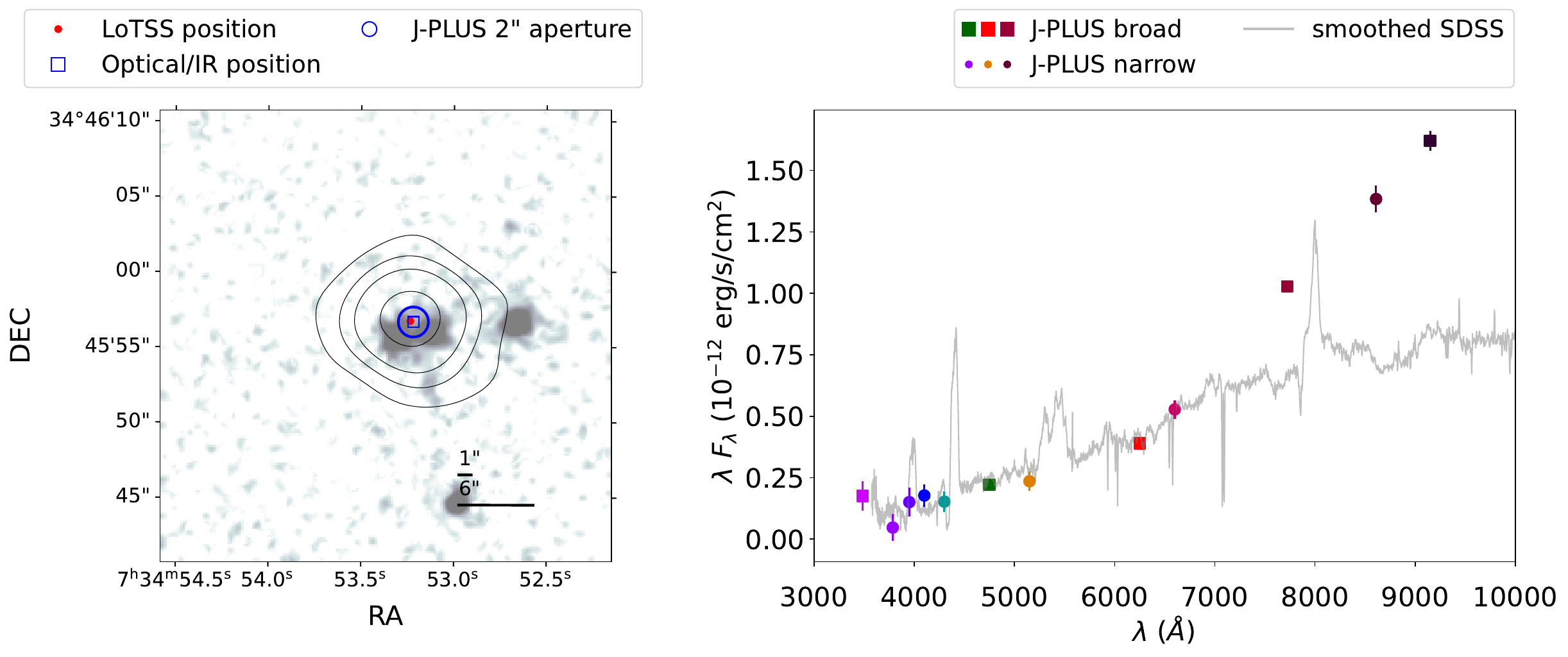}
  \caption{Two objects in very close proximity. Left panel: optical image from J-PLUS in gray scale, with the J-PLUS and LoTSS positions (blue square and red dot respectively), and the radio contours (black lines) from LoTSS drawn at 1, 2, 3, 5 and 7$\sigma$. The blue circle is the J-PLUS 2 arcsec aperture selected to match the size of the SDSS BOSS fiber. The two central objects are a star and a QSO but J-PLUS \texttt{AUTO} photometry blends them into a single identification with \texttt{TILE\_ID} = 87749 and \texttt{NUMBER} = 19089. Right panel: 2 arcsec aperture fluxes and error bars measured in the 12 J-PLUS filters as the colored symbols (squares for broad and circles for narrow bands, some error bars are smaller than the symbol size) and their SDSS DR18 \citep{Almeida2023} spectra in gray, smoothed and re-scaled to the J-PLUS flux in $r$. }\label{fig:close_proximity_objects}
\end{figure*}

Here we outline a few potential issues that can arise with the data. 

\begin{itemize}
     
\item First of all, given that this is a multi-wavelength catalog that combines positions from multiple surveys with different resolutions, there are a few cases where the cross-matches may not be correct. 
The bigger radius used in the J-PLUS/catWISE/SDSS cross-match and the smaller resolution of catWISE can, in some borderline cases, associate the optical/IR position to the wrong J-PLUS object. This can be mitigated if the user filters objects with high \texttt{AngDist} (e.g. $>$3 arcsec). Another problematic case is when several J-PLUS objects are within the 5 arcsec cross-match radius with the optical/IR position. As we only include the closest, in some cases the classification of the object, the redshift and the photometry may be incorrectly associated or partly contaminated by the other objects. Sometimes the objects are so close together that WISE and even J-PLUS cannot resolve them. We show an example in the left panel of Fig.~\ref{fig:close_proximity_objects}. The two central objects are very close to each other and are blended in J-PLUS. SDSS classifies them as a star and a QSO. BANNJOS classifies the blended object as a star, but the optical position is closer to the QSO and the cross-matched gets the spectroscopic redshift of the QSO. In the right panel, the 3 arcsec flux in the red wavelengths may be contaminated by the blended object. These objects are flagged in the catalog with the \texttt{close\_objects} flag. 

\item Additionally, the BANNJOS classifications use the cross-matched {\it Gaia} data to measure parallaxes of the J-PLUS objects. Sometimes, the same object is detected as several components in {\it Gaia} (maybe different regions of the same galaxy). This will make BANNJOS classify the same object several times with different mean probabilities. Most of the time, the variations are very slight and do not change the classification of the object (e.g., it is always classified as a galaxy with mean $>$ 0.9). The catalog includes the different mean probabilities for the same object duplicating the rows of all columns except the BANNJOS values, where the different values will be included. We mark these rows with the \texttt{ambiguous\_classification} flag.

\item Another potential issue may be that, even if the cross-matches are correct, the BANNJOS classification of the object does not match the spectroscopic classification of SDSS or DESI and thus the redshift is incompatible. BANNJOS uses parallaxes while SDSS does not. This problem especially affects QSO and white dwarfs, which can be mistaken at first glance, but galaxies can also be affected. For example, a few of the objects BANNJOS classified as galaxies have a much lower SDSS spec-z (compatible with zero) than the J-PLUS photo-z because their SDSS classification is a star. When adjusting their stellar masses to this spectroscopic redshift, the values become very small. All these cases are flagged with \texttt{incompatible\_spec\_class}. Additionally, sometimes the spectroscopic redshift from SDSS or DESI is incorrect due to a line-misidentification or a source being too weak. In the cases where the spec-z is above 5, or the difference between the photo-z and spec-z from SDSS/DESI is more than 0.2, we flag them with the \texttt{unreliable\_spec\_z}.

\item Finally, there are many objects classified as stars with associated radio emission that could be coming from other nearby or undetected objects. A search within the general J-PLUS catalog identifies 6,391 stars exhibiting extended radio emission with a nearby object that is not classified as a star. We perform some filtering criteria explained in Section \ref{sect:Stars} to find the most reliable radio star candidates, which we flag with \texttt{radio\_star\_candidate} = `reliable', while the rest have `unreliable'. 

\end{itemize}

 All these flags help the user decide which cases to filter in their case study. Many of these flags can overlap as well, e.g., an object classified as a star with extended radio emission, very close to another QSO ($<$ 5 arcsec) and with a SDSS spectral classification of QSO will have the \texttt{close\_objects}, \texttt{candidate\_radio\_star} = `unreliable' and \texttt{incompatible\_spec\_class} flags.

\section{Conclusions and future work}

In this work we introduce \textsc{J-HERTz} (J-PLUS Heritage Exploration of Radio Targets at $z < 5$), a comprehensive multi-wavelength catalog combining photometry in 12 optical broad and narrow bands from J-PLUS DR3, 4 IR bands from \textit{WISE}, and deep low-frequency radio data from LoTSS DR2 at $144\, \rm{MHz}$. The catalog includes 489,897 objects distributed over 2100\,deg$^2$ of the northern sky, with robust Bayesian neural network classifications for 390,000 galaxies (320,000 with r $<$ 21), 31,000 quasars, and 20,000 stars. J-HERTz constitutes a new resource for the study of galaxy and AGN co-evolution, the radio properties of stars and QSOs, and the long-term impact of radio jets on galaxy growth, especially in the low-redshift Universe. It enables population-wide statistical studies with higher precision and completeness than previously available over this area of the sky. 

Significantly improved photometric redshifts are provided for 235,000 galaxies, exploiting the combination of deep broad-band photometry and narrow band observations from the two surveys. These enhanced redshift estimates demonstrate greater accuracy and precision compared to prior estimates from J-PLUS DR3 and LoTSS DR2.  The radio-to-IR ratio based on LOFAR $144\, \rm{MHz}$ and WISE W2 ($4.6\, \rm{\micron}$) flux measurements is introduced as a new radio-loudness indicator, discussed in this work for the first time. Additionally, stellar masses and SFRs based on SED-fitting using narrow-band photometry are provided for the galaxy population. 

An interesting result is the identification of 831 possible radio star candidates showing compact low-frequency radio emission. 
Our filtering process discards the most probable false associations, and the likelihood-ratio analysis using {\it Gaia} DR3 indicates that the number of matches between optical and radio bands is far above what would be expected from random cross-matches. Determining the nature of the radio emission in these objects would require additional data, nevertheless previous studies in the literature suggest that the radio continuum emission may be due to coronal activity or due to coupling between the magnetospheres of the star and a massive orbiting exoplanet.

QSOs show a bimodal distribution in radio-loudness with an approximate division at $\text{log} \, (f_{144}/f_{W2}) = 1.76$, adopted here as threshold between radio-quiet and radio-loud sources (lower right panel of Fig. \ref{fig:radio_properties}). At low redshifts (z $<$ 0.2) 5\% of QSOs are radio-loud and $\sim$95\% are radio-quiet, however the radio-loud fraction increases to $\sim$12\% at z $\sim$ 1. These QSOs are brighter in WISE W2 and show redder W1-W2 colors than galaxies, due to hot dust associated to AGN activity (Fig. \ref{fig:radio_WISE_class_comparison}). In fact, W1-W2 efficiently separates radio-loud QSOs from radio-galaxies, with a threshold around W1-W2 = -0.27. While most radio-quiet QSOs are above this value, radio-quiet galaxies are spread across a wide range in W1-W2 ($\sim$160,000 with W1-W2 $<$ -0.27 vs. $\sim$138,000 with W1-W2 $>$ -0.27 at r $<$ 21, right panel of Fig. \ref{fig:WISE_radio_loudness}).

Most galaxies are radio-quiet and lie in the IR–radio relation (left panel of Fig. \ref{fig:radio_WISE_class_comparison}). The majority of radio-quiet galaxies with blue W1-W2 colors are passive galaxies, showing intermediate masses ($10.78^{+0.44}_{-0.56} \, \text{log} \, \left( \frac{M_*}{M_{\odot}} \right)$) and low sSFRs (0.01-0.25 $\text{Gyr}^{-1}$). Radio-quiet galaxies with red W1-W2 colors have tentatively lower masses ($10.53^{+0.30}_{-0.33} \, \text{log} \, \left( \frac{M_*}{M_{\odot}} \right)$) and highest sSFRs (0.04-0.34 $\text{Gyr}^{-1}$), reaching values compatible with starburst galaxies ($6$\% of them have values larger than 0.5 $\text{Gyr}^{-1}$). This suggests that the red IR colors are associated with ongoing starbursts activity. However, around half of galaxies with red IR colors (W1-W2 $>$ 0.25) and available SDSS spectroscopy are classified as QSOs by \citet{Almeida2023} and might be AGN with a resolved host.

Notably, there are $\sim$14,000 radio-loud galaxies (at r $<$ 21) with no clear optical or IR signs of nuclear activity. Most of these galaxies have extended radio emission at low frequencies likely tracing past jet activity. These are also the most massive ($11.24^{+0.28}_{-0.38} \, \text{log} \, \left( \frac{M_*}{M_{\odot}} \right)$) and most quenched galaxies (1$\sigma$ spread of the sSFR of 0.01-0.09 $\text{Gyr}^{-1}$, Fig. \ref{fig:stellar_mass_SFR_galaxies}). About half of them are compatible with low excitation radio galaxies (LERGs) according to the spectral classification provided by \citet{Drake2024}, while we estimate that $\gtrsim$20\% could be optically quiescent radio galaxies with residual or absent nuclear activity, and extended relic radio lobes at low frequencies being the only evidence of the past activity. Further studies of their radio morphologies, spectral indices, surface brightness and core prominences are necessary to completely characterize these sources.

The J-HERTz catalog is available as part of the J-PLUS database. In the near future, the catalog will be updated with the upcoming J-PLUS DR4, adding $\sim$2,000 $\rm{deg^{2}}$ with improved photo-z estimates (Hernán-Caballero in prep.) We will use J-HERTz to select a clean sample of star-forming galaxies and compare SFRs and radio luminosities for this population, to characterize the radio-SFR correlation (Fernández Gil et al. in prep). This will allow us to determine SFRs of radio detected galaxies missing precise optical photometry and also establish a selection criteria between star-forming and AGN radio galaxies. Additional radio surveys at higher frequencies such as the Arecibo Legacy Fast ALFA \ion{H}{1} \citep[ALFALFA;][]{Haynes2018} or the first data release of the APERture Tile In Focus array \citep[Apertif;][]{vanCappellen2022} will trace the HI emission in these sources. Finally, we aim to further characterize the host galactic properties, estimate the black hole accretion rates using emission line ratios thanks to the precise narrow band photometry of J-PLUS, and compare them with past AGN activity seen in older and extended jets detected with low-frequency radio. 

\begin{acknowledgements}
We thank the anonymous referee for their constructive comments and suggestions, which helped improve the quality of this manuscript. DFG acknowledges financial support by the Predoctoral Gobierno de Aragón fellowship 2023-2027. DFG, JAFO, FAB, AHC, and ALC acknowledge financial support by the Spanish Ministry of Science and Innovation (MCIN/AEI/10.13039/501100011033), by ``ERDF A way of making Europe'' and by ``European Union NextGenerationEU/PRTR'' through the grants PID2021-124918NB-C44 and CNS2023-145339. JAFO acknowledges MCIN and the European Union -- NextGenerationEU through the Recovery and Resilience Facility project ICTS-MRR-2021-03-CEFCA. AdP acknowledges financial support from the Severo Ochoa grant CEX2021-001131-S funded by MCIN/AEI/ 10.13039/501100011033. 
Based on observations made with the JAST80 telescope and T80Cam camera for the J-PLUS project at the Observatorio Astrof\'{\i}sico de Javalambre (OAJ), in Teruel, owned, managed, and operated by the Centro de Estudios de F\'{\i}sica del  Cosmos de Arag\'on (CEFCA). We acknowledge the OAJ Data Processing and Archiving Unit (UPAD; Cristobal-Hornillos et al. 2012) for reducing the OAJ data used in this work.
Funding for the J-PLUS Project has been provided by the Governments of Spain and Arag\'on through the Fondo de Inversiones de Teruel; the Aragonese Government through the Research Groups E96, E103, E16\_17R, E16\_20R, and E16\_23R; the Spanish Ministry of Science and Innovation (MCIN/AEI/10.13039/501100011033 y FEDER, Una manera de hacer Europa) with grants PID2021-124918NB-C41, PID2021-124918NB-C42, PID2021-124918NA-C43, and PID2021-124918NB-C44; the Spanish Ministry of Science, Innovation and Universities (MCIU/AEI/FEDER, UE) with grants PGC2018-097585-B-C21 and PGC2018-097585-B-C22; the Spanish Ministry of Economy and Competitiveness (MINECO) under AYA2015-66211-C2-1-P, AYA2015-66211-C2-2, AYA2012-30789, and ICTS-2009-14; and European FEDER funding (FCDD10-4E-867, FCDD13-4E-2685). The Brazilian agencies FINEP, FAPESP, and the National Observatory of Brazil have also contributed to this project. 
LOFAR is the Low Frequency Array, designed and constructed by ASTRON. It has observing, data processing, and data storage facilities in several countries, which are owned by various parties (each with their own funding sources), and which are collectively operated by the ILT foundation under a joint scientific policy. The ILT resources have benefited from the following recent major funding sources: CNRS-INSU, Observatoire de Paris and Université d’Orléans, France; BMBF, MIWF-NRW, MPG, Germany; Science Foundation Ireland (SFI), Department of Business, Enterprise and Innovation (DBEI), Ireland; NWO, The Netherlands; The Science and Technology Facilities Council, UK; Ministry of Science and Higher Education, Poland; The Istituto Nazionale di Astrofisica (INAF), Italy. 
This publication makes use of data products from the Wide-field Infrared Survey Explorer, which is a joint project of the University of California, Los Angeles, and the Jet Propulsion Laboratory/California Institute of Technology, and NEOWISE, which is a project of the Jet Propulsion Laboratory/California Institute of Technology. WISE and NEOWISE are funded by the National Aeronautics and Space Administration.
\end{acknowledgements}

\bibliography{jhertz}{}

@ARTICLE{Abolfathi2018,
       author = {{Abolfathi}, Bela and {Aguado}, D.~S. and {Aguilar}, Gabriela and {Allende Prieto}, Carlos and {Almeida}, Andres and {Ananna}, Tonima Tasnim and {Anders}, Friedrich and {Anderson}, Scott F. and {Andrews}, Brett H. and {Anguiano}, Borja and {Arag{\'o}n-Salamanca}, Alfonso and {Argudo-Fern{\'a}ndez}, Maria and {Armengaud}, Eric and {Ata}, Metin and {Aubourg}, Eric and {Avila-Reese}, Vladimir and {Badenes}, Carles and {Bailey}, Stephen and {Balland}, Christophe and {Barger}, Kathleen A. and {Barrera-Ballesteros}, Jorge and {Bartosz}, Curtis and {Bastien}, Fabienne and {Bates}, Dominic and {Baumgarten}, Falk and {Bautista}, Julian and {Beaton}, Rachael and {Beers}, Timothy C. and {Belfiore}, Francesco and {Bender}, Chad F. and {Bernardi}, Mariangela and {Bershady}, Matthew A. and {Beutler}, Florian and {Bird}, Jonathan C. and {Bizyaev}, Dmitry and {Blanc}, Guillermo A. and {Blanton}, Michael R. and {Blomqvist}, Michael and {Bolton}, Adam S. and {Boquien}, M{\'e}d{\'e}ric and {Borissova}, Jura and {Bovy}, Jo and {Bradna Diaz}, Christian Andres and {Brandt}, William Nielsen and {Brinkmann}, Jonathan and {Brownstein}, Joel R. and {Bundy}, Kevin and {Burgasser}, Adam J. and {Burtin}, Etienne and {Busca}, Nicol{\'a}s G. and {Ca{\~n}as}, Caleb I. and {Cano-D{\'\i}az}, Mariana and {Cappellari}, Michele and {Carrera}, Ricardo and {Casey}, Andrew R. and {Cervantes Sodi}, Bernardo and {Chen}, Yanping and {Cherinka}, Brian and {Chiappini}, Cristina and {Choi}, Peter Doohyun and {Chojnowski}, Drew and {Chuang}, Chia-Hsun and {Chung}, Haeun and {Clerc}, Nicolas and {Cohen}, Roger E. and {Comerford}, Julia M. and {Comparat}, Johan and {Correa do Nascimento}, Janaina and {da Costa}, Luiz and {Cousinou}, Marie-Claude and {Covey}, Kevin and {Crane}, Jeffrey D. and {Cruz-Gonzalez}, Irene and {Cunha}, Katia and {da Silva Ilha}, Gabriele and {Damke}, Guillermo J. and {Darling}, Jeremy and {Davidson}, Jr., James W. and {Dawson}, Kyle and {de Icaza Lizaola}, Miguel Angel C. and {de la Macorra}, Axel and {de la Torre}, Sylvain and {De Lee}, Nathan and {de Sainte Agathe}, Victoria and {Deconto Machado}, Alice and {Dell'Agli}, Flavia and {Delubac}, Timoth{\'e}e and {Diamond-Stanic}, Aleksandar M. and {Donor}, John and {Downes}, Juan Jos{\'e} and {Drory}, Niv and {du Mas des Bourboux}, H{\'e}lion and {Duckworth}, Christopher J. and {Dwelly}, Tom and {Dyer}, Jamie and {Ebelke}, Garrett and {Davis Eigenbrot}, Arthur and {Eisenstein}, Daniel J. and {Elsworth}, Yvonne P. and {Emsellem}, Eric and {Eracleous}, Michael and {Erfanianfar}, Ghazaleh and {Escoffier}, Stephanie and {Fan}, Xiaohui and {Fern{\'a}ndez Alvar}, Emma and {Fernandez-Trincado}, J.~G. and {Fernando Cirolini}, Rafael and {Feuillet}, Diane and {Finoguenov}, Alexis and {Fleming}, Scott W. and {Font-Ribera}, Andreu and {Freischlad}, Gordon and {Frinchaboy}, Peter and {Fu}, Hai and {G{\'o}mez Maqueo Chew}, Yilen and {Galbany}, Llu{\'\i}s and {Garc{\'\i}a P{\'e}rez}, Ana E. and {Garcia-Dias}, R. and {Garc{\'\i}a-Hern{\'a}ndez}, D.~A. and {Garma Oehmichen}, Luis Alberto and {Gaulme}, Patrick and {Gelfand}, Joseph and {Gil-Mar{\'\i}n}, H{\'e}ctor and {Gillespie}, Bruce A. and {Goddard}, Daniel and {Gonz{\'a}lez Hern{\'a}ndez}, Jonay I. and {Gonzalez-Perez}, Violeta and {Grabowski}, Kathleen and {Green}, Paul J. and {Grier}, Catherine J. and {Gueguen}, Alain and {Guo}, Hong and {Guy}, Julien and {Hagen}, Alex and {Hall}, Patrick and {Harding}, Paul and {Hasselquist}, Sten and {Hawley}, Suzanne and {Hayes}, Christian R. and {Hearty}, Fred and {Hekker}, Saskia and {Hernandez}, Jesus and {Hernandez Toledo}, Hector and {Hogg}, David W. and {Holley-Bockelmann}, Kelly and {Holtzman}, Jon A. and {Hou}, Jiamin and {Hsieh}, Bau-Ching and {Hunt}, Jason A.~S. and {Hutchinson}, Timothy A. and {Hwang}, Ho Seong and {Jimenez Angel}, Camilo Eduardo and {Johnson}, Jennifer A. and {Jones}, Amy and {J{\"o}nsson}, Henrik and {Jullo}, Eric and {Khan}, Fahim Sakil and {Kinemuchi}, Karen and {Kirkby}, David and {Kirkpatrick}, IV, Charles C. and {Kitaura}, Francisco-Shu and {Knapp}, Gillian R. and {Kneib}, Jean-Paul and {Kollmeier}, Juna A. and {Lacerna}, Ivan and {Lane}, Richard R. and {Lang}, Dustin and {Law}, David R. and {Le Goff}, Jean-Marc and {Lee}, Young-Bae and {Li}, Hongyu and {Li}, Cheng and {Lian}, Jianhui and {Liang}, Yu and {Lima}, Marcos and {Lin}, Lihwai and {Long}, Dan and {Lucatello}, Sara and {Lundgren}, Britt and {Mackereth}, J. Ted and {MacLeod}, Chelsea L. and {Mahadevan}, Suvrath and {Maia}, Marcio Antonio Geimba and {Majewski}, Steven and {Manchado}, Arturo and {Maraston}, Claudia and {Mariappan}, Vivek and {Marques-Chaves}, Rui and {Masseron}, Thomas and {Masters}, Karen L. and {McDermid}, Richard M. and {McGreer}, Ian D. and {Melendez}, Matthew and {Meneses-Goytia}, Sofia and {Merloni}, Andrea and {Merrifield}, Michael R. and {Meszaros}, Szabolcs and {Meza}, Andres and {Minchev}, Ivan and {Minniti}, Dante},
        title = "{The Fourteenth Data Release of the Sloan Digital Sky Survey: First Spectroscopic Data from the Extended Baryon Oscillation Spectroscopic Survey and from the Second Phase of the Apache Point Observatory Galactic Evolution Experiment}",
      journal = {\apjs},
         year = 2018,
        month = apr,
       volume = {235},
       number = {2},
          eid = {42},
        pages = {42},
          doi = {10.3847/1538-4365/aa9e8a},
archivePrefix = {arXiv},
       eprint = {1707.09322},
 primaryClass = {astro-ph.GA},
       adsurl = {https://ui.adsabs.harvard.edu/abs/2018ApJS..235...42A}
}

@ARTICLE{Ahumada2020,
       author = {{Ahumada}, Romina and {Allende Prieto}, Carlos and {Almeida}, Andr{\'e}s and {Anders}, Friedrich and {Anderson}, Scott F. and {Andrews}, Brett H. and {Anguiano}, Borja and {Arcodia}, Riccardo and {Armengaud}, Eric and {Aubert}, Marie and {Avila}, Santiago and {Avila-Reese}, Vladimir and {Badenes}, Carles and {Balland}, Christophe and {Barger}, Kat and {Barrera-Ballesteros}, Jorge K. and {Basu}, Sarbani and {Bautista}, Julian and {Beaton}, Rachael L. and {Beers}, Timothy C. and {Benavides}, B. Izamar T. and {Bender}, Chad F. and {Bernardi}, Mariangela and {Bershady}, Matthew and {Beutler}, Florian and {Bidin}, Christian Moni and {Bird}, Jonathan and {Bizyaev}, Dmitry and {Blanc}, Guillermo A. and {Blanton}, Michael R. and {Boquien}, M{\'e}d{\'e}ric and {Borissova}, Jura and {Bovy}, Jo and {Brandt}, W.~N. and {Brinkmann}, Jonathan and {Brownstein}, Joel R. and {Bundy}, Kevin and {Bureau}, Martin and {Burgasser}, Adam and {Burtin}, Etienne and {Cano-D{\'\i}az}, Mariana and {Capasso}, Raffaella and {Cappellari}, Michele and {Carrera}, Ricardo and {Chabanier}, Sol{\`e}ne and {Chaplin}, William and {Chapman}, Michael and {Cherinka}, Brian and {Chiappini}, Cristina and {Doohyun Choi}, Peter and {Chojnowski}, S. Drew and {Chung}, Haeun and {Clerc}, Nicolas and {Coffey}, Damien and {Comerford}, Julia M. and {Comparat}, Johan and {da Costa}, Luiz and {Cousinou}, Marie-Claude and {Covey}, Kevin and {Crane}, Jeffrey D. and {Cunha}, Katia and {Ilha}, Gabriele da Silva and {Dai}, Yu Sophia and {Damsted}, Sanna B. and {Darling}, Jeremy and {Davidson}, Jr., James W. and {Davies}, Roger and {Dawson}, Kyle and {De}, Nikhil and {de la Macorra}, Axel and {De Lee}, Nathan and {Queiroz}, Anna B{\'a}rbara de Andrade and {Deconto Machado}, Alice and {de la Torre}, Sylvain and {Dell'Agli}, Flavia and {du Mas des Bourboux}, H{\'e}lion and {Diamond-Stanic}, Aleksandar M. and {Dillon}, Sean and {Donor}, John and {Drory}, Niv and {Duckworth}, Chris and {Dwelly}, Tom and {Ebelke}, Garrett and {Eftekharzadeh}, Sarah and {Davis Eigenbrot}, Arthur and {Elsworth}, Yvonne P. and {Eracleous}, Mike and {Erfanianfar}, Ghazaleh and {Escoffier}, Stephanie and {Fan}, Xiaohui and {Farr}, Emily and {Fern{\'a}ndez-Trincado}, Jos{\'e} G. and {Feuillet}, Diane and {Finoguenov}, Alexis and {Fofie}, Patricia and {Fraser-McKelvie}, Amelia and {Frinchaboy}, Peter M. and {Fromenteau}, Sebastien and {Fu}, Hai and {Galbany}, Llu{\'\i}s and {Garcia}, Rafael A. and {Garc{\'\i}a-Hern{\'a}ndez}, D.~A. and {Garma Oehmichen}, Luis Alberto and {Ge}, Junqiang and {Geimba Maia}, Marcio Antonio and {Geisler}, Doug and {Gelfand}, Joseph and {Goddy}, Julian and {Gonzalez-Perez}, Violeta and {Grabowski}, Kathleen and {Green}, Paul and {Grier}, Catherine J. and {Guo}, Hong and {Guy}, Julien and {Harding}, Paul and {Hasselquist}, Sten and {Hawken}, Adam James and {Hayes}, Christian R. and {Hearty}, Fred and {Hekker}, S. and {Hogg}, David W. and {Holtzman}, Jon A. and {Horta}, Danny and {Hou}, Jiamin and {Hsieh}, Bau-Ching and {Huber}, Daniel and {Hunt}, Jason A.~S. and {Ider Chitham}, J. and {Imig}, Julie and {Jaber}, Mariana and {Jimenez Angel}, Camilo Eduardo and {Johnson}, Jennifer A. and {Jones}, Amy M. and {J{\"o}nsson}, Henrik and {Jullo}, Eric and {Kim}, Yerim and {Kinemuchi}, Karen and {Kirkpatrick}, IV, Charles C. and {Kite}, George W. and {Klaene}, Mark and {Kneib}, Jean-Paul and {Kollmeier}, Juna A. and {Kong}, Hui and {Kounkel}, Marina and {Krishnarao}, Dhanesh and {Lacerna}, Ivan and {Lan}, Ting-Wen and {Lane}, Richard R. and {Law}, David R. and {Le Goff}, Jean-Marc and {Leung}, Henry W. and {Lewis}, Hannah and {Li}, Cheng and {Lian}, Jianhui and {Lin}, Lihwai and {Long}, Dan and {Longa-Pe{\~n}a}, Pen{\'e}lope and {Lundgren}, Britt and {Lyke}, Brad W. and {Mackereth}, J. Ted and {MacLeod}, Chelsea L. and {Majewski}, Steven R. and {Manchado}, Arturo and {Maraston}, Claudia and {Martini}, Paul and {Masseron}, Thomas and {Masters}, Karen L. and {Mathur}, Savita and {McDermid}, Richard M. and {Merloni}, Andrea and {Merrifield}, Michael and {M{\'e}sz{\'a}ros}, Szabolcs and {Miglio}, Andrea and {Minniti}, Dante and {Minsley}, Rebecca and {Miyaji}, Takamitsu and {Mohammad}, Faizan Gohar and {Mosser}, Benoit and {Mueller}, Eva-Maria and {Muna}, Demitri and {Mu{\~n}oz-Guti{\'e}rrez}, Andrea and {Myers}, Adam D. and {Nadathur}, Seshadri and {Nair}, Preethi and {Nandra}, Kirpal and {Correa do Nascimento}, Janaina and {Nevin}, Rebecca Jean and {Newman}, Jeffrey A. and {Nidever}, David L. and {Nitschelm}, Christian and {Noterdaeme}, Pasquier and {O'Connell}, Julia E. and {Olmstead}, Matthew D. and {Oravetz}, Daniel and {Oravetz}, Audrey and {Osorio}, Yeisson and {Pace}, Zachary J. and {Padilla}, Nelson and {Palanque-Delabrouille}, Nathalie and {Palicio}, Pedro A.},
        title = "{The 16th Data Release of the Sloan Digital Sky Surveys: First Release from the APOGEE-2 Southern Survey and Full Release of eBOSS Spectra}",
      journal = {\apjs},
         year = 2020,
        month = jul,
       volume = {249},
       number = {1},
          eid = {3},
        pages = {3},
          doi = {10.3847/1538-4365/ab929e},
archivePrefix = {arXiv},
       eprint = {1912.02905},
 primaryClass = {astro-ph.GA},
       adsurl = {https://ui.adsabs.harvard.edu/abs/2020ApJS..249....3A}
}

@ARTICLE{Alam2015,
       author = {{Alam}, Shadab and {Albareti}, Franco D. and {Allende Prieto}, Carlos and {Anders}, F. and {Anderson}, Scott F. and {Anderton}, Timothy and {Andrews}, Brett H. and {Armengaud}, Eric and {Aubourg}, {\'E}ric and {Bailey}, Stephen and {Basu}, Sarbani and {Bautista}, Julian E. and {Beaton}, Rachael L. and {Beers}, Timothy C. and {Bender}, Chad F. and {Berlind}, Andreas A. and {Beutler}, Florian and {Bhardwaj}, Vaishali and {Bird}, Jonathan C. and {Bizyaev}, Dmitry and {Blake}, Cullen H. and {Blanton}, Michael R. and {Blomqvist}, Michael and {Bochanski}, John J. and {Bolton}, Adam S. and {Bovy}, Jo and {Shelden Bradley}, A. and {Brandt}, W.~N. and {Brauer}, D.~E. and {Brinkmann}, J. and {Brown}, Peter J. and {Brownstein}, Joel R. and {Burden}, Angela and {Burtin}, Etienne and {Busca}, Nicol{\'a}s G. and {Cai}, Zheng and {Capozzi}, Diego and {Carnero Rosell}, Aurelio and {Carr}, Michael A. and {Carrera}, Ricardo and {Chambers}, K.~C. and {Chaplin}, William James and {Chen}, Yen-Chi and {Chiappini}, Cristina and {Chojnowski}, S. Drew and {Chuang}, Chia-Hsun and {Clerc}, Nicolas and {Comparat}, Johan and {Covey}, Kevin and {Croft}, Rupert A.~C. and {Cuesta}, Antonio J. and {Cunha}, Katia and {da Costa}, Luiz N. and {Da Rio}, Nicola and {Davenport}, James R.~A. and {Dawson}, Kyle S. and {De Lee}, Nathan and {Delubac}, Timoth{\'e}e and {Deshpande}, Rohit and {Dhital}, Saurav and {Dutra-Ferreira}, Let{\'\i}cia and {Dwelly}, Tom and {Ealet}, Anne and {Ebelke}, Garrett L. and {Edmondson}, Edward M. and {Eisenstein}, Daniel J. and {Ellsworth}, Tristan and {Elsworth}, Yvonne and {Epstein}, Courtney R. and {Eracleous}, Michael and {Escoffier}, Stephanie and {Esposito}, Massimiliano and {Evans}, Michael L. and {Fan}, Xiaohui and {Fern{\'a}ndez-Alvar}, Emma and {Feuillet}, Diane and {Filiz Ak}, Nurten and {Finley}, Hayley and {Finoguenov}, Alexis and {Flaherty}, Kevin and {Fleming}, Scott W. and {Font-Ribera}, Andreu and {Foster}, Jonathan and {Frinchaboy}, Peter M. and {Galbraith-Frew}, J.~G. and {Garc{\'\i}a}, Rafael A. and {Garc{\'\i}a-Hern{\'a}ndez}, D.~A. and {Garc{\'\i}a P{\'e}rez}, Ana E. and {Gaulme}, Patrick and {Ge}, Jian and {G{\'e}nova-Santos}, R. and {Georgakakis}, A. and {Ghezzi}, Luan and {Gillespie}, Bruce A. and {Girardi}, L{\'e}o and {Goddard}, Daniel and {Gontcho}, Satya Gontcho A. and {Gonz{\'a}lez Hern{\'a}ndez}, Jonay I. and {Grebel}, Eva K. and {Green}, Paul J. and {Grieb}, Jan Niklas and {Grieves}, Nolan and {Gunn}, James E. and {Guo}, Hong and {Harding}, Paul and {Hasselquist}, Sten and {Hawley}, Suzanne L. and {Hayden}, Michael and {Hearty}, Fred R. and {Hekker}, Saskia and {Ho}, Shirley and {Hogg}, David W. and {Holley-Bockelmann}, Kelly and {Holtzman}, Jon A. and {Honscheid}, Klaus and {Huber}, Daniel and {Huehnerhoff}, Joseph and {Ivans}, Inese I. and {Jiang}, Linhua and {Johnson}, Jennifer A. and {Kinemuchi}, Karen and {Kirkby}, David and {Kitaura}, Francisco and {Klaene}, Mark A. and {Knapp}, Gillian R. and {Kneib}, Jean-Paul and {Koenig}, Xavier P. and {Lam}, Charles R. and {Lan}, Ting-Wen and {Lang}, Dustin and {Laurent}, Pierre and {Le Goff}, Jean-Marc and {Leauthaud}, Alexie and {Lee}, Khee-Gan and {Lee}, Young Sun and {Licquia}, Timothy C. and {Liu}, Jian and {Long}, Daniel C. and {L{\'o}pez-Corredoira}, Mart{\'\i}n and {Lorenzo-Oliveira}, Diego and {Lucatello}, Sara and {Lundgren}, Britt and {Lupton}, Robert H. and {Mack}, III, Claude E. and {Mahadevan}, Suvrath and {Maia}, Marcio A.~G. and {Majewski}, Steven R. and {Malanushenko}, Elena and {Malanushenko}, Viktor and {Manchado}, A. and {Manera}, Marc and {Mao}, Qingqing and {Maraston}, Claudia and {Marchwinski}, Robert C. and {Margala}, Daniel and {Martell}, Sarah L. and {Martig}, Marie and {Masters}, Karen L. and {Mathur}, Savita and {McBride}, Cameron K. and {McGehee}, Peregrine M. and {McGreer}, Ian D. and {McMahon}, Richard G. and {M{\'e}nard}, Brice and {Menzel}, Marie-Luise and {Merloni}, Andrea and {M{\'e}sz{\'a}ros}, Szabolcs and {Miller}, Adam A. and {Miralda-Escud{\'e}}, Jordi and {Miyatake}, Hironao and {Montero-Dorta}, Antonio D. and {More}, Surhud and {Morganson}, Eric and {Morice-Atkinson}, Xan and {Morrison}, Heather L. and {Mosser}, Ben{\^o}it and {Muna}, Demitri and {Myers}, Adam D. and {Nandra}, Kirpal and {Newman}, Jeffrey A. and {Neyrinck}, Mark and {Nguyen}, Duy Cuong and {Nichol}, Robert C. and {Nidever}, David L. and {Noterdaeme}, Pasquier and {Nuza}, Sebasti{\'a}n E. and {O'Connell}, Julia E. and {O'Connell}, Robert W. and {O'Connell}, Ross and {Ogando}, Ricardo L.~C. and {Olmstead}, Matthew D. and {Oravetz}, Audrey E. and {Oravetz}, Daniel J. and {Osumi}, Keisuke and {Owen}, Russell and {Padgett}, Deborah L. and {Padmanabhan}, Nikhil and {Paegert}, Martin and {Palanque-Delabrouille}, Nathalie and {Pan}, Kaike},
        title = "{The Eleventh and Twelfth Data Releases of the Sloan Digital Sky Survey: Final Data from SDSS-III}",
      journal = {\apjs},
         year = 2015,
        month = jul,
       volume = {219},
       number = {1},
          eid = {12},
        pages = {12},
          doi = {10.1088/0067-0049/219/1/12},
archivePrefix = {arXiv},
       eprint = {1501.00963},
 primaryClass = {astro-ph.IM},
       adsurl = {https://ui.adsabs.harvard.edu/abs/2015ApJS..219...12A}
}

@ARTICLE{Almeida2023,
       author = {{Almeida}, Andr{\'e}s and {Anderson}, Scott F. and {Argudo-Fern{\'a}ndez}, Maria and {Badenes}, Carles and {Barger}, Kat and {Barrera-Ballesteros}, Jorge K. and {Bender}, Chad F. and {Benitez}, Erika and {Besser}, Felipe and {Bird}, Jonathan C. and {Bizyaev}, Dmitry and {Blanton}, Michael R. and {Bochanski}, John and {Bovy}, Jo and {Brandt}, William Nielsen and {Brownstein}, Joel R. and {Buchner}, Johannes and {Bulbul}, Esra and {Burchett}, Joseph N. and {Cano D{\'\i}az}, Mariana and {Carlberg}, Joleen K. and {Casey}, Andrew R. and {Chandra}, Vedant and {Cherinka}, Brian and {Chiappini}, Cristina and {Coker}, Abigail A. and {Comparat}, Johan and {Conroy}, Charlie and {Contardo}, Gabriella and {Cortes}, Arlin and {Covey}, Kevin and {Crane}, Jeffrey D. and {Cunha}, Katia and {Dabbieri}, Collin and {Davidson}, James W. and {Davis}, Megan C. and {de Andrade Queiroz}, Anna Barbara and {De Lee}, Nathan and {M{\'e}ndez Delgado}, Jos{\'e} Eduardo and {Demasi}, Sebastian and {Di Mille}, Francesco and {Donor}, John and {Dow}, Peter and {Dwelly}, Tom and {Eracleous}, Mike and {Eriksen}, Jamey and {Fan}, Xiaohui and {Farr}, Emily and {Frederick}, Sara and {Fries}, Logan and {Frinchaboy}, Peter and {G{\"a}nsicke}, Boris T. and {Ge}, Junqiang and {Gonz{\'a}lez {\'A}vila}, Consuelo and {Grabowski}, Katie and {Grier}, Catherine and {Guiglion}, Guillaume and {Gupta}, Pramod and {Hall}, Patrick and {Hawkins}, Keith and {Hayes}, Christian R. and {Hermes}, J.~J. and {Hern{\'a}ndez-Garc{\'\i}a}, Lorena and {Hogg}, David W. and {Holtzman}, Jon A. and {Ibarra-Medel}, Hector Javier and {Ji}, Alexander and {Jofre}, Paula and {Johnson}, Jennifer A. and {Jones}, Amy M. and {Kinemuchi}, Karen and {Kluge}, Matthias and {Koekemoer}, Anton and {Kollmeier}, Juna A. and {Kounkel}, Marina and {Krishnarao}, Dhanesh and {Krumpe}, Mirko and {Lacerna}, Ivan and {Lago}, Paulo Jakson Assuncao and {Laporte}, Chervin and {Liu}, Chao and {Liu}, Ang and {Liu}, Xin and {Lopes}, Alexandre Roman and {Macktoobian}, Matin and {Majewski}, Steven R. and {Malanushenko}, Viktor and {Maoz}, Dan and {Masseron}, Thomas and {Masters}, Karen L. and {Matijevic}, Gal and {McBride}, Aidan and {Medan}, Ilija and {Merloni}, Andrea and {Morrison}, Sean and {Myers}, Natalie and {M{\'e}sz{\'a}ros}, Szabolcs and {Negrete}, C. Alenka and {Nidever}, David L. and {Nitschelm}, Christian and {Oravetz}, Daniel and {Oravetz}, Audrey and {Pan}, Kaike and {Peng}, Yingjie and {Pinsonneault}, Marc H. and {Pogge}, Rick and {Qiu}, Dan and {Ramirez}, Solange V. and {Rix}, Hans-Walter and {Fern{\'a}ndez Rosso}, Daniela and {Runnoe}, Jessie and {Salvato}, Mara and {Sanchez}, Sebastian F. and {Santana}, Felipe A. and {Saydjari}, Andrew and {Sayres}, Conor and {Schlaufman}, Kevin C. and {Schneider}, Donald P. and {Schwope}, Axel and {Serna}, Javier and {Shen}, Yue and {Sobeck}, Jennifer and {Song}, Ying-Yi and {Souto}, Diogo and {Spoo}, Taylor and {Stassun}, Keivan G. and {Steinmetz}, Matthias and {Straumit}, Ilya and {Stringfellow}, Guy and {S{\'a}nchez-Gallego}, Jos{\'e} and {Taghizadeh-Popp}, Manuchehr and {Tayar}, Jamie and {Thakar}, Ani and {Tissera}, Patricia B. and {Tkachenko}, Andrew and {Hernandez Toledo}, Hector and {Trakhtenbrot}, Benny and {Fern{\'a}ndez-Trincado}, Jos{\'e} G. and {Troup}, Nicholas and {Trump}, Jonathan R. and {Tuttle}, Sarah and {Ulloa}, Natalie and {Vazquez-Mata}, Jose Antonio and {Vera Alfaro}, Pablo and {Villanova}, Sandro and {Wachter}, Stefanie and {Weijmans}, Anne-Marie and {Wheeler}, Adam and {Wilson}, John and {Wojno}, Leigh and {Wolf}, Julien and {Xue}, Xiang-Xiang and {Ybarra}, Jason E. and {Zari}, Eleonora and {Zasowski}, Gail},
        title = "{The Eighteenth Data Release of the Sloan Digital Sky Surveys: Targeting and First Spectra from SDSS-V}",
      journal = {\apjs},
         year = 2023,
        month = aug,
       volume = {267},
       number = {2},
          eid = {44},
        pages = {44},
          doi = {10.3847/1538-4365/acda98},
archivePrefix = {arXiv},
       eprint = {2301.07688},
 primaryClass = {astro-ph.GA},
       adsurl = {https://ui.adsabs.harvard.edu/abs/2023ApJS..267...44A}
}

@ARTICLE{Almosallam2016,
       author = {{Almosallam}, Ibrahim A. and {Lindsay}, Sam N. and {Jarvis}, Matt J. and {Roberts}, Stephen J.},
        title = "{A sparse Gaussian process framework for photometric redshift estimation}",
      journal = {\mnras},
         year = 2016,
        month = jan,
       volume = {455},
       number = {3},
        pages = {2387-2401},
          doi = {10.1093/mnras/stv2425},
archivePrefix = {arXiv},
       eprint = {1505.05489},
 primaryClass = {astro-ph.IM},
       adsurl = {https://ui.adsabs.harvard.edu/abs/2016MNRAS.455.2387A}
}

@ARTICLE{Arizo2025,
       author = {{Arizo-Borillo}, F.~D. and {Lopez-Sanjuan}, C. and {Pintos-Castro}, I. and {Fernandez-Ontiveros}, J.~A. and {Kuutma}, T. and {Lumbreras-Calle}, A. and {Hernan-Caballero}, A. and {Dominguez-Sanchez}, H. and {De Lucia}, G. and {Fontanot}, F. and {Diaz-Garcia}, L.~A. and {Vilchez}, J.~M. and {Rahna}, P.~T. and {Cenarro}, A.~J. and {Cristobal-Hornillos}, D. and {Hernandez-Monteagudo}, C. and {Marin-Franch}, A. and {Moles}, M. and {Varela}, J. and {Vazquez Ramio}, H. and {Alcaniz}, J. and {Dupke}, R.~A. and {Ederoclite}, A. and {Sodre Jr.}, L. and {Angulo}, R.~E.},
        title = "{J-PLUS: The stellar mass function of quiescent and star forming galaxies at 0.05 <= z <= 0.2}",
      journal = {arXiv e-prints},
         year = 2025,
        month = sep,
          eid = {arXiv:2509.03404},
        pages = {arXiv:2509.03404},
archivePrefix = {arXiv},
       eprint = {2509.03404},
 primaryClass = {astro-ph.GA},
       adsurl = {https://ui.adsabs.harvard.edu/abs/2025arXiv250903404A}
}

@ARTICLE{Arnaudova2024,
       author = {{Arnaudova}, M.~I. and {Smith}, D.~J.~B. and {Hardcastle}, M.~J. and {Das}, S. and {Drake}, A. and {Duncan}, K. and {G{\"u}rkan}, G. and {Magliocchetti}, M. and {Morabito}, L.~K. and {Petley}, J.~W. and {Shenoy}, S. and {Tasse}, C.},
        title = "{Exploring the radio loudness of SDSS quasars with spectral stacking}",
      journal = {\mnras},
         year = 2024,
        month = mar,
       volume = {528},
       number = {3},
        pages = {4547-4567},
          doi = {10.1093/mnras/stae233},
archivePrefix = {arXiv},
       eprint = {2401.08774},
 primaryClass = {astro-ph.GA},
       adsurl = {https://ui.adsabs.harvard.edu/abs/2024MNRAS.528.4547A}
}

@ARTICLE{Arnouts1999,
       author = {{Arnouts}, S. and {Cristiani}, S. and {Moscardini}, L. and {Matarrese}, S. and {Lucchin}, F. and {Fontana}, A. and {Giallongo}, E.},
        title = "{Measuring and modelling the redshift evolution of clustering: the Hubble Deep Field North}",
      journal = {\mnras},
         year = 1999,
        month = dec,
       volume = {310},
       number = {2},
        pages = {540-556},
          doi = {10.1046/j.1365-8711.1999.02978.x},
archivePrefix = {arXiv},
       eprint = {astro-ph/9902290},
 primaryClass = {astro-ph},
       adsurl = {https://ui.adsabs.harvard.edu/abs/1999MNRAS.310..540A}
}

@ARTICLE{Baldwin1981,
       author = {{Baldwin}, J.~A. and {Phillips}, M.~M. and {Terlevich}, R.},
        title = "{Classification parameters for the emission-line spectra of extragalactic objects.}",
      journal = {\pasp},
         year = 1981,
        month = feb,
       volume = {93},
        pages = {5-19},
          doi = {10.1086/130766},
       adsurl = {https://ui.adsabs.harvard.edu/abs/1981PASP...93....5B}
}

@ARTICLE{Balokovic2012,
       author = {{Balokovi{\'c}}, M. and {Smol{\v{c}}i{\'c}}, V. and {Ivezi{\'c}}, {\v{Z}}. and {Zamorani}, G. and {Schinnerer}, E. and {Kelly}, B.~C.},
        title = "{Disclosing the Radio Loudness Distribution Dichotomy in Quasars: An Unbiased Monte Carlo Approach Applied to the SDSS-FIRST Quasar Sample}",
      journal = {\apj},
         year = 2012,
        month = nov,
       volume = {759},
       number = {1},
          eid = {30},
        pages = {30},
          doi = {10.1088/0004-637X/759/1/30},
archivePrefix = {arXiv},
       eprint = {1209.1099},
 primaryClass = {astro-ph.GA},
       adsurl = {https://ui.adsabs.harvard.edu/abs/2012ApJ...759...30B}
}

@dataset{BailerJones2021,
       author = {{Bailer-Jones}, C.~A.~L. and {Rybizki}, J. and {Fouesneau}, M. and {Demleitner}, M. and {Andrae}, R.},
        title = "{VizieR Online Data Catalog: Distances to 1.47 billion stars in Gaia EDR3 (Bailer-Jones+, 2021)}",
 howpublished = {VizieR On-line Data Catalog: I/352.  Originally published in: 2021AJ....161..147B},
         year = 2021,
        month = feb,
          eid = {I/352},
       adsurl = {https://ui.adsabs.harvard.edu/abs/2021yCat.1352....0B}
}

@INPROCEEDINGS{Becker1994,
       author = {{Becker}, Robert H. and {White}, Richard L. and {Helfand}, David J.},
        title = "{The VLA's FIRST Survey}",
    booktitle = {Astronomical Data Analysis Software and Systems III},
         year = 1994,
       editor = {{Crabtree}, D.~R. and {Hanisch}, R.~J. and {Barnes}, J.},
       series = {Astronomical Society of the Pacific Conference Series},
       volume = {61},
        month = jan,
        pages = {165},
       adsurl = {https://ui.adsabs.harvard.edu/abs/1994ASPC...61..165B}
}

@ARTICLE{Bennett1962,
       author = {{Bennett}, A.~S.},
        title = "{The preparation of the revised 3C catalogue of radio sources}",
      journal = {\mnras},
         year = 1962,
        month = jan,
       volume = {125},
        pages = {75},
          doi = {10.1093/mnras/125.1.75},
       adsurl = {https://ui.adsabs.harvard.edu/abs/1962MNRAS.125...75B}
}

@ARTICLE{Best2005,
       author = {{Best}, P.~N. and {Kauffmann}, G. and {Heckman}, T.~M. and {Brinchmann}, J. and {Charlot}, S. and {Ivezi{\'c}}, {\v{Z}}. and {White}, S.~D.~M.},
        title = "{The host galaxies of radio-loud active galactic nuclei: mass dependences, gas cooling and active galactic nuclei feedback}",
      journal = {\mnras},
         year = 2005,
        month = sep,
       volume = {362},
       number = {1},
        pages = {25-40},
          doi = {10.1111/j.1365-2966.2005.09192.x},
archivePrefix = {arXiv},
       eprint = {astro-ph/0506269},
 primaryClass = {astro-ph},
       adsurl = {https://ui.adsabs.harvard.edu/abs/2005MNRAS.362...25B}
}

@ARTICLE{Boquien2019,
       author = {{Boquien}, M. and {Burgarella}, D. and {Roehlly}, Y. and {Buat}, V. and {Ciesla}, L. and {Corre}, D. and {Inoue}, A.~K. and {Salas}, H.},
        title = "{CIGALE: a python Code Investigating GALaxy Emission}",
      journal = {\aap},
         year = 2019,
        month = feb,
       volume = {622},
          eid = {A103},
        pages = {A103},
          doi = {10.1051/0004-6361/201834156},
archivePrefix = {arXiv},
       eprint = {1811.03094},
 primaryClass = {astro-ph.GA},
       adsurl = {https://ui.adsabs.harvard.edu/abs/2019A&A...622A.103B}
}

@dataset{deBruyn2000,
       author = {{de Bruyn}, G. and {Miley}, G. and {Rengelink}, R. and {Tang}, Y. and {Bremer}, M. and {Rottgering}, H. and {Raimond}, R. and {Bremer}, M. and {Fullagar}, D.},
        title = "{VizieR Online Data Catalog: The Westerbork Northern Sky Survey (Leiden, 1998)}",
 howpublished = {VizieR On-line Data Catalog: VIII/62.  Originally published in: WENSS Collaboration NFRA/ASTRON and Leiden Observatory (1998)},
         year = 2000,
        month = feb,
          eid = {VIII/62},
       adsurl = {https://ui.adsabs.harvard.edu/abs/2000yCat.8062....0D}
}

@ARTICLE{Bruzual2003,
       author = {{Bruzual}, G. and {Charlot}, S.},
        title = "{Stellar population synthesis at the resolution of 2003}",
      journal = {\mnras},
         year = 2003,
        month = oct,
       volume = {344},
       number = {4},
        pages = {1000-1028},
          doi = {10.1046/j.1365-8711.2003.06897.x},
archivePrefix = {arXiv},
       eprint = {astro-ph/0309134},
 primaryClass = {astro-ph},
       adsurl = {https://ui.adsabs.harvard.edu/abs/2003MNRAS.344.1000B}
}

@ARTICLE{Bonoli2021,
       author = {{Bonoli}, S. and {Mar{\'\i}n-Franch}, A. and {Varela}, J. and {V{\'a}zquez Rami{\'o}}, H. and {Abramo}, L.~R. and {Cenarro}, A.~J. and {Dupke}, R.~A. and {V{\'\i}lchez}, J.~M. and {Crist{\'o}bal-Hornillos}, D. and {Gonz{\'a}lez Delgado}, R.~M. and {Hern{\'a}ndez-Monteagudo}, C. and {L{\'o}pez-Sanjuan}, C. and {Muniesa}, D.~J. and {Civera}, T. and {Ederoclite}, A. and {Hern{\'a}n-Caballero}, A. and {Marra}, V. and {Baqui}, P.~O. and {Cortesi}, A. and {Cypriano}, E.~S. and {Daflon}, S. and {de Amorim}, A.~L. and {D{\'\i}az-Garc{\'\i}a}, L.~A. and {Diego}, J.~M. and {Mart{\'\i}nez-Solaeche}, G. and {P{\'e}rez}, E. and {Placco}, V.~M. and {Prada}, F. and {Queiroz}, C. and {Alcaniz}, J. and {Alvarez-Candal}, A. and {Cepa}, J. and {Maroto}, A.~L. and {Roig}, F. and {Siffert}, B.~B. and {Taylor}, K. and {Benitez}, N. and {Moles}, M. and {Sodr{\'e}}, L. and {Carneiro}, S. and {Mendes de Oliveira}, C. and {Abdalla}, E. and {Angulo}, R.~E. and {Aparicio Resco}, M. and {Balaguera-Antol{\'\i}nez}, A. and {Ballesteros}, F.~J. and {Brito-Silva}, D. and {Broadhurst}, T. and {Carrasco}, E.~R. and {Castro}, T. and {Cid Fernandes}, R. and {Coelho}, P. and {de Melo}, R.~B. and {Doubrawa}, L. and {Fernandez-Soto}, A. and {Ferrari}, F. and {Finoguenov}, A. and {Garc{\'\i}a-Benito}, R. and {Iglesias-P{\'a}ramo}, J. and {Jim{\'e}nez-Teja}, Y. and {Kitaura}, F.~S. and {Laur}, J. and {Lopes}, P.~A.~A. and {Lucatelli}, G. and {Mart{\'\i}nez}, V.~J. and {Maturi}, M. and {Overzier}, R.~A. and {Pigozzo}, C. and {Quartin}, M. and {Rodr{\'\i}guez-Mart{\'\i}n}, J.~E. and {Salzano}, V. and {Tamm}, A. and {Tempel}, E. and {Umetsu}, K. and {Valdivielso}, L. and {von Marttens}, R. and {Zitrin}, A. and {D{\'\i}az-Mart{\'\i}n}, M.~C. and {L{\'o}pez-Alegre}, G. and {L{\'o}pez-Sainz}, A. and {Yanes-D{\'\i}az}, A. and {Rueda-Teruel}, F. and {Rueda-Teruel}, S. and {Abril Iba{\~n}ez}, J. and {L Ant{\'o}n Bravo}, J. and {Bello Ferrer}, R. and {Bielsa}, S. and {Casino}, J.~M. and {Castillo}, J. and {Chueca}, S. and {Cuesta}, L. and {Garzar{\'a}n Calderaro}, J. and {Iglesias-Marzoa}, R. and {{\'I}niguez}, C. and {Lamadrid Gutierrez}, J.~L. and {Lopez-Martinez}, F. and {Lozano-P{\'e}rez}, D. and {Ma{\'\i}cas Sacrist{\'a}n}, N. and {Molina-Ib{\'a}{\~n}ez}, E.~L. and {Moreno-Signes}, A. and {Rodr{\'\i}guez Llano}, S. and {Royo Navarro}, M. and {Tilve Rua}, V. and {Andrade}, U. and {Alfaro}, E.~J. and {Akras}, S. and {Arnalte-Mur}, P. and {Ascaso}, B. and {Barbosa}, C.~E. and {Beltr{\'a}n Jim{\'e}nez}, J. and {Benetti}, M. and {Bengaly}, C.~A.~P. and {Bernui}, A. and {Blanco-Pillado}, J.~J. and {Borges Fernandes}, M. and {Bregman}, J.~N. and {Bruzual}, G. and {Calderone}, G. and {Carvano}, J.~M. and {Casarini}, L. and {Chaves-Montero}, J. and {Chies-Santos}, A.~L. and {Coutinho de Carvalho}, G. and {Dimauro}, P. and {Duarte Puertas}, S. and {Figueruelo}, D. and {Gonz{\'a}lez-Serrano}, J.~I. and {Guerrero}, M.~A. and {Gurung-L{\'o}pez}, S. and {Herranz}, D. and {Huertas-Company}, M. and {Irwin}, J.~A. and {Izquierdo-Villalba}, D. and {Kanaan}, A. and {Kehrig}, C. and {Kirkpatrick}, C.~C. and {Lim}, J. and {Lopes}, A.~R. and {Lopes de Oliveira}, R. and {Marcos-Caballero}, A. and {Mart{\'\i}nez-Delgado}, D. and {Mart{\'\i}nez-Gonz{\'a}lez}, E. and {Mart{\'\i}nez-Somonte}, G. and {Oliveira}, N. and {Orsi}, A.~A. and {Penna-Lima}, M. and {Reis}, R.~R.~R. and {Spinoso}, D. and {Tsujikawa}, S. and {Vielva}, P. and {Vitorelli}, A.~Z. and {Xia}, J.~Q. and {Yuan}, H.~B. and {Arroyo-Polonio}, A. and {Dantas}, M.~L.~L. and {Galarza}, C.~A. and {Gon{\c{c}}alves}, D.~R. and {Gon{\c{c}}alves}, R.~S. and {Gonzalez}, J.~E. and {Gonzalez}, A.~H. and {Greisel}, N. and {Jim{\'e}nez-Esteban}, F. and {Landim}, R.~G. and {Lazzaro}, D. and {Magris}, G. and {Monteiro-Oliveira}, R. and {Pereira}, C.~B. and {Rebou{\c{c}}as}, M.~J. and {Rodriguez-Espinosa}, J.~M. and {Santos da Costa}, S. and {Telles}, E.},
        title = "{The miniJPAS survey: A preview of the Universe in 56 colors}",
      journal = {\aap},
         year = 2021,
        month = sep,
       volume = {653},
          eid = {A31},
        pages = {A31},
          doi = {10.1051/0004-6361/202038841},
archivePrefix = {arXiv},
       eprint = {2007.01910},
 primaryClass = {astro-ph.CO},
       adsurl = {https://ui.adsabs.harvard.edu/abs/2021A&A...653A..31B}
}

@ARTICLE{Callingham2019,
       author = {{Callingham}, J.~R. and {Vedantham}, H.~K. and {Pope}, B.~J.~S. and {Shimwell}, T.~W. and {LoTSS Team}},
        title = "{LoTSS-HETDEX and Gaia: Blind Search for Radio Emission from Stellar Systems Dominated by False Positives}",
      journal = {Research Notes of the American Astronomical Society},
         year = 2019,
        month = feb,
       volume = {3},
       number = {2},
          eid = {37},
        pages = {37},
          doi = {10.3847/2515-5172/ab07c3},
       adsurl = {https://ui.adsabs.harvard.edu/abs/2019RNAAS...3...37C}
}

@ARTICLE{Callingham2021,
       author = {{Callingham}, J.~R. and {Vedantham}, H.~K. and {Shimwell}, T.~W. and {Pope}, B.~J.~S. and {Davis}, I.~E. and {Best}, P.~N. and {Hardcastle}, M.~J. and {R{\"o}ttgering}, H.~J.~A. and {Sabater}, J. and {Tasse}, C. and {van Weeren}, R.~J. and {Williams}, W.~L. and {Zarka}, P. and {de Gasperin}, F. and {Drabent}, A.},
        title = "{The population of M dwarfs observed at low radio frequencies}",
      journal = {Nature Astronomy},
         year = 2021,
        month = dec,
       volume = {5},
        pages = {1233-1239},
          doi = {10.1038/s41550-021-01483-0},
archivePrefix = {arXiv},
       eprint = {2110.03713},
 primaryClass = {astro-ph.SR},
       adsurl = {https://ui.adsabs.harvard.edu/abs/2021NatAs...5.1233C}
}

@ARTICLE{vanCappellen2022,
       author = {{van Cappellen}, W.~A. and {Oosterloo}, T.~A. and {Verheijen}, M.~A.~W. and {Adams}, E.~A.~K. and {Adebahr}, B. and {Braun}, R. and {Hess}, K.~M. and {Holties}, H. and {van der Hulst}, J.~M. and {Hut}, B. and {Kooistra}, E. and {van Leeuwen}, J. and {Loose}, G.~M. and {Morganti}, R. and {Moss}, V.~A. and {Orr{\'u}}, E. and {Ruiter}, M. and {Schoenmakers}, A.~P. and {Vermaas}, N.~J. and {Wijnholds}, S.~J. and {van Amesfoort}, A.~S. and {Arts}, M.~J. and {Attema}, J.~J. and {Bakker}, L. and {Bassa}, C.~G. and {Bast}, J.~E. and {Benthem}, P. and {Beukema}, R. and {Blaauw}, R. and {de Blok}, W.~J.~G. and {Bouwhuis}, M. and {van den Brink}, R.~H. and {Connor}, L. and {Coolen}, A.~H.~W.~M. and {Damstra}, S. and {van Diepen}, G.~N.~J. and {de Goei}, R. and {D{\'e}nes}, H. and {Drost}, M. and {Ebbendorf}, N. and {Frank}, B.~S. and {Gardenier}, D.~W. and {Gerbers}, M. and {Grange}, Y.~G. and {Grit}, T. and {Gunst}, A.~W. and {Gupta}, N. and {Ivashina}, M.~V. and {J{\'o}zsa}, G.~I.~G. and {Janssen}, G.~H. and {Koster}, A. and {Kruithof}, G.~H. and {Kuindersma}, S.~J. and {Kutkin}, A. and {Lucero}, D.~M. and {Maan}, Y. and {Maccagni}, F.~M. and {van der Marel}, J. and {Mika}, A. and {Morawietz}, J. and {Mulder}, H. and {Mulder}, E. and {Norden}, M.~J. and {Offringa}, A.~R. and {Oostrum}, L.~C. and {Overeem}, R.~E. and {Paragi}, Z. and {Pepping}, H.~J. and {Petroff}, E. and {Pisano}, D.~J. and {Polatidis}, A.~G. and {Prasad}, P. and {de Reijer}, J.~P.~R. and {Romein}, J.~W. and {Schaap}, J. and {Schoonderbeek}, G.~W. and {Schulz}, R. and {van der Schuur}, D. and {Sclocco}, A. and {Sluman}, J.~J. and {Smits}, R. and {Stappers}, B.~W. and {Straal}, S.~M. and {Stuurwold}, K.~J.~C. and {Verstappen}, J. and {Vohl}, D. and {Wierenga}, K.~J. and {Woestenburg}, E.~E.~M. and {Zanting}, A.~W. and {Ziemke}, J.},
        title = "{Apertif: Phased array feeds for the Westerbork Synthesis Radio Telescope. System overview and performance characteristics}",
      journal = {\aap},
         year = 2022,
        month = feb,
       volume = {658},
          eid = {A146},
        pages = {A146},
          doi = {10.1051/0004-6361/202141739},
archivePrefix = {arXiv},
       eprint = {2109.14234},
 primaryClass = {astro-ph.IM},
       adsurl = {https://ui.adsabs.harvard.edu/abs/2022A&A...658A.146V}
}

@INPROCEEDINGS{Cenarro14,
   author = {{Cenarro}, A.~J. and {Moles}, M. and {Mar{\'{\i}}n-Franch}, A. and 
	{Crist{\'o}bal-Hornillos}, D. and {Yanes D{\'{\i}}az}, A. and 
	{Ederoclite}, A. and {Varela}, J. and {V{\'a}zquez-Rami{\'o}}, H. and 
	{Valdivielso}, L. and {Ben{\'{\i}}tez}, N. and {Cepa}, J. and 
	{Dupke}, R. and {Fern{\'a}ndez-Soto}, A. and {Mendes de Oliveira}, C. and 
	{Sodr{\'e}}, L. and {Taylor}, K. and {Rueda-Teruel}, S. and 
	{Rueda-Teruel}, F. and {Luis-Simoes}, R. and {Chueca}, S. and 
	{Ant{\'o}n}, J.~L. and {Bello}, R. and {D{\'{\i}}az-Mart{\'{\i}}n}, M.~C. and 
	{Guill{\'e}n-Civera}, L. and {Hern{\'a}ndez-Fuertes}, J. and 
	{Iglesias-Marzoa}, R. and {Jim{\'e}nez-Mej{\'{\i}}as}, D. and 
	{Lasso-Cabrera}, N.~M. and {L{\'o}pez-Alegre}, G. and {L{\'o}pez-Sainz}, A. and 
	{Rodr{\'{\i}}guez-Hern{\'a}ndez}, M.~A.~C. and {Su{\'a}rez}, O. and 
	{Lamadrid}, J.~L. and {Ma{\'{\i}}cas}, N. and {Abril-Iba{\~n}ez}, J. and 
	{Tilve}, V. and {Rodr{\'{\i}}guez-Llano}, S.},
    title = "{The Observatorio Astrof{\'{\i}}sico de Javalambre: current status, developments, operations, and strategies}",
booktitle = {Observatory Operations: Strategies, Processes, and Systems V},
     year = 2014,
   series = {\procspie},
   volume = 9149,
    month = aug,
      eid = {91491I},
    pages = {91491I},
      doi = {10.1117/12.2055455},
   adsurl = {http://esoads.eso.org/abs/2014SPIE.9149E..1IC}
}

@ARTICLE{Cenarro19,
       author = {{Cenarro}, A.~J. and {Moles}, M. and {Crist{\'o}bal-Hornillos}, D. and {Mar{\'\i}n-Franch}, A. and {Ederoclite}, A. and {Varela}, J. and {L{\'o}pez-Sanjuan}, C. and {Hern{\'a}ndez-Monteagudo}, C. and {Angulo}, R.~E. and {V{\'a}zquez Rami{\'o}}, H. and {Viironen}, K. and {Bonoli}, S. and {Orsi}, A.~A. and {Hurier}, G. and {San Roman}, I. and {Greisel}, N. and {Vilella-Rojo}, G. and {D{\'\i}az-Garc{\'\i}a}, L.~A. and {Logro{\~n}o-Garc{\'\i}a}, R. and {Gurung-L{\'o}pez}, S. and {Spinoso}, D. and {Izquierdo-Villalba}, D. and {Aguerri}, J.~A.~L. and {Allende Prieto}, C. and {Bonatto}, C. and {Carvano}, J.~M. and {Chies-Santos}, A.~L. and {Daflon}, S. and {Dupke}, R.~A. and {Falc{\'o}n-Barroso}, J. and {Gon{\c{c}}alves}, D.~R. and {Jim{\'e}nez-Teja}, Y. and {Molino}, A. and {Placco}, V.~M. and {Solano}, E. and {Whitten}, D.~D. and {Abril}, J. and {Ant{\'o}n}, J.~L. and {Bello}, R. and {Bielsa de Toledo}, S. and {Castillo-Ram{\'\i}rez}, J. and {Chueca}, S. and {Civera}, T. and {D{\'\i}az-Mart{\'\i}n}, M.~C. and {Dom{\'\i}nguez-Mart{\'\i}nez}, M. and {Garzar{\'a}n-Calderaro}, J. and {Hern{\'a}ndez-Fuertes}, J. and {Iglesias-Marzoa}, R. and {I{\~n}iguez}, C. and {Jim{\'e}nez Ruiz}, J.~M. and {Kruuse}, K. and {Lamadrid}, J.~L. and {Lasso-Cabrera}, N. and {L{\'o}pez-Alegre}, G. and {L{\'o}pez-Sainz}, A. and {Ma{\'\i}cas}, N. and {Moreno-Signes}, A. and {Muniesa}, D.~J. and {Rodr{\'\i}guez-Llano}, S. and {Rueda-Teruel}, F. and {Rueda-Teruel}, S. and {Soriano-Lagu{\'\i}a}, I. and {Tilve}, V. and {Valdivielso}, L. and {Yanes-D{\'\i}az}, A. and {Alcaniz}, J.~S. and {Mendes de Oliveira}, C. and {Sodr{\'e}}, L. and {Coelho}, P. and {Lopes de Oliveira}, R. and {Tamm}, A. and {Xavier}, H.~S. and {Abramo}, L.~R. and {Akras}, S. and {Alfaro}, E.~J. and {Alvarez-Candal}, A. and {Ascaso}, B. and {Beasley}, M.~A. and {Beers}, T.~C. and {Borges Fernandes}, M. and {Bruzual}, G.~R. and {Buzzo}, M.~L. and {Carrasco}, J.~M. and {Cepa}, J. and {Cortesi}, A. and {Costa-Duarte}, M.~V. and {De Pr{\'a}}, M. and {Favole}, G. and {Galarza}, A. and {Galbany}, L. and {Garcia}, K. and {Gonz{\'a}lez Delgado}, R.~M. and {Gonz{\'a}lez-Serrano}, J.~I. and {Guti{\'e}rrez-Soto}, L.~A. and {Hernandez-Jimenez}, J.~A. and {Kanaan}, A. and {Kuncarayakti}, H. and {Landim}, R.~C.~G. and {Laur}, J. and {Licandro}, J. and {Lima Neto}, G.~B. and {Lyman}, J.~D. and {Ma{\'\i}z Apell{\'a}niz}, J. and {Miralda-Escud{\'e}}, J. and {Morate}, D. and {Nogueira-Cavalcante}, J.~P. and {Novais}, P.~M. and {Oncins}, M. and {Oteo}, I. and {Overzier}, R.~A. and {Pereira}, C.~B. and {Rebassa-Mansergas}, A. and {Reis}, R.~R.~R. and {Roig}, F. and {Sako}, M. and {Salvador-Rusi{\~n}ol}, N. and {Sampedro}, L. and {S{\'a}nchez-Bl{\'a}zquez}, P. and {Santos}, W.~A. and {Schmidtobreick}, L. and {Siffert}, B.~B. and {Telles}, E. and {Vilchez}, J.~M.},
        title = "{J-PLUS: The Javalambre Photometric Local Universe Survey}",
      journal = {\aap},
         year = 2019,
        month = feb,
       volume = {622},
          eid = {A176},
        pages = {A176},
          doi = {10.1051/0004-6361/201833036},
archivePrefix = {arXiv},
       eprint = {1804.02667},
 primaryClass = {astro-ph.GA},
       adsurl = {https://ui.adsabs.harvard.edu/abs/2019A&A...622A.176C}
}

@ARTICLE{Chabrier2003,
       author = {{Chabrier}, Gilles},
        title = "{Galactic Stellar and Substellar Initial Mass Function}",
      journal = {\pasp},
         year = 2003,
        month = jul,
       volume = {115},
       number = {809},
        pages = {763-795},
          doi = {10.1086/376392},
archivePrefix = {arXiv},
       eprint = {astro-ph/0304382},
 primaryClass = {astro-ph},
       adsurl = {https://ui.adsabs.harvard.edu/abs/2003PASP..115..763C}
}

@ARTICLE{Chen2023,
       author = {{Chen}, Sina and {Laor}, Ari and {Behar}, Ehud and {Baldi}, Ranieri D. and {Gelfand}, Joseph D.},
        title = "{The radio emission in radio-quiet quasars: the VLBA perspective}",
      journal = {\mnras},
         year = 2023,
        month = oct,
       volume = {525},
       number = {1},
        pages = {164-182},
          doi = {10.1093/mnras/stad2289},
archivePrefix = {arXiv},
       eprint = {2307.13599},
 primaryClass = {astro-ph.GA},
       adsurl = {https://ui.adsabs.harvard.edu/abs/2023MNRAS.525..164C}
}

@ARTICLE{Ching2017,
       author = {{Ching}, John H.~Y. and {Sadler}, Elaine M. and {Croom}, Scott M. and {Johnston}, Helen M. and {Pracy}, Michael B. and {Couch}, Warrick J. and {Hopkins}, A.~M. and {Jurek}, Russell J. and {Pimbblet}, K.~A.},
        title = "{The Large Area Radio Galaxy Evolution Spectroscopic Survey (LARGESS): survey design, data catalogue and GAMA/WiggleZ spectroscopy}",
      journal = {\mnras},
         year = 2017,
        month = jan,
       volume = {464},
       number = {2},
        pages = {1306-1332},
          doi = {10.1093/mnras/stw2396},
archivePrefix = {arXiv},
       eprint = {1609.05578},
 primaryClass = {astro-ph.GA},
       adsurl = {https://ui.adsabs.harvard.edu/abs/2017MNRAS.464.1306C}
}

@ARTICLE{Ching2017b,
       author = {{Ching}, J.~H.~Y. and {Croom}, S.~M. and {Sadler}, E.~M. and {Robotham}, A.~S.~G. and {Brough}, S. and {Baldry}, I.~K. and {Bland-Hawthorn}, J. and {Colless}, M. and {Driver}, S.~P. and {Holwerda}, B.~W. and {Hopkins}, A.~M. and {Jarvis}, M.~J. and {Johnston}, H.~M. and {Kelvin}, L.~S. and {Liske}, J. and {Loveday}, J. and {Norberg}, P. and {Pracy}, M.~B. and {Steele}, O. and {Thomas}, D. and {Wang}, L.},
        title = "{Galaxy And Mass Assembly (GAMA): the environments of high- and low-excitation radio galaxies}",
      journal = {\mnras},
         year = 2017,
        month = aug,
       volume = {469},
       number = {4},
        pages = {4584-4599},
          doi = {10.1093/mnras/stx1173},
archivePrefix = {arXiv},
       eprint = {1705.04502},
 primaryClass = {astro-ph.GA},
       adsurl = {https://ui.adsabs.harvard.edu/abs/2017MNRAS.469.4584C}
}

@ARTICLE{Cirasuolo2003,
       author = {{Cirasuolo}, M. and {Magliocchetti}, M. and {Celotti}, A. and {Danese}, L.},
        title = "{The radio-loud/radio-quiet dichotomy: news from the 2dF QSO Redshift Survey}",
      journal = {\mnras},
         year = 2003,
        month = may,
       volume = {341},
       number = {3},
        pages = {993-1004},
          doi = {10.1046/j.1365-8711.2003.06485.x},
archivePrefix = {arXiv},
       eprint = {astro-ph/0301526},
 primaryClass = {astro-ph},
       adsurl = {https://ui.adsabs.harvard.edu/abs/2003MNRAS.341..993C}
}

@ARTICLE{Cohen2007,
       author = {{Cohen}, A.~S. and {Lane}, W.~M. and {Cotton}, W.~D. and {Kassim}, N.~E. and {Lazio}, T.~J.~W. and {Perley}, R.~A. and {Condon}, J.~J. and {Erickson}, W.~C.},
        title = "{The VLA Low-Frequency Sky Survey}",
      journal = {\aj},
         year = 2007,
        month = sep,
       volume = {134},
       number = {3},
        pages = {1245-1262},
          doi = {10.1086/520719},
archivePrefix = {arXiv},
       eprint = {0706.1191},
 primaryClass = {astro-ph},
       adsurl = {https://ui.adsabs.harvard.edu/abs/2007AJ....134.1245C}
}

@ARTICLE{Condon1998,
       author = {{Condon}, J.~J. and {Cotton}, W.~D. and {Greisen}, E.~W. and {Yin}, Q.~F. and {Perley}, R.~A. and {Taylor}, G.~B. and {Broderick}, J.~J.},
        title = "{The NRAO VLA Sky Survey}",
      journal = {\aj},
         year = 1998,
        month = may,
       volume = {115},
       number = {5},
        pages = {1693-1716},
          doi = {10.1086/300337},
       adsurl = {https://ui.adsabs.harvard.edu/abs/1998AJ....115.1693C}
}

@BOOK{Condon2016,
       author = {{Condon}, James J. and {Ransom}, Scott M.},
        title = "{Essential Radio Astronomy}",
         year = 2016,
    publisher = {Princeton University Press},
       adsurl = {https://ui.adsabs.harvard.edu/abs/2016era..book.....C}
}

@ARTICLE{Davis2023,
       author = {{Davis}, Dustin and {Gebhardt}, Karl and {Cooper}, Erin Mentuch and {Bowman}, William P. and {Garcia Castanheira}, Barbara and {Chisholm}, John and {Ciardullo}, Robin and {Fabricius}, Maximilian and {Farrow}, Daniel J. and {Finkelstein}, Steven L. and {Gronwall}, Caryl and {Gawiser}, Eric and {Hill}, Gary J. and {Hopp}, Ulrich and {House}, Lindsay R. and {Jeong}, Donghui and {Kollatschny}, Wolfram and {Komatsu}, Eiichiro and {Liu}, Chenxu and {Niemeyer}, Maja Lujan and {Saldana-Lopez}, Alberto and {Saito}, Shun and {Schneider}, Donald P. and {Snigula}, Jan and {Tuttle}, Sarah and {Weiss}, Laurel H. and {Wisotzki}, Lutz and {Zeimann}, Gregory},
        title = "{HETDEX Public Source Catalog 1-Stacking 50,000 Lyman Alpha Emitters}",
      journal = {\apj},
         year = 2023,
        month = sep,
       volume = {954},
       number = {2},
          eid = {209},
        pages = {209},
          doi = {10.3847/1538-4357/ace4c2},
archivePrefix = {arXiv},
       eprint = {2307.03096},
 primaryClass = {astro-ph.GA},
       adsurl = {https://ui.adsabs.harvard.edu/abs/2023ApJ...954..209D}
}

@ARTICLE{DESI2025,
       author = {{DESI Collaboration} and {Abdul-Karim}, M. and {Adame}, A.~G. and {Aguado}, D. and {Aguilar}, J. and {Ahlen}, S. and {Alam}, S. and {Aldering}, G. and {Alexander}, D.~M. and {Alfarsy}, R. and {Allen}, L. and {Allende Prieto}, C. and {Alves}, O. and {Anand}, A. and {Andrade}, U. and {Armengaud}, E. and {Avila}, S. and {Aviles}, A. and {Awan}, H. and {Bailey}, S. and {Baleato Lizancos}, A. and {Ballester}, O. and {Bault}, A. and {Bautista}, J. and {BenZvi}, S. and {Beraldo e Silva}, L. and {Bermejo-Climent}, J.~R. and {Beutler}, F. and {Bianchi}, D. and {Blake}, C. and {Blum}, R. and {Bolton}, A.~S. and {Bonici}, M. and {Brieden}, S. and {Brodzeller}, A. and {Brooks}, D. and {Buckley-Geer}, E. and {Burtin}, E. and {Canning}, R. and {Carnero Rosell}, A. and {Carr}, A. and {Carrilho}, P. and {Casas}, L. and {Castander}, F.~J. and {Cereskaite}, R. and {Cervantes-Cota}, J.~L. and {Chaussidon}, E. and {Chaves-Montero}, J. and {Chen}, S. and {Chen}, X. and {Claybaugh}, T. and {Cole}, S. and {Cooper}, A.~P. and {Cousinou}, M. -C. and {Cuceu}, A. and {Davis}, T.~M. and {Dawson}, K.~S. and {de Belsunce}, R. and {de la Cruz}, R. and {de la Macorra}, A. and {de Mattia}, A. and {Deiosso}, N. and {Della Costa}, J. and {Demina}, R. and {Demirbozan}, U. and {DeRose}, J. and {Dey}, A. and {Dey}, B. and {Ding}, J. and {Ding}, Z. and {Doel}, P. and {Douglass}, K. and {Dowicz}, M. and {Ebina}, H. and {Edelstein}, J. and {Eisenstein}, D.~J. and {Elbers}, W. and {Emas}, N. and {Escoffier}, S. and {Fagrelius}, P. and {Fan}, X. and {Fanning}, K. and {Fawcett}, V.~A. and {Fern\textbackslash'andez-Garc\textbackslash'ia}, E. and {Ferraro}, S. and {Findlay}, N. and {Font-Ribera}, A. and {Forero-Romero}, J.~E. and {Forero-S\textbackslash'anchez}, D. and {Frenk}, C.~S. and {G\textbackslash''ansicke}, B.~T. and {Galbany}, L. and {Garc\textbackslash'ia-Bellido}, J. and {Garcia-Quintero}, C. and {Garrison}, L.~H. and {Gazta\textbackslash}, E. and {Gil-Mar\textbackslash'in}, H. and {Gnedin}, O.~Y. and {Gontcho}, S. Gontcho A and {Gonzalez-Morales}, A.~X. and {Gonzalez-Perez}, V. and {Gordon}, C. and {Graur}, O. and {Green}, D. and {Gruen}, D. and {Gsponer}, R. and {Guandalin}, C. and {Gutierrez}, G. and {Guy}, J. and {Hahn}, C. and {Han}, J.~J. and {Han}, J. and {He}, S. and {Herrera-Alcantar}, H.~K. and {Honscheid}, K. and {Hou}, J. and {Howlett}, C. and {Huterer}, D. and {Irsic}, V. and {Ishak}, M. and {Jacques}, A. and {Jimenez}, J. and {Jing}, Y.~P. and {Joachimi}, B. and {Joudaki}, S. and {Joyce}, R. and {Jullo}, E. and {Juneau}, S. and {Karacayli}, N.~G. and {Karim}, T. and {Kehoe}, R. and {Kent}, S. and {Khederlarian}, A. and {Kirkby}, D. and {Kisner}, T. and {Kitaura}, F. -S. and {Kizhuprakkat}, N. and {Kong}, H. and {Koposov}, S.~E. and {Kremin}, A. and {Krolewski}, A. and {Lahav}, O. and {Lai}, Y. and {Lamman}, C. and {Lan}, T. -W. and {Landriau}, M. and {Lang}, D. and {Lange}, J.~U. and {Lasker}, J. and {Le Goff}, J.~M. and {Le Guillou}, L. and {Leauthaud}, A. and {Levi}, M.~E. and {Li}, S. and {Li}, T.~S. and {Lodha}, K. and {Lokken}, M. and {Luo}, Y. and {Magneville}, C. and {Manera}, M. and {Manser}, C.~J. and {Margala}, D. and {Martini}, P. and {Maus}, M. and {McCullough}, J. and {McDonald}, P. and {Medina}, G.~E. and {Medina-Varela}, L. and {Meisner}, A. and {Mena-Fern\textbackslash'andez}, J. and {Menegas}, A. and {Mezcua}, M. and {Miquel}, R. and {Montero-Camacho}, P. and {Moon}, J. and {Moustakas}, J. and {Mu\~noz-Guti\textbackslash'errez}, A. and {Mu\~noz-Santos}, D. and {Myers}, A.~D. and {Myles}, J. and {Nadathur}, S. and {Najita}, J. and {Napolitano}, L. and {Newman}, J.~A. and {Nikakhtar}, F. and {Nikutta}, R. and {Niz}, G. and {Noriega}, H.~E. and {Padmanabhan}, N. and {Paillas}, E. and {Palanque-Delabrouille}, N. and {Palmese}, A. and {Pan}, J. and {Pan}, Z. and {Parkinson}, D. and {Peacock}, J. and {Percival}, W.~J. and {P\textbackslash'erez-Fern\textbackslash'andez}, A. and {P\textbackslash'erez-R\textbackslash`afols}, I. and {Peterson}, P.},
        title = "{Data Release 1 of the Dark Energy Spectroscopic Instrument}",
      journal = {arXiv e-prints},
         year = 2025,
        month = mar,
          eid = {arXiv:2503.14745},
        pages = {arXiv:2503.14745},
          doi = {10.48550/arXiv.2503.14745},
archivePrefix = {arXiv},
       eprint = {2503.14745},
 primaryClass = {astro-ph.CO},
       adsurl = {https://ui.adsabs.harvard.edu/abs/2025arXiv250314745D}
}

@ARTICLE{Dey2019,
       author = {{Dey}, Arjun and {Schlegel}, David J. and {Lang}, Dustin and {Blum}, Robert and {Burleigh}, Kaylan and {Fan}, Xiaohui and {Findlay}, Joseph R. and {Finkbeiner}, Doug and {Herrera}, David and {Juneau}, St{\'e}phanie and {Landriau}, Martin and {Levi}, Michael and {McGreer}, Ian and {Meisner}, Aaron and {Myers}, Adam D. and {Moustakas}, John and {Nugent}, Peter and {Patej}, Anna and {Schlafly}, Edward F. and {Walker}, Alistair R. and {Valdes}, Francisco and {Weaver}, Benjamin A. and {Y{\`e}che}, Christophe and {Zou}, Hu and {Zhou}, Xu and {Abareshi}, Behzad and {Abbott}, T.~M.~C. and {Abolfathi}, Bela and {Aguilera}, C. and {Alam}, Shadab and {Allen}, Lori and {Alvarez}, A. and {Annis}, James and {Ansarinejad}, Behzad and {Aubert}, Marie and {Beechert}, Jacqueline and {Bell}, Eric F. and {BenZvi}, Segev Y. and {Beutler}, Florian and {Bielby}, Richard M. and {Bolton}, Adam S. and {Brice{\~n}o}, C{\'e}sar and {Buckley-Geer}, Elizabeth J. and {Butler}, Karen and {Calamida}, Annalisa and {Carlberg}, Raymond G. and {Carter}, Paul and {Casas}, Ricard and {Castander}, Francisco J. and {Choi}, Yumi and {Comparat}, Johan and {Cukanovaite}, Elena and {Delubac}, Timoth{\'e}e and {DeVries}, Kaitlin and {Dey}, Sharmila and {Dhungana}, Govinda and {Dickinson}, Mark and {Ding}, Zhejie and {Donaldson}, John B. and {Duan}, Yutong and {Duckworth}, Christopher J. and {Eftekharzadeh}, Sarah and {Eisenstein}, Daniel J. and {Etourneau}, Thomas and {Fagrelius}, Parker A. and {Farihi}, Jay and {Fitzpatrick}, Mike and {Font-Ribera}, Andreu and {Fulmer}, Leah and {G{\"a}nsicke}, Boris T. and {Gaztanaga}, Enrique and {George}, Koshy and {Gerdes}, David W. and {Gontcho}, Satya Gontcho A. and {Gorgoni}, Claudio and {Green}, Gregory and {Guy}, Julien and {Harmer}, Diane and {Hernandez}, M. and {Honscheid}, Klaus and {Huang}, Lijuan Wendy and {James}, David J. and {Jannuzi}, Buell T. and {Jiang}, Linhua and {Joyce}, Richard and {Karcher}, Armin and {Karkar}, Sonia and {Kehoe}, Robert and {Kneib}, Jean-Paul and {Kueter-Young}, Andrea and {Lan}, Ting-Wen and {Lauer}, Tod R. and {Le Guillou}, Laurent and {Le Van Suu}, Auguste and {Lee}, Jae Hyeon and {Lesser}, Michael and {Perreault Levasseur}, Laurence and {Li}, Ting S. and {Mann}, Justin L. and {Marshall}, Robert and {Mart{\'\i}nez-V{\'a}zquez}, C.~E. and {Martini}, Paul and {du Mas des Bourboux}, H{\'e}lion and {McManus}, Sean and {Meier}, Tobias Gabriel and {M{\'e}nard}, Brice and {Metcalfe}, Nigel and {Mu{\~n}oz-Guti{\'e}rrez}, Andrea and {Najita}, Joan and {Napier}, Kevin and {Narayan}, Gautham and {Newman}, Jeffrey A. and {Nie}, Jundan and {Nord}, Brian and {Norman}, Dara J. and {Olsen}, Knut A.~G. and {Paat}, Anthony and {Palanque-Delabrouille}, Nathalie and {Peng}, Xiyan and {Poppett}, Claire L. and {Poremba}, Megan R. and {Prakash}, Abhishek and {Rabinowitz}, David and {Raichoor}, Anand and {Rezaie}, Mehdi and {Robertson}, A.~N. and {Roe}, Natalie A. and {Ross}, Ashley J. and {Ross}, Nicholas P. and {Rudnick}, Gregory and {Safonova}, Sasha and {Saha}, Abhijit and {S{\'a}nchez}, F. Javier and {Savary}, Elodie and {Schweiker}, Heidi and {Scott}, Adam and {Seo}, Hee-Jong and {Shan}, Huanyuan and {Silva}, David R. and {Slepian}, Zachary and {Soto}, Christian and {Sprayberry}, David and {Staten}, Ryan and {Stillman}, Coley M. and {Stupak}, Robert J. and {Summers}, David L. and {Sien Tie}, Suk and {Tirado}, H. and {Vargas-Maga{\~n}a}, Mariana and {Vivas}, A. Katherina and {Wechsler}, Risa H. and {Williams}, Doug and {Yang}, Jinyi and {Yang}, Qian and {Yapici}, Tolga and {Zaritsky}, Dennis and {Zenteno}, A. and {Zhang}, Kai and {Zhang}, Tianmeng and {Zhou}, Rongpu and {Zhou}, Zhimin},
        title = "{Overview of the DESI Legacy Imaging Surveys}",
      journal = {\aj},
         year = 2019,
        month = may,
       volume = {157},
       number = {5},
          eid = {168},
        pages = {168},
          doi = {10.3847/1538-3881/ab089d},
archivePrefix = {arXiv},
       eprint = {1804.08657},
 primaryClass = {astro-ph.IM},
       adsurl = {https://ui.adsabs.harvard.edu/abs/2019AJ....157..168D}
}

@ARTICLE{Drake2024,
       author = {{Drake}, A.~B. and {Smith}, D.~J.~B. and {Hardcastle}, M.~J. and {Best}, P.~N. and {Kondapally}, R. and {Arnaudova}, M.~I. and {Das}, S. and {Shenoy}, S. and {Duncan}, K.~J. and {R{\"o}ttgering}, H.~J.~A. and {Tasse}, C.},
        title = "{The LOFAR two metre sky survey data release 2: probabilistic spectral source classifications and faint radio source demographics}",
      journal = {\mnras},
         year = 2024,
        month = oct,
       volume = {534},
       number = {2},
        pages = {1107-1126},
          doi = {10.1093/mnras/stae2117},
archivePrefix = {arXiv},
       eprint = {2409.11465},
 primaryClass = {astro-ph.GA},
       adsurl = {https://ui.adsabs.harvard.edu/abs/2024MNRAS.534.1107D}
}

@ARTICLE{Driessen2024,
       author = {{Driessen}, Laura Nicole and {Pritchard}, Joshua and {Murphy}, Tara and {Heald}, George and {Robrade}, Jan and {Das}, Barnali and {Duchesne}, Stefan William and {Kaplan}, David L. and {Lenc}, Emil and {Lynch}, Christene R. and {Mitchell-Bolton}, Jackson and {Pope}, Benjamin J.~S. and {Rose}, Kovi and {Stelzer}, Beate and {Wang}, Yuanming and {Zic}, Andrew},
        title = "{The Sydney Radio Star Catalogue: Properties of radio stars at megahertz to gigahertz frequencies}",
      journal = {\pasa},
         year = 2024,
        month = nov,
       volume = {41},
          eid = {e084},
        pages = {e084},
          doi = {10.1017/pasa.2024.72},
archivePrefix = {arXiv},
       eprint = {2404.07418},
 primaryClass = {astro-ph.SR},
       adsurl = {https://ui.adsabs.harvard.edu/abs/2024PASA...41...84D}
}

@ARTICLE{Duncan2022,
       author = {{Duncan}, Kenneth J.},
        title = "{All-purpose, all-sky photometric redshifts for the Legacy Imaging Surveys Data Release 8}",
      journal = {\mnras},
         year = 2022,
        month = may,
       volume = {512},
       number = {3},
        pages = {3662-3683},
          doi = {10.1093/mnras/stac608},
archivePrefix = {arXiv},
       eprint = {2203.01949},
 primaryClass = {astro-ph.GA},
       adsurl = {https://ui.adsabs.harvard.edu/abs/2022MNRAS.512.3662D}
}

@ARTICLE{Dutta2023,
       author = {{Dutta}, Sushant and {Singh}, Veeresh and {Chandra}, C.~H. Ishwara and {Wadadekar}, Yogesh and {Kayal}, Abhijit and {Heywood}, Ian},
        title = "{Search and Characterization of Remnant Radio Galaxies in the XMM-LSS Deep Field}",
      journal = {\apj},
         year = 2023,
        month = feb,
       volume = {944},
       number = {2},
          eid = {176},
        pages = {176},
          doi = {10.3847/1538-4357/acaf01},
archivePrefix = {arXiv},
       eprint = {2212.10133},
 primaryClass = {astro-ph.GA},
       adsurl = {https://ui.adsabs.harvard.edu/abs/2023ApJ...944..176D}
}

@ARTICLE{Elvis1994,
       author = {{Elvis}, Martin and {Wilkes}, Belinda J. and {McDowell}, Jonathan C. and {Green}, Richard F. and {Bechtold}, Jill and {Willner}, S.~P. and {Oey}, M.~S. and {Polomski}, Elisha and {Cutri}, Roc},
        title = "{Atlas of Quasar Energy Distributions}",
      journal = {\apjs},
         year = 1994,
        month = nov,
       volume = {95},
        pages = {1},
          doi = {10.1086/192093},
       adsurl = {https://ui.adsabs.harvard.edu/abs/1994ApJS...95....1E}
}

@ARTICLE{FaranoffRiley1974,
       author = {{Fanaroff}, B.~L. and {Riley}, J.~M.},
        title = "{The morphology of extragalactic radio sources of high and low luminosity}",
      journal = {\mnras},
         year = 1974,
        month = may,
       volume = {167},
        pages = {31P-36P},
          doi = {10.1093/mnras/167.1.31P},
       adsurl = {https://ui.adsabs.harvard.edu/abs/1974MNRAS.167P..31F}
}

@ARTICLE{Ferrari2008,
       author = {{Ferrari}, C. and {Govoni}, F. and {Schindler}, S. and {Bykov}, A.~M. and {Rephaeli}, Y.},
        title = "{Observations of Extended Radio Emission in Clusters}",
      journal = {\ssr},
         year = 2008,
        month = feb,
       volume = {134},
       number = {1-4},
        pages = {93-118},
          doi = {10.1007/s11214-008-9311-x},
archivePrefix = {arXiv},
       eprint = {0801.0985},
 primaryClass = {astro-ph},
       adsurl = {https://ui.adsabs.harvard.edu/abs/2008SSRv..134...93F}
}

@ARTICLE{Hale2021,
       author = {{Hale}, Catherine L. and {McConnell}, D. and {Thomson}, A.~J.~M. and {Lenc}, E. and {Heald}, G.~H. and {Hotan}, A.~W. and {Leung}, J.~K. and {Moss}, V.~A. and {Murphy}, T. and {Pritchard}, J. and {Sadler}, E.~M. and {Stewart}, A.~J. and {Whiting}, M.~T.},
        title = "{The Rapid ASKAP Continuum Survey Paper II: First Stokes I Source Catalogue Data Release}",
      journal = {\pasa},
         year = 2021,
        month = nov,
       volume = {38},
          eid = {e058},
        pages = {e058},
          doi = {10.1017/pasa.2021.47},
archivePrefix = {arXiv},
       eprint = {2109.00956},
 primaryClass = {astro-ph.GA},
       adsurl = {https://ui.adsabs.harvard.edu/abs/2021PASA...38...58H}
}

@ARTICLE{LeFevre2005,
       author = {{Le F{\`e}vre}, O. and {Vettolani}, G. and {Garilli}, B. and {Tresse}, L. and {Bottini}, D. and {Le Brun}, V. and {Maccagni}, D. and {Picat}, J.~P. and {Scaramella}, R. and {Scodeggio}, M. and {Zanichelli}, A. and {Adami}, C. and {Arnaboldi}, M. and {Arnouts}, S. and {Bardelli}, S. and {Bolzonella}, M. and {Cappi}, A. and {Charlot}, S. and {Ciliegi}, P. and {Contini}, T. and {Foucaud}, S. and {Franzetti}, P. and {Gavignaud}, I. and {Guzzo}, L. and {Ilbert}, O. and {Iovino}, A. and {McCracken}, H.~J. and {Marano}, B. and {Marinoni}, C. and {Mathez}, G. and {Mazure}, A. and {Meneux}, B. and {Merighi}, R. and {Paltani}, S. and {Pell{\`o}}, R. and {Pollo}, A. and {Pozzetti}, L. and {Radovich}, M. and {Zamorani}, G. and {Zucca}, E. and {Bondi}, M. and {Bongiorno}, A. and {Busarello}, G. and {Lamareille}, F. and {Mellier}, Y. and {Merluzzi}, P. and {Ripepi}, V. and {Rizzo}, D.},
        title = "{The VIMOS VLT deep survey. First epoch VVDS-deep survey: 11 564 spectra with 17.5 {\ensuremath{\leq}} IAB {\ensuremath{\leq}} 24, and the redshift distribution over 0 {\ensuremath{\leq}} z {\ensuremath{\leq}} 5}",
      journal = {\aap},
         year = 2005,
        month = sep,
       volume = {439},
       number = {3},
        pages = {845-862},
          doi = {10.1051/0004-6361:20041960},
archivePrefix = {arXiv},
       eprint = {astro-ph/0409133},
 primaryClass = {astro-ph},
       adsurl = {https://ui.adsabs.harvard.edu/abs/2005A&A...439..845L}
}

@ARTICLE{Gaia2023,
       author = {{Gaia Collaboration} and {Vallenari}, A. and {Brown}, A.~G.~A. and {Prusti}, T. and {de Bruijne}, J.~H.~J. and {Arenou}, F. and {Babusiaux}, C. and {Biermann}, M. and {Creevey}, O.~L. and {Ducourant}, C. and {Evans}, D.~W. and {Eyer}, L. and {Guerra}, R. and {Hutton}, A. and {Jordi}, C. and {Klioner}, S.~A. and {Lammers}, U.~L. and {Lindegren}, L. and {Luri}, X. and {Mignard}, F. and {Panem}, C. and {Pourbaix}, D. and {Randich}, S. and {Sartoretti}, P. and {Soubiran}, C. and {Tanga}, P. and {Walton}, N.~A. and {Bailer-Jones}, C.~A.~L. and {Bastian}, U. and {Drimmel}, R. and {Jansen}, F. and {Katz}, D. and {Lattanzi}, M.~G. and {van Leeuwen}, F. and {Bakker}, J. and {Cacciari}, C. and {Casta{\~n}eda}, J. and {De Angeli}, F. and {Fabricius}, C. and {Fouesneau}, M. and {Fr{\'e}mat}, Y. and {Galluccio}, L. and {Guerrier}, A. and {Heiter}, U. and {Masana}, E. and {Messineo}, R. and {Mowlavi}, N. and {Nicolas}, C. and {Nienartowicz}, K. and {Pailler}, F. and {Panuzzo}, P. and {Riclet}, F. and {Roux}, W. and {Seabroke}, G.~M. and {Sordo}, R. and {Th{\'e}venin}, F. and {Gracia-Abril}, G. and {Portell}, J. and {Teyssier}, D. and {Altmann}, M. and {Andrae}, R. and {Audard}, M. and {Bellas-Velidis}, I. and {Benson}, K. and {Berthier}, J. and {Blomme}, R. and {Burgess}, P.~W. and {Busonero}, D. and {Busso}, G. and {C{\'a}novas}, H. and {Carry}, B. and {Cellino}, A. and {Cheek}, N. and {Clementini}, G. and {Damerdji}, Y. and {Davidson}, M. and {de Teodoro}, P. and {Nu{\~n}ez Campos}, M. and {Delchambre}, L. and {Dell'Oro}, A. and {Esquej}, P. and {Fern{\'a}ndez-Hern{\'a}ndez}, J. and {Fraile}, E. and {Garabato}, D. and {Garc{\'\i}a-Lario}, P. and {Gosset}, E. and {Haigron}, R. and {Halbwachs}, J. -L. and {Hambly}, N.~C. and {Harrison}, D.~L. and {Hern{\'a}ndez}, J. and {Hestroffer}, D. and {Hodgkin}, S.~T. and {Holl}, B. and {Jan{\ss}en}, K. and {Jevardat de Fombelle}, G. and {Jordan}, S. and {Krone-Martins}, A. and {Lanzafame}, A.~C. and {L{\"o}ffler}, W. and {Marchal}, O. and {Marrese}, P.~M. and {Moitinho}, A. and {Muinonen}, K. and {Osborne}, P. and {Pancino}, E. and {Pauwels}, T. and {Recio-Blanco}, A. and {Reyl{\'e}}, C. and {Riello}, M. and {Rimoldini}, L. and {Roegiers}, T. and {Rybizki}, J. and {Sarro}, L.~M. and {Siopis}, C. and {Smith}, M. and {Sozzetti}, A. and {Utrilla}, E. and {van Leeuwen}, M. and {Abbas}, U. and {{\'A}brah{\'a}m}, P. and {Abreu Aramburu}, A. and {Aerts}, C. and {Aguado}, J.~J. and {Ajaj}, M. and {Aldea-Montero}, F. and {Altavilla}, G. and {{\'A}lvarez}, M.~A. and {Alves}, J. and {Anders}, F. and {Anderson}, R.~I. and {Anglada Varela}, E. and {Antoja}, T. and {Baines}, D. and {Baker}, S.~G. and {Balaguer-N{\'u}{\~n}ez}, L. and {Balbinot}, E. and {Balog}, Z. and {Barache}, C. and {Barbato}, D. and {Barros}, M. and {Barstow}, M.~A. and {Bartolom{\'e}}, S. and {Bassilana}, J. -L. and {Bauchet}, N. and {Becciani}, U. and {Bellazzini}, M. and {Berihuete}, A. and {Bernet}, M. and {Bertone}, S. and {Bianchi}, L. and {Binnenfeld}, A. and {Blanco-Cuaresma}, S. and {Blazere}, A. and {Boch}, T. and {Bombrun}, A. and {Bossini}, D. and {Bouquillon}, S. and {Bragaglia}, A. and {Bramante}, L. and {Breedt}, E. and {Bressan}, A. and {Brouillet}, N. and {Brugaletta}, E. and {Bucciarelli}, B. and {Burlacu}, A. and {Butkevich}, A.~G. and {Buzzi}, R. and {Caffau}, E. and {Cancelliere}, R. and {Cantat-Gaudin}, T. and {Carballo}, R. and {Carlucci}, T. and {Carnerero}, M.~I. and {Carrasco}, J.~M. and {Casamiquela}, L. and {Castellani}, M. and {Castro-Ginard}, A. and {Chaoul}, L. and {Charlot}, P. and {Chemin}, L. and {Chiaramida}, V. and {Chiavassa}, A. and {Chornay}, N. and {Comoretto}, G. and {Contursi}, G. and {Cooper}, W.~J. and {Cornez}, T. and {Cowell}, S. and {Crifo}, F. and {Cropper}, M. and {Crosta}, M. and {Crowley}, C. and {Dafonte}, C. and {Dapergolas}, A. and {David}, M. and {David}, P. and {de Laverny}, P. and {De Luise}, F. and {De March}, R.},
        title = "{Gaia Data Release 3. Summary of the content and survey properties}",
      journal = {\aap},
         year = 2023,
        month = jun,
       volume = {674},
          eid = {A1},
        pages = {A1},
          doi = {10.1051/0004-6361/202243940},
archivePrefix = {arXiv},
       eprint = {2208.00211},
 primaryClass = {astro-ph.GA},
       adsurl = {https://ui.adsabs.harvard.edu/abs/2023A&A...674A...1G}
}

@ARTICLE{deGasperin2021,
       author = {{de Gasperin}, F. and {Williams}, W.~L. and {Best}, P. and {Br{\"u}ggen}, M. and {Brunetti}, G. and {Cuciti}, V. and {Dijkema}, T.~J. and {Hardcastle}, M.~J. and {Norden}, M.~J. and {Offringa}, A. and {Shimwell}, T. and {van Weeren}, R. and {Bomans}, D. and {Bonafede}, A. and {Botteon}, A. and {Callingham}, J.~R. and {Cassano}, R. and {Chy{\.z}y}, K.~T. and {Emig}, K.~L. and {Edler}, H. and {Haverkorn}, M. and {Heald}, G. and {Heesen}, V. and {Iacobelli}, M. and {Intema}, H.~T. and {Kadler}, M. and {Ma{\l}ek}, K. and {Mevius}, M. and {Miley}, G. and {Mingo}, B. and {Morabito}, L.~K. and {Sabater}, J. and {Morganti}, R. and {Orr{\'u}}, E. and {Pizzo}, R. and {Prandoni}, I. and {Shulevski}, A. and {Tasse}, C. and {Vaccari}, M. and {Zarka}, P. and {R{\"o}ttgering}, H.},
        title = "{The LOFAR LBA Sky Survey. I. Survey description and preliminary data release}",
      journal = {\aap},
         year = 2021,
        month = apr,
       volume = {648},
          eid = {A104},
        pages = {A104},
          doi = {10.1051/0004-6361/202140316},
archivePrefix = {arXiv},
       eprint = {2102.09238},
 primaryClass = {astro-ph.IM},
       adsurl = {https://ui.adsabs.harvard.edu/abs/2021A&A...648A.104D}
}

@ARTICLE{Gregory1996,
       author = {{Gregory}, P.~C. and {Scott}, W.~K. and {Douglas}, K. and {Condon}, J.~J.},
        title = "{The GB6 Catalog of Radio Sources}",
      journal = {\apjs},
         year = 1996,
        month = apr,
       volume = {103},
        pages = {427},
          doi = {10.1086/192282},
       adsurl = {https://ui.adsabs.harvard.edu/abs/1996ApJS..103..427G}
}

@ARTICLE{Gurkan2014,
       author = {{G{\"u}rkan}, G. and {Hardcastle}, M.~J. and {Jarvis}, M.~J.},
        title = "{The Wide-field Infrared Survey Explorer properties of complete samples of radio-loud active galactic nucleus}",
      journal = {\mnras},
         year = 2014,
        month = feb,
       volume = {438},
       number = {2},
        pages = {1149-1161},
          doi = {10.1093/mnras/stt2264},
archivePrefix = {arXiv},
       eprint = {1308.4843},
 primaryClass = {astro-ph.CO},
       adsurl = {https://ui.adsabs.harvard.edu/abs/2014MNRAS.438.1149G}
}

@ARTICLE{Haarlem2013,
       author = {{van Haarlem}, M.~P. and {Wise}, M.~W. and {Gunst}, A.~W. and {Heald}, G. and {McKean}, J.~P. and {Hessels}, J.~W.~T. and {de Bruyn}, A.~G. and {Nijboer}, R. and {Swinbank}, J. and {Fallows}, R. and {Brentjens}, M. and {Nelles}, A. and {Beck}, R. and {Falcke}, H. and {Fender}, R. and {H{\"o}randel}, J. and {Koopmans}, L.~V.~E. and {Mann}, G. and {Miley}, G. and {R{\"o}ttgering}, H. and {Stappers}, B.~W. and {Wijers}, R.~A.~M.~J. and {Zaroubi}, S. and {van den Akker}, M. and {Alexov}, A. and {Anderson}, J. and {Anderson}, K. and {van Ardenne}, A. and {Arts}, M. and {Asgekar}, A. and {Avruch}, I.~M. and {Batejat}, F. and {B{\"a}hren}, L. and {Bell}, M.~E. and {Bell}, M.~R. and {van Bemmel}, I. and {Bennema}, P. and {Bentum}, M.~J. and {Bernardi}, G. and {Best}, P. and {B{\^\i}rzan}, L. and {Bonafede}, A. and {Boonstra}, A. -J. and {Braun}, R. and {Bregman}, J. and {Breitling}, F. and {van de Brink}, R.~H. and {Broderick}, J. and {Broekema}, P.~C. and {Brouw}, W.~N. and {Br{\"u}ggen}, M. and {Butcher}, H.~R. and {van Cappellen}, W. and {Ciardi}, B. and {Coenen}, T. and {Conway}, J. and {Coolen}, A. and {Corstanje}, A. and {Damstra}, S. and {Davies}, O. and {Deller}, A.~T. and {Dettmar}, R. -J. and {van Diepen}, G. and {Dijkstra}, K. and {Donker}, P. and {Doorduin}, A. and {Dromer}, J. and {Drost}, M. and {van Duin}, A. and {Eisl{\"o}ffel}, J. and {van Enst}, J. and {Ferrari}, C. and {Frieswijk}, W. and {Gankema}, H. and {Garrett}, M.~A. and {de Gasperin}, F. and {Gerbers}, M. and {de Geus}, E. and {Grie{\ss}meier}, J. -M. and {Grit}, T. and {Gruppen}, P. and {Hamaker}, J.~P. and {Hassall}, T. and {Hoeft}, M. and {Holties}, H.~A. and {Horneffer}, A. and {van der Horst}, A. and {van Houwelingen}, A. and {Huijgen}, A. and {Iacobelli}, M. and {Intema}, H. and {Jackson}, N. and {Jelic}, V. and {de Jong}, A. and {Juette}, E. and {Kant}, D. and {Karastergiou}, A. and {Koers}, A. and {Kollen}, H. and {Kondratiev}, V.~I. and {Kooistra}, E. and {Koopman}, Y. and {Koster}, A. and {Kuniyoshi}, M. and {Kramer}, M. and {Kuper}, G. and {Lambropoulos}, P. and {Law}, C. and {van Leeuwen}, J. and {Lemaitre}, J. and {Loose}, M. and {Maat}, P. and {Macario}, G. and {Markoff}, S. and {Masters}, J. and {McFadden}, R.~A. and {McKay-Bukowski}, D. and {Meijering}, H. and {Meulman}, H. and {Mevius}, M. and {Middelberg}, E. and {Millenaar}, R. and {Miller-Jones}, J.~C.~A. and {Mohan}, R.~N. and {Mol}, J.~D. and {Morawietz}, J. and {Morganti}, R. and {Mulcahy}, D.~D. and {Mulder}, E. and {Munk}, H. and {Nieuwenhuis}, L. and {van Nieuwpoort}, R. and {Noordam}, J.~E. and {Norden}, M. and {Noutsos}, A. and {Offringa}, A.~R. and {Olofsson}, H. and {Omar}, A. and {Orr{\'u}}, E. and {Overeem}, R. and {Paas}, H. and {Pandey-Pommier}, M. and {Pandey}, V.~N. and {Pizzo}, R. and {Polatidis}, A. and {Rafferty}, D. and {Rawlings}, S. and {Reich}, W. and {de Reijer}, J. -P. and {Reitsma}, J. and {Renting}, G.~A. and {Riemers}, P. and {Rol}, E. and {Romein}, J.~W. and {Roosjen}, J. and {Ruiter}, M. and {Scaife}, A. and {van der Schaaf}, K. and {Scheers}, B. and {Schellart}, P. and {Schoenmakers}, A. and {Schoonderbeek}, G. and {Serylak}, M. and {Shulevski}, A. and {Sluman}, J. and {Smirnov}, O. and {Sobey}, C. and {Spreeuw}, H. and {Steinmetz}, M. and {Sterks}, C.~G.~M. and {Stiepel}, H. -J. and {Stuurwold}, K. and {Tagger}, M. and {Tang}, Y. and {Tasse}, C. and {Thomas}, I. and {Thoudam}, S. and {Toribio}, M.~C. and {van der Tol}, B. and {Usov}, O. and {van Veelen}, M. and {van der Veen}, A. -J. and {ter Veen}, S. and {Verbiest}, J.~P.~W. and {Vermeulen}, R. and {Vermaas}, N. and {Vocks}, C. and {Vogt}, C. and {de Vos}, M. and {van der Wal}, E. and {van Weeren}, R. and {Weggemans}, H. and {Weltevrede}, P. and {White}, S. and {Wijnholds}, S.~J. and {Wilhelmsson}, T. and {Wucknitz}, O. and {Yatawatta}, S. and {Zarka}, P. and {Zensus}, A. and {van Zwieten}, J.},
        title = "{LOFAR: The LOw-Frequency ARray}",
      journal = {\aap},
         year = 2013,
        month = aug,
       volume = {556},
          eid = {A2},
        pages = {A2},
          doi = {10.1051/0004-6361/201220873},
archivePrefix = {arXiv},
       eprint = {1305.3550},
 primaryClass = {astro-ph.IM},
       adsurl = {https://ui.adsabs.harvard.edu/abs/2013A&A...556A...2V}
}

@ARTICLE{Hardcastle2023,
       author = {{Hardcastle}, M.~J. and {Horton}, M.~A. and {Williams}, W.~L. and {Duncan}, K.~J. and {Alegre}, L. and {Barkus}, B. and {Croston}, J.~H. and {Dickinson}, H. and {Osinga}, E. and {R{\"o}ttgering}, H.~J.~A. and {Sabater}, J. and {Shimwell}, T.~W. and {Smith}, D.~J.~B. and {Best}, P.~N. and {Botteon}, A. and {Br{\"u}ggen}, M. and {Drabent}, A. and {de Gasperin}, F. and {G{\"u}rkan}, G. and {Hajduk}, M. and {Hale}, C.~L. and {Hoeft}, M. and {Jamrozy}, M. and {Kunert-Bajraszewska}, M. and {Kondapally}, R. and {Magliocchetti}, M. and {Mahatma}, V.~H. and {Mostert}, R.~I.~J. and {O'Sullivan}, S.~P. and {Pajdosz-{\'S}mierciak}, U. and {Petley}, J. and {Pierce}, J.~C.~S. and {Prandoni}, I. and {Schwarz}, D.~J. and {Shulewski}, A. and {Siewert}, T.~M. and {Stott}, J.~P. and {Tang}, H. and {Vaccari}, M. and {Zheng}, X. and {Bailey}, T. and {Desbled}, S. and {Goyal}, A. and {Gonano}, V. and {Hanset}, M. and {Kurtz}, W. and {Lim}, S.~M. and {Mielle}, L. and {Molloy}, C.~S. and {Roth}, R. and {Terentev}, I.~A. and {Torres}, M.},
        title = "{The LOFAR Two-Metre Sky Survey. VI. Optical identifications for the second data release}",
      journal = {\aap},
         year = 2023,
        month = oct,
       volume = {678},
          eid = {A151},
        pages = {A151},
          doi = {10.1051/0004-6361/202347333},
archivePrefix = {arXiv},
       eprint = {2309.00102},
 primaryClass = {astro-ph.GA},
       adsurl = {https://ui.adsabs.harvard.edu/abs/2023A&A...678A.151H}
}

@ARTICLE{Hardcastle2025,
       author = {{Hardcastle}, M.~J. and {Pierce}, J.~C.~S. and {Duncan}, K.~J. and {G{\"u}rkan}, G. and {Gong}, Y. and {Horton}, M.~A. and {Mingo}, B. and {R{\"o}ttgering}, H.~J.~A. and {Smith}, D.~J.~B.},
        title = "{Radio AGN selection in LoTSS DR2}",
      journal = {\mnras},
         year = 2025,
        month = may,
       volume = {539},
       number = {2},
        pages = {1856-1878},
          doi = {10.1093/mnras/staf622},
archivePrefix = {arXiv},
       eprint = {2504.09303},
 primaryClass = {astro-ph.GA},
       adsurl = {https://ui.adsabs.harvard.edu/abs/2025MNRAS.539.1856H}
}

@ARTICLE{Haynes2018,
       author = {{Haynes}, Martha P. and {Giovanelli}, Riccardo and {Kent}, Brian R. and {Adams}, Elizabeth A.~K. and {Balonek}, Thomas J. and {Craig}, David W. and {Fertig}, Derek and {Finn}, Rose and {Giovanardi}, Carlo and {Hallenbeck}, Gregory and {Hess}, Kelley M. and {Hoffman}, G. Lyle and {Huang}, Shan and {Jones}, Michael G. and {Koopmann}, Rebecca A. and {Kornreich}, David A. and {Leisman}, Lukas and {Miller}, Jeffrey and {Moorman}, Crystal and {O'Connor}, Jessica and {O'Donoghue}, Aileen and {Papastergis}, Emmanouil and {Troischt}, Parker and {Stark}, David and {Xiao}, Li},
        title = "{The Arecibo Legacy Fast ALFA Survey: The ALFALFA Extragalactic H I Source Catalog}",
      journal = {\apj},
         year = 2018,
        month = jul,
       volume = {861},
       number = {1},
          eid = {49},
        pages = {49},
          doi = {10.3847/1538-4357/aac956},
archivePrefix = {arXiv},
       eprint = {1805.11499},
 primaryClass = {astro-ph.GA},
       adsurl = {https://ui.adsabs.harvard.edu/abs/2018ApJ...861...49H}
}

@ARTICLE{Helou1985,
       author = {{Helou}, G. and {Soifer}, B.~T. and {Rowan-Robinson}, M.},
        title = "{Thermal infrared and nonthermal radio : remarkable correlation in disks of galaxies.}",
      journal = {\apjl},
         year = 1985,
        month = nov,
       volume = {298},
        pages = {L7-L11},
          doi = {10.1086/184556},
       adsurl = {https://ui.adsabs.harvard.edu/abs/1985ApJ...298L...7H}
}

@ARTICLE{Hernan2021,
       author = {{Hern{\'a}n-Caballero}, A. and {Varela}, J. and {L{\'o}pez-Sanjuan}, C. and {Muniesa}, D. and {Civera}, T. and {Chaves-Montero}, J. and {D{\'\i}az-Garc{\'\i}a}, L.~A. and {Laur}, J. and {Hern{\'a}ndez-Monteagudo}, C. and {Abramo}, R. and {Angulo}, R. and {Crist{\'o}bal-Hornillos}, D. and {Gonz{\'a}lez Delgado}, R.~M. and {Greisel}, N. and {Orsi}, A. and {Queiroz}, C. and {Sobral}, D. and {Tamm}, A. and {Tempel}, E. and {V{\'a}zquez Rami{\'o}}, H. and {Alcaniz}, J. and {Ben{\'\i}tez}, N. and {Bonoli}, S. and {Carneiro}, S. and {Cenarro}, J. and {Dupke}, R. and {Ederoclite}, A. and {Mar{\'\i}n-Franch}, A. and {Mendes de Oliveira}, C. and {Moles}, M. and {Sodr{\'e}}, L. and {Taylor}, K. and {Cypriano}, E.~S. and {Mart{\'\i}nez-Solaeche}, G.},
        title = "{The miniJPAS survey: Photometric redshift catalogue}",
      journal = {\aap},
         year = 2021,
        month = oct,
       volume = {654},
          eid = {A101},
        pages = {A101},
          doi = {10.1051/0004-6361/202141236},
archivePrefix = {arXiv},
       eprint = {2108.03271},
 primaryClass = {astro-ph.GA},
       adsurl = {https://ui.adsabs.harvard.edu/abs/2021A&A...654A.101H}
}

@ARTICLE{Heckman2014,
       author = {{Heckman}, Timothy M. and {Best}, Philip N.},
        title = "{The Coevolution of Galaxies and Supermassive Black Holes: Insights from Surveys of the Contemporary Universe}",
      journal = {\araa},
         year = 2014,
        month = aug,
       volume = {52},
        pages = {589-660},
          doi = {10.1146/annurev-astro-081913-035722},
archivePrefix = {arXiv},
       eprint = {1403.4620},
 primaryClass = {astro-ph.GA},
       adsurl = {https://ui.adsabs.harvard.edu/abs/2014ARA&A..52..589H}
}

@ARTICLE{Hernan2024,
       author = {{Hern{\'a}n-Caballero}, A. and {Akhlaghi}, M. and {L{\'o}pez-Sanjuan}, C. and {V{\'a}zquez Rami{\'o}}, H. and {Laur}, J. and {Varela}, J. and {Civera}, T. and {Muniesa}, D. and {Finoguenov}, A. and {Fern{\'a}ndez-Ontiveros}, J.~A. and {Dom{\'\i}nguez S{\'a}nchez}, H. and {Chaves-Montero}, J. and {Fern{\'a}ndez-Soto}, A. and {Lumbreras-Calle}, A. and {D{\'\i}az-Garc{\'\i}a}, L.~A. and {del Pino}, A. and {Gonz{\'a}lez Delgado}, R.~M. and {Hern{\'a}ndez-Monteagudo}, C. and {Coelho}, P. and {Jim{\'e}nez-Teja}, Y. and {Lopes}, P.~A.~A. and {Marra}, V. and {Tempel}, E. and {V{\'\i}lchez}, J.~M. and {Abramo}, R. and {Alcaniz}, J. and {Ben{\'\i}tez}, N. and {Bonoli}, S. and {Carneiro}, S. and {Cenarro}, J. and {Crist{\'o}bal-Hornillos}, D. and {Dupke}, R. and {Ederoclite}, A. and {Mar{\'\i}n-Franch}, A. and {Mendes de Oliveira}, C. and {Moles}, M. and {Sodr{\'e}}, L. and {Taylor}, K.},
        title = "{The miniJPAS survey: Maximising the photo-z accuracy from multi-survey datasets with probability conflation}",
      journal = {\aap},
         year = 2024,
        month = apr,
       volume = {684},
          eid = {A61},
        pages = {A61},
          doi = {10.1051/0004-6361/202348513},
archivePrefix = {arXiv},
       eprint = {2311.04220},
 primaryClass = {astro-ph.GA},
       adsurl = {https://ui.adsabs.harvard.edu/abs/2024A&A...684A..61H}
}

@ARTICLE{Ho2002,
       author = {{Ho}, Luis C.},
        title = "{On the Relationship between Radio Emission and Black Hole Mass in Galactic Nuclei}",
      journal = {\apj},
         year = 2002,
        month = jan,
       volume = {564},
       number = {1},
        pages = {120-132},
          doi = {10.1086/324399},
archivePrefix = {arXiv},
       eprint = {astro-ph/0110440},
 primaryClass = {astro-ph},
       adsurl = {https://ui.adsabs.harvard.edu/abs/2002ApJ...564..120H}
}

@ARTICLE{Hurley2017,
       author = {{Hurley-Walker}, N. and {Callingham}, J.~R. and {Hancock}, P.~J. and {Franzen}, T.~M.~O. and {Hindson}, L. and {Kapi{\'n}ska}, A.~D. and {Morgan}, J. and {Offringa}, A.~R. and {Wayth}, R.~B. and {Wu}, C. and {Zheng}, Q. and {Murphy}, T. and {Bell}, M.~E. and {Dwarakanath}, K.~S. and {For}, B. and {Gaensler}, B.~M. and {Johnston-Hollitt}, M. and {Lenc}, E. and {Procopio}, P. and {Staveley-Smith}, L. and {Ekers}, R. and {Bowman}, J.~D. and {Briggs}, F. and {Cappallo}, R.~J. and {Deshpande}, A.~A. and {Greenhill}, L. and {Hazelton}, B.~J. and {Kaplan}, D.~L. and {Lonsdale}, C.~J. and {McWhirter}, S.~R. and {Mitchell}, D.~A. and {Morales}, M.~F. and {Morgan}, E. and {Oberoi}, D. and {Ord}, S.~M. and {Prabu}, T. and {Shankar}, N. Udaya and {Srivani}, K.~S. and {Subrahmanyan}, R. and {Tingay}, S.~J. and {Webster}, R.~L. and {Williams}, A. and {Williams}, C.~L.},
        title = "{GaLactic and Extragalactic All-sky Murchison Widefield Array (GLEAM) survey - I. A low-frequency extragalactic catalogue}",
      journal = {\mnras},
         year = 2017,
        month = jan,
       volume = {464},
       number = {1},
        pages = {1146-1167},
          doi = {10.1093/mnras/stw2337},
archivePrefix = {arXiv},
       eprint = {1610.08318},
 primaryClass = {astro-ph.GA},
       adsurl = {https://ui.adsabs.harvard.edu/abs/2017MNRAS.464.1146H}
}

@ARTICLE{Ivezic2002,
       author = {{Ivezi{\'c}}, {\v{Z}}eljko and {Menou}, Kristen and {Knapp}, Gillian R. and {Strauss}, Michael A. and {Lupton}, Robert H. and {Vanden Berk}, Daniel E. and {Richards}, Gordon T. and {Tremonti}, Christy and {Weinstein}, Michael A. and {Anderson}, Scott and {Bahcall}, Neta A. and {Becker}, Robert H. and {Bernardi}, Mariangela and {Blanton}, Michael and {Eisenstein}, Daniel and {Fan}, Xiaohui and {Finkbeiner}, Douglas and {Finlator}, Kristian and {Frieman}, Joshua and {Gunn}, James E. and {Hall}, Pat B. and {Kim}, Rita S.~J. and {Kinkhabwala}, Ali and {Narayanan}, Vijay K. and {Rockosi}, Constance M. and {Schlegel}, David and {Schneider}, Donald P. and {Strateva}, Iskra and {SubbaRao}, Mark and {Thakar}, Aniruddha R. and {Voges}, Wolfgang and {White}, Richard L. and {Yanny}, Brian and {Brinkmann}, Jonathan and {Doi}, Mamoru and {Fukugita}, Masataka and {Hennessy}, Gregory S. and {Munn}, Jeffrey A. and {Nichol}, Robert C. and {York}, Donald G.},
        title = "{Optical and Radio Properties of Extragalactic Sources Observed by the FIRST Survey and the Sloan Digital Sky Survey}",
      journal = {\aj},
         year = 2002,
        month = nov,
       volume = {124},
       number = {5},
        pages = {2364-2400},
          doi = {10.1086/344069},
archivePrefix = {arXiv},
       eprint = {astro-ph/0202408},
 primaryClass = {astro-ph},
       adsurl = {https://ui.adsabs.harvard.edu/abs/2002AJ....124.2364I}
}

@ARTICLE{Jansky1933,
       author = {{Jansky}, Karl G.},
        title = "{Radio Waves from Outside the Solar System}",
      journal = {\nat},
         year = 1933,
        month = jul,
       volume = {132},
       number = {3323},
        pages = {66},
          doi = {10.1038/132066a0},
       adsurl = {https://ui.adsabs.harvard.edu/abs/1933Natur.132...66J}
}

@ARTICLE{Kellermann1989,
       author = {{Kellermann}, K.~I. and {Sramek}, R. and {Schmidt}, M. and {Shaffer}, D.~B. and {Green}, R.},
        title = "{VLA Observations of Objects in the Palomar Bright Quasar Survey}",
      journal = {\aj},
         year = 1989,
        month = oct,
       volume = {98},
        pages = {1195},
          doi = {10.1086/115207},
       adsurl = {https://ui.adsabs.harvard.edu/abs/1989AJ.....98.1195K}
}

@ARTICLE{Kellermann2016,
       author = {{Kellermann}, K.~I. and {Condon}, J.~J. and {Kimball}, A.~E. and {Perley}, R.~A. and {Ivezi{\'c}}, {\v{Z}}eljko},
        title = "{Radio-loud and Radio-quiet QSOs}",
      journal = {\apj},
         year = 2016,
        month = nov,
       volume = {831},
       number = {2},
          eid = {168},
        pages = {168},
          doi = {10.3847/0004-637X/831/2/168},
archivePrefix = {arXiv},
       eprint = {1608.04586},
 primaryClass = {astro-ph.GA},
       adsurl = {https://ui.adsabs.harvard.edu/abs/2016ApJ...831..168K}
}

@ARTICLE{Kewley2001,
       author = {{Kewley}, L.~J. and {Dopita}, M.~A. and {Sutherland}, R.~S. and {Heisler}, C.~A. and {Trevena}, J.},
        title = "{Theoretical Modeling of Starburst Galaxies}",
      journal = {\apj},
         year = 2001,
        month = jul,
       volume = {556},
       number = {1},
        pages = {121-140},
          doi = {10.1086/321545},
archivePrefix = {arXiv},
       eprint = {astro-ph/0106324},
 primaryClass = {astro-ph},
       adsurl = {https://ui.adsabs.harvard.edu/abs/2001ApJ...556..121K}
}

@ARTICLE{Kuzmicz2018,
       author = {{Ku{\'z}micz}, Agnieszka and {Jamrozy}, Marek and {Bronarska}, Katarzyna and {Janda-Boczar}, Katarzyna and {Saikia}, D.~J.},
        title = "{An Updated Catalog of Giant Radio Sources}",
      journal = {\apjs},
         year = 2018,
        month = sep,
       volume = {238},
       number = {1},
          eid = {9},
        pages = {9},
          doi = {10.3847/1538-4365/aad9ff},
archivePrefix = {arXiv},
       eprint = {1809.09008},
 primaryClass = {astro-ph.GA},
       adsurl = {https://ui.adsabs.harvard.edu/abs/2018ApJS..238....9K}
}

@ARTICLE{Lang2014,
       author = {{Lang}, Dustin},
        title = "{unWISE: Unblurred Coadds of the WISE Imaging}",
      journal = {\aj},
         year = 2014,
        month = may,
       volume = {147},
       number = {5},
          eid = {108},
        pages = {108},
          doi = {10.1088/0004-6256/147/5/108},
archivePrefix = {arXiv},
       eprint = {1405.0308},
 primaryClass = {astro-ph.IM},
       adsurl = {https://ui.adsabs.harvard.edu/abs/2014AJ....147..108L}
}

@INPROCEEDINGS{Laing1994,
       author = {{Laing}, R.~A. and {Jenkins}, C.~R. and {Wall}, J.~V. and {Unger}, S.~W.},
        title = "{Spectrophotometry of a Complete Sample of 3CR Radio Sources: Implications for Unified Models}",
    booktitle = {The Physics of Active Galaxies},
         year = 1994,
       editor = {{Bicknell}, Geoffrey V. and {Dopita}, Michael A. and {Quinn}, Peter J.},
       series = {Astronomical Society of the Pacific Conference Series},
       volume = {54},
        month = jan,
        pages = {201},
       adsurl = {https://ui.adsabs.harvard.edu/abs/1994ASPC...54..201L}
}

@ARTICLE{Laor2000,
       author = {{Laor}, Ari},
        title = "{On Black Hole Masses and Radio Loudness in Active Galactic Nuclei}",
      journal = {\apjl},
         year = 2000,
        month = nov,
       volume = {543},
       number = {2},
        pages = {L111-L114},
          doi = {10.1086/317280},
archivePrefix = {arXiv},
       eprint = {astro-ph/0009192},
 primaryClass = {astro-ph},
       adsurl = {https://ui.adsabs.harvard.edu/abs/2000ApJ...543L.111L}
}

@ARTICLE{LopezSanJuan2019,
       author = {{L{\'o}pez-Sanjuan}, C. and {V{\'a}zquez Rami{\'o}}, H. and {Varela}, J. and {Spinoso}, D. and {Angulo}, R.~E. and {Muniesa}, D. and {Viironen}, K. and {Crist{\'o}bal-Hornillos}, D. and {Cenarro}, A.~J. and {Ederoclite}, A. and {Mar{\'\i}n-Franch}, A. and {Moles}, M. and {Ascaso}, B. and {Bonoli}, S. and {Chies-Santos}, A.~L. and {Coelho}, P.~R.~T. and {Costa-Duarte}, M.~V. and {Cortesi}, A. and {D{\'\i}az-Garc{\'\i}a}, L.~A. and {Dupke}, R.~A. and {Galbany}, L. and {Hern{\'a}ndez-Monteagudo}, C. and {Logro{\~n}o-Garc{\'\i}a}, R. and {Molino}, A. and {Orsi}, A. and {Placco}, V.~M. and {Sampedro}, L. and {San Roman}, I. and {Vilella-Rojo}, G. and {Whitten}, D.~D. and {Mendes de Oliveira}, C.~L. and {Sodr{\'e}}, L.},
        title = "{J-PLUS: Morphological star/galaxy classification by PDF analysis}",
      journal = {\aap},
         year = 2019,
        month = feb,
       volume = {622},
          eid = {A177},
        pages = {A177},
          doi = {10.1051/0004-6361/201732480},
archivePrefix = {arXiv},
       eprint = {1804.02673},
 primaryClass = {astro-ph.GA},
       adsurl = {https://ui.adsabs.harvard.edu/abs/2019A&A...622A.177L}
}

@ARTICLE{LopezSanJuan24,
       author = {{L{\'o}pez-Sanjuan}, C. and {V{\'a}zquez Rami{\'o}}, H. and {Xiao}, K. and {Yuan}, H. and {Carrasco}, J.~M. and {Varela}, J. and {Crist{\'o}bal-Hornillos}, D. and {Tremblay}, P. -E. and {Ederoclite}, A. and {Mar{\'\i}n-Franch}, A. and {Cenarro}, A.~J. and {Coelho}, P.~R.~T. and {Daflon}, S. and {del Pino}, A. and {Dom{\'\i}nguez S{\'a}nchez}, H. and {Fern{\'a}ndez-Ontiveros}, J.~A. and {Hern{\'a}n-Caballero}, A. and {Jim{\'e}nez-Esteban}, F.~M. and {Alcaniz}, J. and {Angulo}, R.~E. and {Dupke}, R.~A. and {Hern{\'a}ndez-Monteagudo}, C. and {Moles}, M. and {Sodr{\'e}}, L.},
        title = "{J-PLUS: Toward a homogeneous photometric calibration using Gaia BP/RP low-resolution spectra}",
      journal = {\aap},
         year = 2024,
        month = mar,
       volume = {683},
          eid = {A29},
        pages = {A29},
          doi = {10.1051/0004-6361/202346012},
archivePrefix = {arXiv},
       eprint = {2301.12395},
 primaryClass = {astro-ph.SR},
       adsurl = {https://ui.adsabs.harvard.edu/abs/2024A&A...683A..29L}
}

@ARTICLE{Lumbreras2022,
       author = {{Lumbreras-Calle}, A. and {L{\'o}pez-Sanjuan}, C. and {Sobral}, D. and {Fern{\'a}ndez-Ontiveros}, J.~A. and {V{\'\i}lchez}, J.~M. and {Hern{\'a}n-Caballero}, A. and {Akhlaghi}, M. and {D{\'\i}az-Garc{\'\i}a}, L.~A. and {Alcaniz}, J. and {Angulo}, R.~E. and {Cenarro}, A.~J. and {Crist{\'o}bal-Hornillos}, D. and {Dupke}, R.~A. and {Ederoclite}, A. and {Hern{\'a}ndez-Monteagudo}, C. and {Mar{\'\i}n-Franch}, A. and {Moles}, M. and {Sodr{\'e}}, L. and {V{\'a}zquez Rami{\'o}}, H. and {Varela}, J.},
        title = "{J-PLUS: Uncovering a large population of extreme [OIII] emitters in the local Universe}",
      journal = {\aap},
         year = 2022,
        month = dec,
       volume = {668},
          eid = {A60},
        pages = {A60},
          doi = {10.1051/0004-6361/202142898},
archivePrefix = {arXiv},
       eprint = {2112.06938},
 primaryClass = {astro-ph.GA},
       adsurl = {https://ui.adsabs.harvard.edu/abs/2022A&A...668A..60L}
}

@ARTICLE{MadauDickinson2014,
       author = {{Madau}, Piero and {Dickinson}, Mark},
        title = "{Cosmic Star-Formation History}",
      journal = {\araa},
         year = 2014,
        month = aug,
       volume = {52},
        pages = {415-486},
          doi = {10.1146/annurev-astro-081811-125615},
archivePrefix = {arXiv},
       eprint = {1403.0007},
 primaryClass = {astro-ph.CO},
       adsurl = {https://ui.adsabs.harvard.edu/abs/2014ARA&A..52..415M}
}

@ARTICLE{Mahatma2018,
       author = {{Mahatma}, V.~H. and {Hardcastle}, M.~J. and {Williams}, W.~L. and {Brienza}, M. and {Br{\"u}ggen}, M. and {Croston}, J.~H. and {Gurkan}, G. and {Harwood}, J.~J. and {Kunert-Bajraszewska}, M. and {Morganti}, R. and {R{\"o}ttgering}, H.~J.~A. and {Shimwell}, T.~W. and {Tasse}, C.},
        title = "{Remnant radio-loud AGN in the Herschel-ATLAS field}",
      journal = {\mnras},
         year = 2018,
        month = apr,
       volume = {475},
       number = {4},
        pages = {4557-4578},
          doi = {10.1093/mnras/sty025},
archivePrefix = {arXiv},
       eprint = {1801.01067},
 primaryClass = {astro-ph.GA},
       adsurl = {https://ui.adsabs.harvard.edu/abs/2018MNRAS.475.4557M}
}

@ARTICLE{Mahony2012,
       author = {{Mahony}, Elizabeth K. and {Sadler}, Elaine M. and {Croom}, Scott M. and {Ekers}, Ronald D. and {Feain}, Ilana J. and {Murphy}, Tara},
        title = "{Is the Observed High-frequency Radio Luminosity Distribution of QSOs Bimodal?}",
      journal = {\apj},
         year = 2012,
        month = jul,
       volume = {754},
       number = {1},
          eid = {12},
        pages = {12},
          doi = {10.1088/0004-637X/754/1/12},
archivePrefix = {arXiv},
       eprint = {1205.2233},
 primaryClass = {astro-ph.CO},
       adsurl = {https://ui.adsabs.harvard.edu/abs/2012ApJ...754...12M}
}

@INPROCEEDINGS{MarinFranch15,
       author = {{Mar{\'{\i}}n-Franch}, Antonio and {Taylor}, Keith and {Cenarro}, Javier and
         {Cristobal-Hornillos}, David and {Moles}, Mariano},
        title = "{T80Cam: a wide field camera for the J-PLUS survey}",
    booktitle = {IAU General Assembly},
         year = "2015",
       volume = {29},
        month = "Aug",
          eid = {2257381},
        pages = {2257381},
       adsurl = {https://ui.adsabs.harvard.edu/abs/2015IAUGA..2257381M}
}

@dataset{Marocco2020,
       author = {{Marocco}, F. and {Eisenhardt}, P.~R.~M. and {Fowler}, J.~W. and {Kirkpatrick}, J.~D. and {Meisner}, A.~M. and {Schlafly}, E.~F. and {Stanford}, S.~A. and {Garcia}, N. and {Caselden}, D. and {Cushing}, M.~C. and {Cutri}, R.~M. and {Faherty}, J.~K. and {Gelino}, C.~R. and {Gonzalez}, A.~H. and {Jarrett}, T.~H. and {Koontz}, R. and {Mainzer}, A. and {Marchese}, E.~J. and {Mobasher}, B. and {Schlegel}, D.~J. and {Stern}, D. and {Teplitz}, H.~I. and {Wright E.~L. (The Catwise Team)}},
        title = "{VizieR Online Data Catalog: The CatWISE2020 catalog (updated version 28-Jan-2021) (Marocco+, 2021)}",
 howpublished = {VizieR On-line Data Catalog: II/365.  Originally published in: 2021ApJS..253....8M; 2020arXiv201213084M; 2020ApJS..247...69E},
         year = 2020,
        month = jul,
          eid = {II/365},
       adsurl = {https://ui.adsabs.harvard.edu/abs/2020yCat.2365....0M}
}

@ARTICLE{Mauch2007,
       author = {{Mauch}, Tom and {Sadler}, Elaine M.},
        title = "{Radio sources in the 6dFGS: local luminosity functions at 1.4 GHz for star-forming galaxies and radio-loud AGN}",
      journal = {\mnras},
         year = 2007,
        month = mar,
       volume = {375},
       number = {3},
        pages = {931-950},
          doi = {10.1111/j.1365-2966.2006.11353.x},
archivePrefix = {arXiv},
       eprint = {astro-ph/0612018},
 primaryClass = {astro-ph},
       adsurl = {https://ui.adsabs.harvard.edu/abs/2007MNRAS.375..931M}
}

@dataset{Mauch2013,
       author = {{Mauch}, T. and {Murphy}, T. and {Buttery}, H.~J. and {Curran}, J. and {Hunstead}, R.~W. and {Piestrzynski}, B. and {Robertson}, J.~G. and {Sadler}, E.~M.},
        title = "{VizieR Online Data Catalog: Sydney University Molonglo Sky Survey (SUMSS V2.1) (Mauch+ 2008)}",
 howpublished = {VizieR On-line Data Catalog: VIII/81B.  Originally published in: 2003MNRAS.342.1117M; 2013yCat.8081....0M},
         year = 2013,
        month = jun,
          eid = {VIII/81B},
       adsurl = {https://ui.adsabs.harvard.edu/abs/2013yCat.8081....0M}
}

@INCOLLECTION{Melrose1985,
       author = {{Melrose}, D.~B.},
        title = "{Plasma emission mechanisms.}",
    booktitle = {Solar Radiophysics: Studies of Emission from the Sun at Metre Wavelengths},
         year = 1985,
    publisher = {Cambridge University Press},
       editor = {{McLean}, D.~J. and {Labrum}, N.~R.},
        pages = {177-210},
       adsurl = {https://ui.adsabs.harvard.edu/abs/1985srph.book..177M}
}

@ARTICLE{Merloni2007,
       author = {{Merloni}, Andrea and {Heinz}, Sebastian},
        title = "{Measuring the kinetic power of active galactic nuclei in the radio mode}",
      journal = {\mnras},
         year = 2007,
        month = oct,
       volume = {381},
       number = {2},
        pages = {589-601},
          doi = {10.1111/j.1365-2966.2007.12253.x},
archivePrefix = {arXiv},
       eprint = {0707.3356},
 primaryClass = {astro-ph},
       adsurl = {https://ui.adsabs.harvard.edu/abs/2007MNRAS.381..589M}
}

@ARTICLE{Mingo2016,
       author = {{Mingo}, B. and {Watson}, M.~G. and {Rosen}, S.~R. and {Hardcastle}, M.~J. and {Ruiz}, A. and {Blain}, A. and {Carrera}, F.~J. and {Mateos}, S. and {Pineau}, F. -X. and {Stewart}, G.~C.},
        title = "{The MIXR sample: AGN activity versus star formation across the cross-correlation of WISE, 3XMM, and FIRST/NVSS}",
      journal = {\mnras},
         year = 2016,
        month = nov,
       volume = {462},
       number = {3},
        pages = {2631-2667},
          doi = {10.1093/mnras/stw1826},
archivePrefix = {arXiv},
       eprint = {1607.06471},
 primaryClass = {astro-ph.GA},
       adsurl = {https://ui.adsabs.harvard.edu/abs/2016MNRAS.462.2631M}
}

@ARTICLE{Molino2019,
       author = {{Molino}, A. and {Costa-Duarte}, M.~V. and {Mendes de Oliveira}, C. and {Cenarro}, A.~J. and {Lima Neto}, G.~B. and {Cypriano}, E.~S. and {Sodr{\'e}}, L. and {Coelho}, P. and {Chow-Mart{\'\i}nez}, M. and {Monteiro-Oliveira}, R. and {Sampedro}, L. and {Cristobal-Hornillos}, D. and {Varela}, J. and {Ederoclite}, A. and {Chies-Santos}, A.~L. and {Schoenell}, W. and {Ribeiro}, T. and {Mar{\'\i}n-Franch}, A. and {L{\'o}pez-Sanjuan}, C. and {Hern{\'a}ndez-Fern{\'a}ndez}, J.~D. and {Cortesi}, A. and {V{\'a}zquez Rami{\'o}}, H. and {Santos}, W. and {Cibirka}, N. and {Novais}, P. and {Pereira}, E. and {Hern{\'a}ndez-Jimenez}, J.~A. and {Jimenez-Teja}, Y. and {Moles}, M. and {Ben{\'\i}tez}, N. and {Dupke}, R.},
        title = "{J-PLUS: On the identification of new cluster members in the double galaxy cluster A2589 and A2593 using PDFs}",
      journal = {\aap},
         year = 2019,
        month = feb,
       volume = {622},
          eid = {A178},
        pages = {A178},
          doi = {10.1051/0004-6361/201731348},
archivePrefix = {arXiv},
       eprint = {1804.03640},
 primaryClass = {astro-ph.CO},
       adsurl = {https://ui.adsabs.harvard.edu/abs/2019A&A...622A.178M}
}

@ARTICLE{Mostert2023,
       author = {{Mostert}, Rafa{\"e}l I.~J. and {Morganti}, Raffaella and {Brienza}, Marisa and {Duncan}, Kenneth J. and {Oei}, Martijn S.~S.~L. and {R{\"o}ttgering}, Huub J.~A. and {Alegre}, Lara and {Hardcastle}, Martin J. and {Jurlin}, Nika},
        title = "{Finding AGN remnant candidates based on radio morphology with machine learning}",
      journal = {\aap},
         year = 2023,
        month = jun,
       volume = {674},
          eid = {A208},
        pages = {A208},
          doi = {10.1051/0004-6361/202346035},
archivePrefix = {arXiv},
       eprint = {2304.05813},
 primaryClass = {astro-ph.GA},
       adsurl = {https://ui.adsabs.harvard.edu/abs/2023A&A...674A.208M}
}

@ARTICLE{Rahna2025,
       author = {{Rahna}, P.~T. and {Akhlaghi}, M. and {L{\'o}pez-Sanjuan}, C. and {Logro{\~n}o-Garc{\'\i}a}, R. and {Muniesa}, D.~J. and {Dom{\'\i}nguez-S{\'a}nchez}, H. and {Fern{\'a}ndez-Ontiveros}, J.~A. and {Sobral}, David and {Lumbreras-Calle}, A. and {Chies-Santos}, A.~L. and {Rodr{\'\i}guez-Mart{\'\i}n}, J.~E. and {Eskandarlou}, S. and {Ederoclite}, A. and {Alvarez-Candal}, A. and {V{\'a}zquez Rami{\'o}}, H. and {Cenarro}, A.~J. and {Mar{\'\i}n-Franch}, A. and {Alcaniz}, J. and {Angulo}, R.~E. and {Crist{\'o}bal-Hornillos}, D. and {Dupke}, R.~A. and {Hern{\'a}ndez-Monteagudo}, C. and {Moles}, M. and {Sodr{\'e}}, Jr., L. and {Varela}, J.},
        title = "{J-PLUS: Spectroscopic validation of H$\alpha$ emission line maps in spatially resolved galaxies}",
      journal = {arXiv e-prints},
         year = 2025,
        month = feb,
          eid = {arXiv:2502.05830},
        pages = {arXiv:2502.05830},
          doi = {10.48550/arXiv.2502.05830},
archivePrefix = {arXiv},
       eprint = {2502.05830},
 primaryClass = {astro-ph.GA},
       adsurl = {https://ui.adsabs.harvard.edu/abs/2025arXiv250205830R}
}

@ARTICLE{Rees1990,
       author = {{Rees}, Nick},
        title = "{A deep 38-MHz radio survey of the area delta >+60.}",
      journal = {\mnras},
         year = 1990,
        month = may,
       volume = {244},
        pages = {233-246},
       adsurl = {https://ui.adsabs.harvard.edu/abs/1990MNRAS.244..233R}
}
\bibliographystyle{aasjournalv7}

\section*{Catalog parameters}\label{app:catalog}

\startlongtable
\begin{deluxetable}{lllll}\label{tab:statistics}
\tablecaption{Provided parameters in the catalog with their units and a short description.}
\centering
\setlength{\tabcolsep}{1.pt}
\tablehead{   \text{Parameter name}       & & \text{Description} &  & \text{Units}}
\startdata
   \texttt{TILE\_ID}  & & \text{Tile id of the reference $r$ band image of the object in J-PLUS.} &  & \\
   \texttt{NUMBER}  & & \text{Number id assigned by \texttt{SExtractor} to the object in J-PLUS.} & & \\
   \texttt{Source\_name} & & \text{Object Identifier (ILT name) in LoTSS.} & & \\
   \texttt{R.A.} & & \text{Radio right ascension from LoTSS.} & & \text{deg} \\
   \texttt{Decl.} & & \text{Radio declination from LoTSS.} & & \text{deg} \\
   \texttt{Opt\_RA} & & \text{Right ascension of the optical/IR counterpart from WISE/SDSS.} & & \text{deg} \\
   \texttt{Opt\_DEC} & & \text{Declination of the optical/IR counterpart from WISE/SDSS.} & & \text{deg} \\
    \texttt{ang\_Dist} & & \text{Angular distance between the optical/IR counterpart and the J-PLUS matching object.} & & \text{arcsec} \\  
   \texttt{Total\_flux} & & \text{144-MHz total flux density from LoTSS.} & & \text{mJy} \\
   \texttt{E\_Total\_flux} & & \text{Error on total flux density from LoTSS.} & & \text{mJy} \\
    \texttt{LAS} & & \text{Radio angular size estimate from LoTSS.} & & \text{arcsec} \\
   \texttt{Resolved} & & \text{Resolved flag from LoTSS.} & & \\ \texttt{CLASS\_GALAXY\_mean} & & \text{Mean probability of being a galaxy by BANNJOS.} & & \\
   \texttt{CLASS\_QSO\_mean} & & \text{Mean probability of being a QSO by BANNJOS.} & & \\
   \texttt{CLASS\_STAR\_mean} & & \text{Mean probability of being a star by BANNJOS.} & & \\
   \texttt{CLASS\_GALAXY\_std} & & \text{Standard deviation of probability of being a galaxy by BANNJOS.} & & \\
   \texttt{CLASS\_QSO\_std} & & \text{Standard deviation of probability of being a QSO by BANNJOS.} & & \\
   \texttt{CLASS\_STAR\_std} & & \text{Standard deviation of probability of being a star by BANNJOS.} & & \\
   \texttt{LoTSS\_zphot} & & \text{Photo-z estimate from LoTSS.} & & \\
   \texttt{LoTSS\_zphot\_err} & & \text{Photo-z error from LoTSS.} & & \\
   \texttt{flag\_qual} & & \text{Predicted photo-z quality of LoTSS, 0 if bad, 1 if good.} & &  \\
   \texttt{JPLUS\_zphot} & & \text{Photo-z estimate from J-PLUS.} & & \\
    \texttt{JPLUS\_zphot\_err} & & \text{Photo-z error from J-PLUS.} & & \\
    \texttt{ODDS} & & \text{Integral of P(z) within \texttt{JPLUS\_zphot} ± 0.03*(1+\texttt{JPLUS\_zphot}).} & & \\
    \texttt{Combined\_zphot} & & \text{Combined photo-z from J-PLUS and LoTSS.} & & \\\texttt{Combined\_zphot\_err} & & \text{Combined photo-z error.} & & \\
    \texttt{zspec} & & \text{Spectroscopic redshift from SDSS, DESI} or HETDEX. & & \\
    \texttt{zspec\_err} & & \text{Spectroscopic redshift error when available}. & & \\
    \texttt{z\_best} & & \text{Best available redshift.} & & \\
    \texttt{z\_best\_err} & & \text{Best available redshift error.} & & \\
    \texttt{z\_source} & & \text{Source of the best available redshift.} & & \\
    \texttt{Size} & & \text{Radio linear size obtained from \text{LAS} using \texttt{z\_best} for the cosmology.} & & \text{kpc} \\
    \texttt{L\_144} & & \text{Radio luminosity obtained from \texttt{Total\_flux} using \texttt{z\_best} for the cosmology.} & & \text{W/Hz} \\
    \texttt{Radio\_loudness} & & \text{Radio-to-IR ratio as $\text{log} \, \left( \frac{f_{144}}{f_{W2}} \right)$} & & dex  \\
    \texttt{FLUX\_AUTO\_Lambda} & & \text{\texttt{AUTO} flux densities for the 12 J-PLUS bands.} & & $\frac{10^{-19}\text{erg}}{\text{s}\,\text{cm}^{2} \text{\AA}}$ \\
    \texttt{FLUX\_RELERR\_AUTO\_Lambda} & & \text{Relative errors of the \texttt{AUTO} flux density values for the 12 J-PLUS bands.} & &  \\
    \texttt{FLAGS} & & \text{Sextractor flags for the 12 J-PLUS bands.} & &  \\
    \texttt{MASK\_FLAGS} & & \text{Mask flags for the 12 J-PLUS bands.} & &  \\
    \texttt{mag\_w1} & & \text{Magnitude in WISE band 1 (taken from LoTSS).} & & mag  \\
    \texttt{magerr\_w1} & & \text{Magnitude error in WISE band 1, or blank for upper limit.} & & mag  \\
    \texttt{mag\_w2} & & \text{Magnitude in WISE band 2 (taken from LoTSS).} & & mag  \\
    \texttt{magerr\_w2} & & \text{Magnitude error in WISE band 2, or blank for upper limit.} & & mag  \\
    \texttt{mag\_w3} & & \text{Magnitude in WISE band 3 (taken from LoTSS).} & & mag  \\
    \texttt{magerr\_w3} & & \text{Magnitude error in WISE band 3, or blank for upper limit.} & & mag  \\
    \texttt{mag\_w4} & & \text{Magnitude in WISE band 4 (taken from LoTSS).} & & mag  \\
    \texttt{magerr\_w4} & & \text{Magnitude error in WISE band 4, or blank for upper limit.} & & mag  \\
    \texttt{Stellar\_mass} & & \text{Stellar mass from CIGALE for \texttt{z\_best}.} & & $M_{\odot}$ \\
    \texttt{Stellar\_mass\_err} & & \text{Stellar mass error from CIGALE for \texttt{z\_best}.} & & $M_{\odot}$ \\
    \texttt{SFR} & & \text{SFR from CIGALE for \texttt{z\_best}.} & & $M_{\odot}$/yr \\
    \texttt{SFR\_err} & & \text{SFR\_err from CIGALE for \texttt{z\_best}.} & & $M_{\odot}$/yr \\
    \texttt{SFR10Myrs} & & \text{SFR10Myrs from CIGALE for \texttt{z\_best}.} & & $M_{\odot}$/yr \\
    \texttt{SFR10Myrs\_err} & & \text{SFR10Myrs\_err from CIGALE for \texttt{z\_best}.} & & $M_{\odot}$/yr \\
    \texttt{SFR100Myrs} & & \text{SFR100Myrs from CIGALE for \texttt{z\_best}.} & & $M_{\odot}$/yr \\
    \texttt{SFR100Myrs\_err} & & \text{SFR100Myrs\_err from CIGALE for \texttt{z\_best}.} & & $M_{\odot}$/yr \\
    \texttt{close\_objects} & & \text{Flag indicating an object closer than 5 arcsec.} & & \\
    & & \text{0 if not present, 1 if present.} & & \\
    \texttt{incompatible\_spec\_class} & & \text{Flag indicating a SDSS/DESI spectroscopic class different from the BANNJOS classification.} & & \\
    & & \text{0 if no, 1 if yes.} & & \\
    \texttt{unreliable\_spec\_z} & & \text{Flag indicating an unreliable SDSS/DESI spectroscopic redshift.} & & \\
    & & \text{0 if no, 1 if yes.} & & \\
    \texttt{candidate\_radio\_star} & & \text{Flag indicating a potential radio star candidate, with values `reliable' or `unreliable'} & & \\
    \texttt{ambiguous\_classification} & & \text{Flag indicating an object with several BANNJOS classifications,} & & \\
    & & \text{mostly due to several detections in GAIA. 0 if not, 1 if yes.} & & \\
    \enddata
\end{deluxetable}

\end{document}